\documentclass[a4paper,11pt]{article}
\pdfoutput=1 

\usepackage[utf8]{inputenc}

\usepackage[no-natbib-sort]{jheppub}
\usepackage{enumitem}

\usepackage{youngtab}

\usepackage{hyperref}
\hypersetup{
    pdftitle={Holographic Bubbles with Jecco: Expanding, Collapsing and Critical},
  }

\usepackage{fancybox}
\usepackage{float}
\usepackage{amsfonts}
\usepackage{amstext}
\usepackage{amsmath}
\usepackage{breqn}
\usepackage{amsbsy}
\usepackage{graphicx}
\usepackage{amssymb}
\usepackage{amsthm}
\usepackage{bm}
\usepackage{comment}
\usepackage{epsfig}
\usepackage{latexsym}
\usepackage{pdflscape}
\usepackage{cancel} 
\usepackage{color}
\usepackage{xcolor}
\usepackage{bm,nicefrac}

\makeatletter
\def\@fpheader{\relax}
\makeatother

\DeclareSymbolFont{AMSa}{U}{msa}{m}{n}
\DeclareSymbolFont{AMSb}{U}{msb}{m}{n}
\DeclareMathSymbol{\fieldR}{\mathalpha}{AMSb}{"52}

\newcommand{\beq}{\begin{eqnarray}}
\newcommand{\eeq}{\end{eqnarray}}
\newcommand{\bea}{\begin{eqnarray}}
\newcommand{\eea}{\end{eqnarray}}
\newcommand{\be}{\begin{equation}}
\newcommand{\ee}{\end{equation}}
\newcommand{\bq}{\begin{equation}}
\newcommand{\eq}{\end{equation}}

\def\6{\partial}

\newcommand{\vwall}{v_{\mathrm{wall}}}

\def\6{\partial}

\usepackage{color}
    \definecolor{darkgreen}{rgb}{0,0.5,0}
    \definecolor{darkblue}{rgb}{0,0,0.6}
    \definecolor{purple}{rgb}{0.4,.2,0.7}

\newcommand{\fig}[1]{Fig.~\ref{#1}}

\newcommand{\nc}{N}

\newcommand{\Sec}[1]{Sec.~\ref{#1}}

\def\E{{\mathcal{E}}}
\def\P{{\mathcal{P}}}
\newcommand{\eqn}[1]{(\ref{#1})}

\newcommand{\Ts}{T_s}

\newcommand{\Es}{\mathcal{E}_s}

\newcommand{\Ehigh}{\mathcal{E}_\mathrm{high}}
\newcommand{\Elow}{\mathcal{E}_\mathrm{low}}
\newcommand{\Ebar}{\bar{\mathcal{E}}}

\newcommand{\sac}{\, , \qquad}

\title{Holographic Bubbles with \texttt{Jecco}: Expanding, Collapsing and Critical}

\author[a,b]{Yago~Bea,}
\author[b]{Jorge~Casalderrey-Solana,}
\author[c]{Thanasis~Giannakopoulos,}
\author[b]{Aron~Jansen,}
\author[b,d]{David~Mateos,}
\author[b]{Mikel~Sanchez-Garitaonandia}
\author[e,c]{and Miguel~Zilh\~ao}

\affiliation[a]{Department of Physics and Helsinki Institute of Physics, PL 64, FI-00014 University of Helsinki, Finland.}
\affiliation[b]{Departament de F\'\i sica Qu\`antica i Astrof\'\i sica and Institut de Ci\`encies del Cosmos (ICC),  Universitat de Barcelona, Mart\'\i\  i Franqu\`es 1, ES-08028, Barcelona, Spain.}
\affiliation[c]{Centro de Astrof\'{\i}sica e Gravita\c c\~ao -- CENTRA,
  Departamento de F\'{\i}sica, Instituto Superior T\'ecnico -- IST, Universidade
  de Lisboa -- UL, Av.\ Rovisco Pais 1, 1049-001 Lisboa, Portugal }
\affiliation[d]{Instituci\'o Catalana de Recerca i Estudis Avan\c cats (ICREA), Passeig Llu\'\i s Companys 23,  ES-08010, Barcelona, Spain.}
\affiliation[e]{Departamento de Matem\'atica da Universidade de Aveiro and
Centre for Research and Development in Mathematics and Applications (CIDMA),
Campus de Santiago,
3810-183 Aveiro, Portugal}

\emailAdd{yago.beabesada@helsinki.fi}
\emailAdd{jorge.casalderrey@ub.edu}
\emailAdd{athanasios.giannakopoulos@tecnico.ulisboa.pt}
\emailAdd{a.p.jansen@icc.ub.edu}
\emailAdd{dmateos@fqa.ub.edu}
\emailAdd{mikeliccub@icc.ub.edu}
\emailAdd{mzilhao@ua.pt}

\abstract{
Cosmological phase transitions can proceed via the nucleation of bubbles that subsequently expand and collide. The resulting gravitational wave spectrum depends crucially on the properties of these bubbles. We extend our previous holographic work on planar bubbles to cylindrical bubbles in 
a strongly-coupled, non-Abelian, four-dimensional gauge theory. This extension brings about two new physical properties. First, the existence of a critical bubble, which we determine. Second, the bubble profile at late times exhibits a richer self-similar structure, which we verify. These results require 
a new 3+1 evolution code called \texttt{Jecco} that solves the Einstein equations in the characteristic formulation in asymptotically AdS spaces. \texttt{Jecco} is written in the Julia programming language and is freely available. We present an outline of the code and the tests performed to assess its robustness and performance.
}

\begin{document} 
\maketitle
\flushbottom

\section{Introduction}
\label{intro}
A first-order, thermal  phase transition in the Early Universe would produce gravitational waves that could be detected in current or future experiments. 
Since the Standard Model of particle physics possesses no first-order transitions \cite{Aoki:2006we,Kajantie:1996mn,Laine:1998vn,Rummukainen:1998as}, the discovery of gravitational waves originating from a cosmological phase transition would amount to the discovery of new physics beyond the Standard Model. The transition  may proceed via bubble nucleation (see e.g.~\cite{Hindmarsh:2020hop} for a review) or via the spinodal instability \cite{Bea:2021zol}. In this paper we will focus on the first case.

Maximising the discovery potential requires an accurate understanding of the bubble properties. These range from the action of the critical bubble that gets nucleated to the terminal velocity of expanding bubbles. The former controls the nucleation rate, whereas the latter controls the characteristic frequency of the produced gravitational waves. Computing these parameters from first principles is challenging even in weakly coupled theories. The former requires knowledge of the  effective potential at finite temperature \cite{Laine:2016hma,Gould:2021ccf}, whereas the latter requires an understanding of out-of-equilibrium physics \cite{Moore:1995ua,Bodeker:2017cim,Hoche:2020ysm}.

In \cite{Bea:2021zsu} we performed the first  holographic calculation of the bubble wall velocity in a strongly-coupled, non-Abelian, four-dimensional gauge theory. 
Because of technical limitations, in this reference we focused on planar bubbles, namely we imposed translational invariance along two of the spatial directions,  in such a way that the dynamics was effectively 1+1 dimensional in the gauge theory and 2+1 dimensional on the gravity side. In this paper we will extend our analysis by imposing translational invariance along only one of the spatial directions. Thus, the effective dynamics will be  2+1 dimensional in the gauge theory and 3+1 dimensional on the gravity side. This will allow us to study  bubbles that have the topology of a cylinder. Since we impose translation invariance along the axis of the cylinder, we will only plot the dependence of physical quantities on the two spatial directions transverse to this axis. We emphasize that we will not impose any  symmetries on these directions,  meaning that the dynamics on the plane transverse to the cylinder axis will be completely general.

The extension from planar to cylindrical bubbles brings about two new physical aspects. The first one is that the surface tension now plays a role. In particular, we will be able to identify a critical bubble in which the inward-pointing force due to the surface tension exactly balances the outward-pointing force coming from the pressure difference between the inside and the outside of the bubble. The second one is that the asymptotic, self-similar profile of an expanding bubble possesses a richer structure than in the planar case. We will verify this by plotting our holographic result for the gauge theory stress tensor at late times as a function of the appropriate scaling variable. We will also compare the holographic result with the hydrodynamic approximation. As expected, we will find that hydrodynamics  provides a good approximation everywhere except at the bubble wall. 

To obtain these results we have developed a new 3+1 evolution code called \texttt{Jecco} that solves the Einstein equations in the characteristic formulation in asymptotically anti de Sitter (AdS)  spaces. The characteristic approach to solving Einstein's equations has a long history.
It dates back to the Bondi-Sachs formalism~\cite{Bondi:1960jsa,Sachs:1962wk},
crucial to the modern understanding of gravitational waves.
For numerical applications, these formulations provide advantages over more
standard spacelike foliations in a number of situations. In the context of
extracting gravitational-wave information, for instance, this approach exploits
the fact that null hypersurfaces reach future null infinity, thereby 
avoiding systematic errors from extrapolation techniques.
Further advantages of such formulations include: the initial data are
free (i.e.~there is no need to solve elliptic equations for the
initial data); no second time derivatives (resulting in fewer
evolution variables); the field equations are conveniently cast as a
set of nested ordinary differential equations (ODEs) which can be
efficiently solved.

Though questions remain about the well-posedness
of these formulations~\cite{Giannakopoulos:2020dih,Giannakopoulos:2021pnh}, characteristic codes have shown
remarkable stability. 
Indeed, the
first ever long-term stable evolutions of moving black holes was
accomplished with a characteristic scheme~\cite{Gomez:1997pd}.
Applications of this approach include the
Cauchy-Characteristic extraction method for the computation of
gravitational waveforms at future null infinity, which
has been
numerically implemented in~\cite{doi:10.1063/1.525904, Bishop:1996gt, Bishop:1997ik,Gomez:1996ge,Reisswig:2009rx,Handmer:2014qha}.
There is an extensive literature on this and related subjects --- see Winicour's
Living Review~\cite{Winicour:2012znc} for an overview.

Despite all the successes and advantages of this approach, one serious
drawback that it faces is the possible formation of caustics, which
typically spoil the numerical simulation. This is particularly severe
when evolving binary black holes, and for this reason the
characteristic approach in solving Einstein's equations lost some
ground in favour of more traditional Cauchy evolution schemes.
More recently, though, the characteristic approach has shown to be particularly
well-adapted for evolutions in the Poincaré patch of AdS
spaces. Crucially for these simulations is the presence of a (non-compact)
planar horizon embedded in the asymptotically AdS space, effectively acting as
an infrared cut-off, which removes caustic formation from the computational
domain.

Through  holography, this approach has facilitated the study of
far-from-equili\-brium dynamics of strongly-coupled gauge theories,
allowing for studies of 
isotropization~\cite{Chesler:2008hg,Heller:2013oxa,Gursoy:2016tgf}, collisions of
gravitational shockwaves (used as models for heavy-ion
collisions)~\cite{Chesler:2010bi,Casalderrey-Solana:2013sxa,Chesler:2015wra},
momentum relaxation~\cite{Balasubramanian:2013yqa},
turbulence~\cite{Adams:2013vsa}, collisions in non-conformal
theories~\cite{Attems:2016tby,Attems:2017zam}, phase transitions and
dynamics of phase
separation~\cite{Attems:2017ezz,Janik:2017ykj,Attems:2019yqn,Bellantuono:2019wbn,Bea:2020ees,Janik:2021jbq,Bea:2021ieq}, 
collisions in theories with phase transitions~\cite{Attems:2018gou}, dynamical instabilities~\cite{Gursoy:2016ggq},
and even applications to gravitational-wave physics~\cite{Ahmadvand:2017xrw,Ahmadvand:2017tue,Bigazzi:2020avc,Ares:2020lbt,Ares:2021nap}
and bubble dynamics~\cite{Bigazzi:2020phm,Bea:2021zsu,Bigazzi:2021ucw,Ares:2021ntv}.
See~\cite{Chesler:2013lia} for more references and a comprehensive overview of
the techniques involved, and
also~\cite{Bantilan:2012vu,Bantilan:2020pay,Bantilan:2020xas} for equivalent
approaches using Cauchy evolutions.

Here we present a new 3+1 code called \texttt{Jecco} (Julia Einstein
Characteristic Code) that solves Einstein's equations in the characteristic
formulation in asymptotically AdS spaces.
\texttt{Jecco} is written in the Julia programming language and comes with
several tools (such as arbitrary-order finite-difference operators as well as
Chebyshev and Fourier differentiation matrices) useful for generic numerical
evolutions. The evolution part of the code would allow for the study of any of
the problems mentioned in the previous paragraph; herein, as mentioned in the
beginning, we will focus on the study of bubble dynamics.
The code is publicly available and can be obtained from
github~\url{https://github.com/mzilhao/Jecco.jl} and Zenodo~\cite{jecco-2022}.
To the best of our knowledge, this is the first such freely available code (see
however the \texttt{PittNull} code~\cite{Bishop:1998uk,Babiuc:2010ze}  for characteristic
evolutions in asymptotically flat spaces, freely available and distributed as
part of the Einstein Toolkit~\cite{EinsteinToolkit:2020_05}).

This paper is organized as follows. In \Sec{sec:model} we introduce the class of models to which our code can be applied, as well as the corresponding equations of motion. In \Sec{sec:code} we discuss the
implementation of these equations in the code and the numerical methods that we use. In \Sec{bubbles} we discuss our new results for cylindrical bubbles. In  \Sec{sec:final} we conclude with some final remarks. The tests of our code are collected in Appendix~\ref{sec:results}. We use $G=c=\hbar=1$ units throughout.

\section{\texttt{Jecco}: a new characteristic code for numerical holography}
\subsection{Equations}
\label{sec:model}

In this section we outline the theoretical background and equations that are
implemented in \texttt{Jecco}. Our approach is similar to that
of~\cite{Chesler:2013lia} and generalises the code presented
in~\cite{Attems:2017zam} to the 3+1 dimensional case. See
also~\cite{Winicour:2012znc} for an overview of the approaches and codes used in
the asymptotically flat setting.

\subsubsection{Equations of motion and characteristic formulation}
We consider a five-dimensional action consisting of gravity coupled to a scalar
field $\phi$ with a non-trivial potential $V(\phi)$. The action for this Einstein-scalar model is
\begin{equation}
  S = \frac{2}{\kappa_5^2}\int d^5x\sqrt{-g}\left[\frac{1}{4}R-\frac{1}{2}\left(\partial\phi\right)^2-V(\phi)\right],
  \label{eqn:Einstein-scalar_model}
\end{equation}
where
$\kappa_5^2$
is the 5D gravitational coupling constant, which in our units takes the value $\kappa_5^2=8\pi$. The resulting dynamical equations of motion read
\begin{equation}
\begin{aligned}
E_{\mu\nu}& \equiv R_{\mu\nu}-\frac{R}{2}g_{\mu\nu}-8\pi T_{\mu\nu}=0,\\
\Phi & \equiv \square\phi -\partial_{\phi}V(\phi)=0,
\label{eq:eom}
\end{aligned}
\end{equation}
where
\[
8\pi T_{\mu\nu} = 2\partial_{\mu}\phi\partial_{\nu}\phi-g_{\mu\nu}\left(g^{\alpha\beta}\partial_{\alpha}\phi\partial_{\beta}\phi+2V(\phi)\right).
\]
Our potential $V(\phi)$ comes from a superpotential $W(\phi)$ with the form
\be
 L W(\phi) = -\frac{3}{2}-\frac{\phi ^2}{2}+ \lambda_4 \, \phi^4
  +\lambda_6 \, \phi^6 \,,
  \label{WW}
\ee
and its explicit expression can be derived via
\[
  V = -\frac{4}{3} W^2 + \frac{1}{2} W'{}^2 ,
\]
resulting in 
\begin{align}
%
L^2 V(\phi) &= -3-\frac{3}{2}\phi^2-\frac{1}{3}\phi^4
          + \left(\frac{4\lambda_4}{3} + 8\lambda_4^2 - 2\lambda_6\right)\phi^6
          +\left(-\frac{4\lambda_4^2}{3} + \frac{4}{3}\lambda_6 + 24\lambda_4 \lambda_6\right)\phi^8 \notag\\
         &\quad{}+\left( 18\lambda_6^2 - \frac{8}{3} \lambda_4 \lambda_6 \right)\phi^{10}-\frac{4}{3}\lambda_6^2 \, \phi^{12}
           \,. \label{eq:potential}
\end{align}
In these equations $\lambda_4$ and $\lambda_6$ are freely specifiable dimensionless parameters related to the parameters $\phi_M$ and $\phi_Q$ used in e.g.~\cite{Bea:2018whf,Bea:2020ees} through
\be
\label{param}
\lambda_4 = - \frac{1}{4\phi_M^2}  \sac \lambda_6 = \frac{1}{\phi_Q}
 \,.
\ee
This potential has a maximum at $\phi=0$, where it admits an exact AdS solution of radius $L$. For numerical purposes we set $L=1$. The holographic dual field theory corresponds to a 3+1 dimensional conformal field theory which is deformed by a source $\Lambda$ for the dimension-three scalar operator $\mathcal{O}_{\phi}$ dual to the scalar field $\phi$.
The thermodynamical and near-equilibrium properties of this model were presented
in~\cite{Attems:2016ugt,Attems:2016tby,Attems:2017ezz} for $\lambda_6 = 0$ and in~\cite{Bea:2018whf,Bea:2020ees} for  $\lambda_6 \neq 0$.

Let us point out that even if here we will always make use of the particular potential~(\ref{eq:potential}), the code implementation is such that more generic potentials can be used
provided that, for low values of the scalar field, they behave as
\begin{equation}
L^2 V(\phi) = -3-\frac{3}{2}\phi^2-\frac{\phi^4}{3}+\mathcal{O}\left(\phi^6\right).
\label{ec:pot_near_bdry}
\end{equation}
The constant term is fixed by the 4+1 dimensional AdS asymptotics and the quadratic one is in correspondence with the scaling dimension of the dual scalar operator $\mathcal{O}_{\phi}$. The quartic term, determined by the other two in our case, ensures the absence of a conformal anomaly, which would give rise to logarithms in the asymptotic expansions. Thence, a change in this near boundary behaviour of the potential would alter the hard-coded asymptotic expansions and variable redefinitions to be introduced in Secs.~\ref{sec:asympt-expansion} and~\ref{sec:redef}.




We now write the following 5D ansatz for the metric in
Eddington-Finkelstein (EF) coordinates
\begin{equation}
\begin{aligned}
  ds^2= g_{\mu\nu} dx^{\mu} dx^{\nu} & = -Adt^2+2dt\left(dr+F_xdx+F_ydy\right) 
   +S^2\Big[e^{-B_1-B_2}\cosh(G)dx^2 \\[2mm]
  & \quad { }+e^{B_1-B_2}\cosh(G)dy^2+2e^{-B_2}\sinh(G)dxdy+e^{2 B_2}dz^2\Big],
  \label{eq:metric}
\end{aligned}
\end{equation}
where all functions depend on the radial coordinate $r$, time $t$ and transverse
directions $x$ and $y$. Nothing depends on the coordinate $z$, so this is
effectively a 3+1 system. Physically, this means that in the gauge theory we impose translation invariance along the $z$-direction together with $z\rightarrow -z$ symmetry. Along the remaining $(t,x,y)$-directions general dynamics is permitted. Note that we denote by $t$ the (ingoing) null bulk coordinate usually labeled $v$ in EF coordinates. At the boundary, $t$ becomes the usual time coordinate. The spatial part of the metric is written such
that $S$ encodes the area of constant $t$ and $r$ slices,
\[
  \sqrt{g|_{dt,dr = 0}}=S^3.
\]
We can recover the 2+1 system of~\cite{Attems:2017zam} by setting
\begin{equation}
\begin{aligned}
&F_y =G=0, & B_1 & =\frac{3}{2}B,   & B_2 & = \frac{1}{2}B, \qquad \mathrm{or}
\\[2mm]
&F_x=G=0,  & B_1 & =-\frac{3}{2}B,  & B_2 & = \frac{1}{2}B,
\end{aligned}
\label{ec:to_2+1}
\end{equation}
for non-trivial dependence only along the $x$ or $y$ direction respectively.
The metric~\eqref{eq:metric} is invariant under
\begin{equation}
\begin{aligned}
r & \rightarrow \bar{r}=r+\xi(t,x,y) \,, \\
S & \rightarrow\bar{S}=S \,, \\
B_1 & \rightarrow\bar{B}_1=B_1 \,, \\
B_2 & \rightarrow\bar{B}_2=B_2 \,, \\
A & \rightarrow\bar{A}=A+2\partial_t\xi(t,x,y) \,, \\
F_x & \rightarrow\bar{F}_x=F_x-\partial_x\xi(t,x,y) \,,\\
F_y & \rightarrow\bar{F}_y=F_y-\partial_y\xi(t,x,y) \,.
\end{aligned}
\label{ec:gauge_trafo}
\end{equation}

Plugging the ansatz~(\ref{eq:metric}) into~\eqref{eq:eom} results in a nested
system of ODEs in the radial (holographic) direction $r$ at each constant $t$
that can be solved sequentially. 
We illustrate this system in Table~\ref{tab:system}.
Each row in the table represents an equation, obtained from the particular combination of the equations of motion (\ref{eq:eom}) as indicated, that takes the form
\begin{equation}
  \left[A_f(t,u,x,y)\, \partial^2_u + B_f(t,u,x,y)\, \partial_u + C_f(t,u,x,y) \right] f(t,u,x,y) = -S_f(t,u,x,y),
  \label{eq:ODEs}
\end{equation}
where $u \equiv 1/r$, $f$ is the corresponding function to be solved for and the coefficients $A_f$, $B_f$, $C_f$ and $S_f$ are fully determined once the preceding equations have been solved.
Dotted functions denote an operation defined as
\begin{equation}
\dot{f} \equiv\left(\partial_t+\frac{A}{2}\partial_r\right)f,
\end{equation}
which are necessary to obtain this nested structure.

\begin{table}[t]
  \begin{center}
    \caption{Nested structure of the equations of motion.}
    \label{tab:system}
\begin{tabular}{c | c}
Function(s) & Combination\\
  \hline\hline
  \small
$S$ & $E_{rr}$\\
\hline
$F_x$ & $E_{rx}-g_{tx}E_{rr}$\\
$F_y$ & $E_{ry}-g_{ty}E_{rr}$\\
\hline
$\dot{S}$ & 	$E_{tr}-\frac{1}{2}g_{	tt}E_{rr}$\\
\hline
$\dot{\phi}$ & $\Phi$\\
\hline
$A$ & $\frac{E_{zz}}{g_{zz}}+\left(g^{ry}g_{ty}+g^{rx}g_{tx}\right)E_{rr}+2g^{rx}\left(E_{rx}-g_{tx}E_{rr}\right)+2g^{ry}\left(E_{ry}-g_{ty}E_{rr}\right)$\\
&$-4\left(E_{tr}-\frac{1}{2}g_{tt}E_{rr}\right)+2\frac{E_{xy}}{g_{xy}}+g_{xx}g^{xx}\left(\frac{E_{yy}}{g_{yy}}+\frac{E_{xx}}{g_{xx}}-2\frac{E_{xy}}{g_{xy}}\right)$\\
\hline
$\dot{B_2}$ & $E_{zz}$\\
\hline
 $\dot{G}$ & $E_{xy}$\\
 $\dot{B_1}$ & $E_{yy}$\\
 \hline
 $\ddot{S}$ & $E_{tt}-\frac{1}{2}g_{tt}E_{tr}-\frac{1}{2}g_{tt}\left(E_{tr}-\frac{1}{2}g_{tt}E_{rr}\right)$\\
 \hline
 $\dot{F_x}$&$E_{tx}-\frac{1}{2}g_{tt}E_{rx}-g_{tx}\left(E_{tr}-\frac{1}{2}g_{tt}E_{rr}\right)$\\
 $\dot{F_y}$&$E_{ty}-\frac{1}{2}g_{tt}E_{ry}-g_{ty}\left(E_{tr}-\frac{1}{2}g_{tt}E_{rr}\right)$\\
 \hline
\end{tabular}
\end{center}
\end{table}

There are three sets of (two) coupled equations, indicated in the table by the absence of a separating line.
These still take the form of (\ref{eq:ODEs}), but now $f$ should be thought of as a vector of the two functions involved, as is the source term $S_f$, while $A_f$, $B_f$ and $C_f$ become $2 \times 2$ matrices.
The equations themselves are lengthy and given in (\ref{ec:nested_first}-\ref{ec:nested_last}).

These equations need to be supplemented with boundary conditions specified at the AdS boundary $u \equiv 1/r = 0$, see \Sec{sec:redef}.
In addition, the functions $B_1(t_0,u,x,y)$,
$B_2(t_0,u,x,y)$, $G(t_0,u,x,y)$ and $\phi(t_0,u,x,y)$ should be thought of as initial data which, unlike for Cauchy-based approaches of solving Einstein's equations,  can be freely specified provided they are consistent with AdS asymptotics.



\subsubsection{Asymptotic expansions}
\label{sec:asympt-expansion}

The study of the near-boundary behaviour ($u\rightarrow 0$) of the functions is relevant for two reasons. The first one is that, as usually for asymptotically AdS (AAdS) spacetimes, some metric components diverge as one approaches the boundary, and their expansion in powers of $u$ is useful to redefine the variables in terms of new, finite ones. The second reason is that it allows us
to understand which boundary conditions to impose on the ODEs~\eqref{eq:ODEs}.


For this purpose, we start with an ansatz that is compatible with the AAdS condition,\footnote{This ansatz must be modified if \eqref{ec:pot_near_bdry} does not hold}
\begin{equation}
\begin{aligned}
  A(t,u,x,y) & = \frac{1}{u^2}+\sum_{n=-1}^{\infty}a_{(2n)}(t,x,y) u^n  \,, &
      B_1(t,u,x,y) & = \sum_{n=1}^{\infty}b_{1n}(t,x,y) u^n \,, \\[2mm]
  B_2(t,u,x,y) & = \sum_{n=1}^{\infty}b_{2n}(t,x,y) u^n \,, &
       G(t,u,x,y) & = \sum_{n=1}^{\infty}g_n(t,x,y) u^n \,, \\[2mm]
  S(t,u,x,y)& =\frac{1}{u} +\sum_{n=0}^{\infty}s_{n}(t,x,y) u^n \,, &
        F_x(t,u,x,y) & = \sum_{n=0}^{\infty}f_{xn}(t,x,y) u^n \,, \\[2mm]
  F_y(t,u,x,y) & = \sum_{n=0}^{\infty}f_{yn}(t,x,y) u^n \,, &
        \phi(t,u,x,y) & = \sum_{n=1}^{\infty}\phi_{n-1}(t,x,y)u^n \,.
\end{aligned}
\label{ec:AAdS_general}
\end{equation}
Substituting into equations
(\ref{ec:nested_first}-\ref{ec:nested_last}) and solving order by order, we obtain
\begingroup
\allowdisplaybreaks
\begin{subequations}
\begin{align}
A(t,u,x,y)&=\frac{1}{u^2}+\frac{2}{u}\xi+\xi^2-2\partial_t\xi-\frac{2\phi_0^2}{3}+u^2a_4 \notag\\[2mm]
&\quad
{}-\frac{2}{3}u^3\left( 3 \xi a_4 + \partial_xf_{x2}+\partial_yf_{y2}+\phi_0\partial_t\phi_2\right)+\mathcal{O}\left(u^4\right),\\[2mm]
B_1(t,u,x,y)&=u^4b_{14}+\mathcal{O}\left(u^5\right), \label{eq:asympt-B1}\\[2mm]
B_2(t,u,x,y)&=u^4b_{24}+\mathcal{O}\left(u^5\right), \label{eq:asympt-B2}\\[2mm]
G(t,u,x,y)&=u^4g_4+\mathcal{O}\left(u^5\right), \label{eq:asympt-G}\\[2mm]
S(t,u,x,y)&=\frac{1}{u}+\xi-\frac{\phi_0^2}{3}u+\frac{1}{3}\xi\phi_0^2u^2+\frac{1}{54}u^3\left(-18\xi^2\phi_0^2+\phi_0^4-18\phi_0\phi_2\right) \notag\\[2mm]
&\quad{}+\frac{\phi_0}{90}u^4\left(30\xi^3\phi_0-5\xi\phi_0^3+90\xi\phi_2-24\partial_t\phi_2\right)+\mathcal{O}\left(u^5\right),\\[2mm]
F_x(t,u,x,y)&=\partial_x\xi+u^2f_{x2} \notag\\[2mm]
&\quad{}-\frac{2}{15}u^3\left(15 \xi f_{x2} + 6\partial_xb_{14}+6\partial_xb_{24}-\partial_yg_4
-2\phi_0\partial_x\phi_2\right)+\mathcal{O}\left(u^4\right),\\[2mm]
F_y(t,u,x,y)&=\partial_y\xi+u^2f_{y2} \notag\\
&\quad{}-\frac{2}{15}u^3\left(15 \xi f_{y2}-6\partial_yb_{14}
  +6\partial_yb_{24}-\partial_xg_4
-2\phi_0\partial_y\phi_2\right)+\mathcal{O}\left(u^4\right),\\[2mm]
\phi(t,u,x,y)&=\phi_0u-\xi\phi_0u^2+u^3\left(\xi^2\phi_0+\phi_2\right)+
u^4\left(\partial_t\phi_2-3\xi\phi_2-\xi^3\phi_0\right)+\mathcal{O}\left(u^5\right),  \label{eq:asympt-phi}
\end{align}
\label{eq:asympt}
\end{subequations}
\endgroup
where $\phi_2$ is \emph{not} the one in (\ref{ec:AAdS_general}), but redefined as
\begin{equation}
\phi_2(t,x,y)\rightarrow\phi_2(t,x,y)+\xi^2(t,x,y)\phi_0.
\end{equation}
Note that $\phi_0$ is a constant, while the remaining variables in this expansion are functions of~$(t,x,y)$. In reality, the near boundary expansions depend on $s_0$ instead of $\xi$. The fact that the former is simply shifted by $\xi$ under \eqref{ec:gauge_trafo} means that we can identify $s_0$ with $\xi$ and exchange them everywhere.

We also need the expansions of ``dotted'' variables, defined
in~\eqref{ec:new_der}, which take the form
\begin{subequations}
\begin{align}
\dot{B_1}(t,u,x,y)&=-2b_{14} u^3+\mathcal{O}\left(u^4\right),\\[2mm]
\dot{B_2}(t,u,x,y)&=-2b_{24} u^3+\mathcal{O}\left(u^4\right),\\[2mm]
\dot{G}(t,u,x,y)&=-2g_{4} u^3+\mathcal{O}\left(u^4\right),\\[2mm]
\dot{S}(t,u,x,y)&=\frac{1}{2u^2}+\frac{\xi }{u}+\frac{\xi^2 }{2}-\frac{\phi_0^2}{6}
+\frac{1}{36}u^2\left(10a_4 -5\phi_0^4+18\phi_0\phi_2 \right)+\mathcal{O}\left(u^3\right),\\[2mm]
\dot{F_x}(t,u,x,y)&=\partial_t\partial_x\xi -uf_{x2} +\mathcal{O}\left(u^2\right),\\[2mm]
\dot{F_y}(t,u,x,y)&=\partial_t\partial_y\xi -uf_{y2} +\mathcal{O}\left(u^2\right),\\[2mm]
\dot{\phi}(t,u,x,y)&=-\frac{\phi_0}{2}+u^2\left(\frac{\phi_0^3}{3}-\frac{3}{2}\phi_2 \right)+\mathcal{O}\left(u^3\right).
\end{align}
\end{subequations}
The function $\xi(t,x,y)$ encodes our residual gauge freedom, and the functions $a_4(t,x,y)$, $f_{x2}(t,x,y)$, $f_{y2}(t,x,y)$ are further constrained to obey
\begin{subequations}
  \label{ec:boundary_evol}
\begin{align}
\partial_ta_4&=-\frac{4}{3}\left(\partial_xf_{x2}+\partial_yf_{y2}+\phi_0\partial_t\phi_2\right),\\[2mm]
\partial_tf_{x2}&=-\frac{1}{4}\partial_xa_4-\partial_xb_{14}-\partial_xb_{24}+\partial_yg_4+\frac{1}{3}\phi_0\partial_x\phi_2,\\[2mm]
\partial_tf_{y2}&=-\frac{1}{4}\partial_ya_4+\partial_yb_{14}-\partial_yb_{24}+\partial_xg_4+\frac{1}{3}\phi_0\partial_y\phi_2,
\end{align}
\end{subequations}
where $b_{14}(t,x,y)$, $b_{24}(t,x,y)$, $g_{4}(t,x,y)$, $\phi_2(t,x,y)$, and $\partial_t\phi_2(t,x,y)$ are understood to be read off from the asymptotic behaviour of $B_{1}(t,r,x,y)$, $B_{2}(t,r,x,y)$, $G(t,r,x,y)$, and $\phi(t,r,x,y)$ in equations~\eqref{eq:asympt-B1}, \eqref{eq:asympt-B2}, \eqref{eq:asympt-G} and \eqref{eq:asympt-phi}. The functions $a_4(t_0,x,y)$, $f_{x2}(t_0,x,y)$, $f_{y2}(t_0,x,y)$, and $\xi(t_0,x,y)$ should also be thought of as initial data, which can be freely specified. $\phi_0$ is a parameter that must also be specified  and  corresponds to the energy scale $\Lambda$ of the dual boundary theory.

\subsubsection{Field redefinitions and boundary conditions}
\label{sec:redef}

For the numerical implementation we find it useful to split the numerical grid
into two parts: the outer grid region (deep bulk) and the inner grid region
(close to the AdS boundary, where boundary conditions are imposed and the
gauge-theory variables are read off).
As mentioned earlier, some of the metric functions diverge at the AdS boundary while others
vanish, being convenient to make some redefinitions inspired by the asymptotic
behaviour of these functions so that the variables employed in the inner grid 
remain of order unity therein.
For the outer grid we choose to make simpler redefinitions, which is helpful for the equation used to fix the gauge variable $\xi$.
Denoting with the $g1$ ($g2$) subscript the variables defined in the inner (outer) grid, the redefinitions that we choose to make are then
\begingroup
\allowdisplaybreaks
\begin{align*}
A(t,u,x,y)&=\frac{1}{u^2}+\frac{2}{u}\xi(t,x,y)+\xi^2(t,x,y)-2\partial_t\xi(t,x,y)-\frac{2\phi_0^2}{3}+u^2A_{g1}(t,u,x,y)\\[2mm]
&=-2\partial_t\xi(t,x,y)+A_{g2}(t,u,x,y)\\[2mm]
B_1(t,u,x,y) &=u^4B_{1g1}(t,u,x,y)\\[2mm]
&=B_{1g2}(t,u,x,y),\\[2mm]
B_2(t,u,x,y) &=u^4B_{2g1}(t,u,x,y)\\[2mm]
&=B_{2g2}(t,u,x,y),\\[2mm]
G(t,u,x,y) &=u^4G_{g1}(t,u,x,y)\\[2mm]
&=G_{g2}(t,u,x,y),\\[2mm]
S(t,u,x,y)&=\frac{1}{u}+\xi(t,x,y)-\frac{\phi_0^2}{3}u+\frac{1}{3}\xi\phi_0^2u^2+u^3S_{g1}(t,u,x,y)\\[2mm]
&=S_{g2}(t,u,x,y),\\[2mm]
F_x(t,u,x,y)&=\partial_x\xi(t,x,y)+u^2F_{xg1}(t,u,x,y)\\[2mm]
&=\partial_x\xi(t,x,y)+F_{xg2}(t,u,x,y),\\[2mm]
F_y(t,u,x,y)&=\partial_y\xi(t,x,y)+u^2F_{yg1}(t,u,x,y)\\[2mm]
&=\partial_y\xi(t,x,y)+F_{yg2}(t,u,x,y),\\[2mm]
\phi(t,u,x,y)&=\phi_0u-\xi(t,x,y)\phi_0u^2+u^3\phi_0^3\phi_{g1}(t,u,x,y)\\[2mm]
&=\phi_{g2}(t,u,x,y),\\[2mm]
\dot{B_1}(t,u,x,y)&=u^3\dot{B}_{1g1}(t,u,x,y)\\[2mm]
&=\dot{B}_{1g2}(t,u,x,y),\\[2mm]
\dot{B_2}(t,u,x,y)&=u^3\dot{B}_{2g1}(t,u,x,y)\\[2mm]
&=\dot{B}_{2g2}(t,u,x,y),\\[2mm]
\dot{G}(t,u,x,y)&=u^3\dot{G}_{g1}(t,u,x,y)\\[2mm]
&=\dot{G}_{g2}(t,u,x,y),\\[2mm]
\dot{S}(t,u,x,y)&=\frac{1}{2u^2}+\frac{\xi(t,x,y)}{u}+\frac{\xi^2(t,x,y)}{2}-\frac{\phi_0^2}{6}+u^2\dot{S}_{g1}(t,u,x,y)\\[2mm]
&=\dot{S}_{g2}(t,u,x,y),\\[2mm]
\dot{\phi}(t,u,x,y)&=-\frac{\phi_0}{2}+u^2\phi_0^3\dot{\phi}_{g1}(t,u,x,y)\\
&=\dot{\phi}_{g2}(t,u,x,y).
\end{align*}
\endgroup

Substituting these redefined variables into the system of equations
(\ref{ec:nested_first}-\ref{ec:nested_last}), we are left with two new versions
of this system, one for the near boundary region (inner grid), and the other one
for the bulk region (outer grid).
The corresponding ODEs can then be integrated in the inner grid ($g1$) by imposing
the following boundary conditions
\begingroup
\allowdisplaybreaks
\begin{subequations}
\begin{align}
S_{g1}|_{u=0}&=\frac{1}{54}\left(-18\xi^2\phi_0^2+\phi_0^4-18\phi_0\phi_2\right),\\[2mm]
\partial_uS_{g1}|_{u=0}&=\frac{\phi_0}{90}\left(30\xi^3\phi_0-5\xi\phi_0^3+90\xi\phi_2-24\partial_t\phi_2\right),\\[2mm]
F_{xg1}|_{u=0}&=f_{x2},\\[2mm]
\partial_uF_{xg1}|_{u=0}&=-\frac{2}{15}\left(15\xi f_{x2}+6\partial_xb_{14}+6\partial_xb_{24}-\partial_yg_4-2\phi_0\partial_x\phi_2\right),\\[2mm]
F_{yg1}|_{u=0}&=f_{y2},\\[2mm]
\partial_uF_{yg1}|_{u=0}&=-\frac{2}{15}\left(15\xi f_{y2}+6\partial_yb_{14}+6\partial_yb_{24}-\partial_xg_4-2\phi_0\partial_y\phi_2\right),\\[2mm]
\dot{S}_{g1}|_{u=0}&=\frac{1}{36}\left(18a_4-5\phi_0^4+18\phi_0\phi_2\right),\\[2mm]
\dot{B}_{1g1}|_{u=0}&=-2b_{14},\\[2mm]
\dot{B}_{2g1}|_{u=0}&=-2b_{24},\\[2mm]
\dot{G}_{g1}|_{u=0}&=-2g_{4},\\[2mm]
\dot{\phi}_{g1}|_{u=0}&=\frac{1}{3}-\frac{3\phi_2}{2\phi_0^3},\\[2mm]
A_{g1}|_{u=0}&=a_4,\\
\partial_uA_{g1}|_{u=0}&=-\frac{2}{3}\left(3\xi a_4+\partial_xf_{x2}+\partial_yf_{y2}+\phi_0\partial_t\phi_2\right).
\end{align}
\label{ec:bc_inner}
\end{subequations}
\endgroup
Once again we note that functions $B_1$, $B_2$, $G$, $\phi$, $a_4$, $f_{x2}$,
$f_{y2}$ and $\xi$ encode the freely-specifiable data.
Once the inner grid ODEs have been solved, we evaluate each function at the
interface with the outer grid to obtain the boundary conditions for the $g2$
variables and integrate the corresponding equations.

\subsubsection{Gauge fixing}
\label{sec:gauge-fixing}

To fully close our system we still need to fix the residual gauge
freedom~(\ref{ec:gauge_trafo}). It is advantageous for the numerical
implementation to have the Apparent Horizon (AH) lie at constant radial slice
$r=r_H$ at all times, so it will be convenient to fix a gauge that enforces this
throughout the numerical evolution.
%
%
We thus want to guarantee that $\Theta|_{r=r_{H}} = 0$ at all times, where
$\Theta$ is the expansion of outgoing null rays. Its explicit expression for the
metric~(\ref{eq:metric}) is shown in Appendix~\ref{sec:AH}.


A simple way to enforce $\Theta|_{r=r_{H}} = 0$ at all times during the numerical evolution is to impose a diffusion-like equation of the form
\begin{equation}
\left(\partial_t\Theta+\kappa\Theta\right)|_{u=u_H}=0 
\label{ec:gauge_fixing}
\end{equation}
with $\kappa>0$,
ensuring that the expansion $\Theta$ is driven towards the fix point
$\Theta|_{u=u_H}=0$ as the time evolution runs, pushing the AH surface to
$u=u_H=\mathrm{constant}$.

The way to proceed is the following. We expand equation (\ref{ec:gauge_fixing}) using~\eqref{ec:expansion} and also the equations of motion for both $\ddot{S}$ and $\dot{F}_{x,y}$. 
Then we substitute all the variables by the outer grid redefinitions, $g2$, and
evaluate them at $u=u_H$. We obtain a linear PDE for $\partial_t\xi$ of the type
\begin{equation}
\left(A^{(\xi)}_{xx}\partial^2_x+A^{(\xi)}_{xy}\partial_x\partial_y+A^{(\xi)}_{yy}\partial^2_y+B^{(\xi)}_x\partial_x+B^{(\xi)}_y\partial_y+C^{(\xi)}\right)\partial_t\xi(t,x,y)=-S^{(\xi)} \,, 
\label{ec:xi_evol}
\end{equation}
which can be readily integrated with periodic boundary conditions in $x$ and
$y$.

\subsubsection{Evolution algorithm}
\label{sec:evol-alg}

Having solved equations (\ref{ec:nested_first}-\ref{ec:nested_last}), we use the definition of the ``dot'' operator, cf.\ equation~\eqref{ec:new_der}, to write
\begin{equation}
  \begin{aligned}
    \partial_t B_1(t,u,x,y) = \dot B_1(t,u,x,y)
    + \frac{u^2}{2} A(t,u,x,y) \partial_u B_1(t,u,x,y) \,,
\end{aligned}
  \label{eq:bulk_t}
\end{equation}
and analogously for $B_2$, $G$ and $\phi$. This tells us how to march these
quantities forward in time.\footnote{In practice we write explicitly the
  evolution equations in terms of the redefined $g1$ and $g2$ functions.}


As outlined in the previous subsections, we decompose our computational grid (in the $u$-direction) into two domains: an inner (near boundary) domain and an outer (bulk) domain. The outer domain can further be split into subdomains. We therefore need to match the evolution variables across these domains. The procedure is outlined in Appendix~A of~\cite{Attems:2017zam} which, for completeness, we here summarize.

The evolution equation for $B_1$ (the case for the remaining evolution variables is analogous) has the generic form
\begin{equation}
\label{eq:interior}
\partial_t B_1(t,u,x,y) = c(t,u,x,y) \partial_u B_1(t,u,x,y) + F_{B_1}(t,u,x,y) \,,
\end{equation}
with
\begin{equation}
\label{eq:g1-mode-speed}
c(t,u,x,y) = \frac{u^2}{2} A(t,u,x,y) \,.
\end{equation}
$c(t,u,x,y)$ is locally the propagation speed, and in the vicinity of some $u=u_0$ lying at the interface between two domains $i$ and $i+1$ we can formally write the solution of this equation (ignoring from now on the $x,y$ dependence) as
\begin{equation*}
B_1(t,u_0) \simeq h(u_0 + c \, t) + \int F_{B_1}
\end{equation*}
for any given function $h$.

Therefore, for $c > 0$ ($c < 0$), information is propagating from domain~$i+1$ to domain~$i$ (domain~$i$ to domain~$i+1$). In order to consistently solve this system, the procedure we employ is to use equation~\eqref{eq:interior} (and corresponding ones for the remaining domains) on all interior points;
at the junction point $u=u_0$ we check the propagation speed $c$ at each $x,y$ point and copy the values according to the propagation direction at the interface junction:
\begin{itemize}
\item $c>0$
\begin{equation}
  \label{eq:plus}
\begin{aligned}
  \partial_t B_1^{(i+1)}|_{u=u_0} & =
    c(u_0) \partial_u B_1^{(i+1)}|_{u=u_0} + F_{B_1}(u_0) \,,
  \\
  \partial_t B_1^{(i)}|_{u=u_0} & = \partial_t B_1^{(i+1)}|_{u=u_0} \,,
\end{aligned}
\end{equation}
i.e., we copy the modes leaving domain~$i+1$ to domain~$i$.

\item $c<0$
\begin{equation}
\label{eq:minus}
\begin{aligned}
\partial_t B_1^{(i)}|_{u=u_0} &   = c(u_0) \partial_u B_1^{(i)}|_{u=u_0} + F_{B_1}(u_0) \,, \\
\partial_t B_1^{(i+1)}|_{u=u_0} & = \partial_t B_1^{(i)}|_{u=u_0} \,,
\end{aligned}
\end{equation}
i.e., we copy the modes leaving domain~$i$ to domain~$i+1$.

\end{itemize}

We can now schematically outline the evolution algorithm, which is as follows.
\begin{enumerate}
\item Initial conditions $B_1(t_0,u,x,y)$, $B_2(t_0,u,x,y)$, $G(t_0,u,x,y)$,
  $\phi(t_0,u,x,y)$, $a_4(t_0,x,y)$, $f_{x2}(t_0,x,y)$, $f_{y2}(t_0,x,y)$ and
  $\xi(t_0,x,y)$ are provided for some initial time $t_0$. \label{enum:1}
\item Equations (\ref{ec:nested_first}-\ref{ec:nested_last}) are solved in
  succession for the redefined variables in the inner grid $g1$, imposing the
  boundary conditions (\ref{ec:bc_inner}), and then the same equations are
  solved for the outer grids $g2$, forcing the variables to match their
  values at grid interfaces. 
\item Equation (\ref{ec:xi_evol}) is solved to find $\partial_t\xi(t_0,x,y)$.
  Expression~\eqref{eq:bulk_t} is then used to evaluate
  $\partial_tB_1(t_0,u,x,y)$, $\partial_tB_2(t_0,u,x,y)$,
  $\partial_tG(t_0,u,x,y)$,
  $\partial_t\phi(t_0,u,x,y)$. 
  These variables communicate
  at domain interfaces through equations~\eqref{eq:plus} and~\eqref{eq:minus}.
%
%
\item Obtain $\partial_ta_4(t_0,x,y)$, $\partial_tf_{x2}(t_0,x,y)$ and
  $\partial_tf_{y2}(t_0,x,y)$ through (\ref{ec:boundary_evol}).
\item Advance $B_1$, $B_2$, $G$, $\phi$, $a_4$, $f_{x2}$, $f_{y2}$ and $\xi$ to
  time $t_1$.
  %

\end{enumerate}

\begin{figure}[tbp]
\centerline{\includegraphics[width=0.7\textwidth]{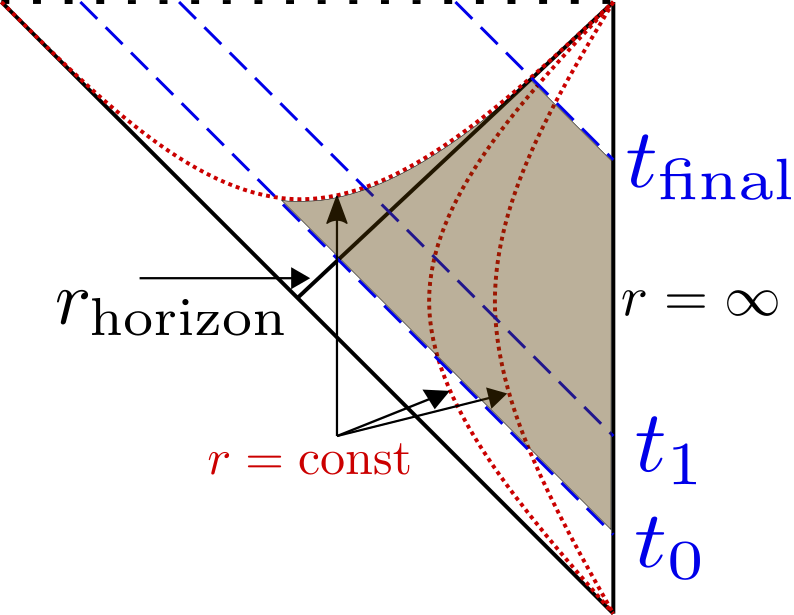}}
\caption{Penrose diagram of the evolution procedure, at constant $x,y$ slices.
  The shaded region represents the region covered by the computational
  domain. \label{fig:penrose} }
\end{figure}

See \fig{fig:penrose} for a cartoon picture of the coordinates used and the evolution scheme (at constant $x,y$).

\subsubsection{Gauge theory expectation values}

The gauge theory expectation values can be obtained from the 
asymptotic behaviour of the bulk variables in a way similar to \cite{Attems:2017zam}.  The result is:
%
%
\begin{equation}
\begin{alignedat}{2}
  \mathcal{E}&=\tfrac{\kappa_5^2}{2L^3} \, \langle T^{tt}\rangle &&
  =-\frac{3}{4}a_4-\phi_0\phi_2+\left(\frac{7}{36}-\lambda_4\right)\phi_0^4,\\[2mm]
  \mathcal{P}_x&= \tfrac{\kappa_5^2}{2L^3} \, \langle T^{xx}\rangle && = 
  -\frac{a_4}{4}-b_{14}-b_{24}+\frac{\phi_0\phi_2}{3}+\left(\frac{-5}{108}+\lambda_4\right)\phi_0^4, \\[2mm]
  \mathcal{P}_{xy}&= \tfrac{\kappa_5^2}{2L^3} \,\langle T^{xy}\rangle && = -g_4,
  \\[2mm]
  \mathcal{P}_y&= \tfrac{\kappa_5^2}{2L^3} \,\langle T^{yy}\rangle  && =
  -\frac{a_4}{4}+b_{14}-b_{24}+\frac{\phi_0\phi_2}{3}+\left(\frac{-5}{108}+\lambda_4\right)\phi_0^4,\\[2mm]
  \mathcal{P}_z&=\tfrac{\kappa_5^2}{2L^3} \,\langle T^{zz}\rangle && =
  -\frac{a_4}{4}+2b_{24}+\frac{\phi_0\phi_2}{3}+\left(\frac{-5}{108}+\lambda_4	\right)\phi_0^4,\\[2mm]
  \mathcal{J}_x&=-\tfrac{\kappa_5^2}{2L^3} \,\langle T^{tx}\rangle && = f_{x2},\\[2mm]
  \mathcal{J}_y&=-\tfrac{\kappa_5^2}{2L^3} \,\langle T^{ty}\rangle && = f_{y2},\\[2mm]
  \mathcal{V}&=  \tfrac{\kappa_5^2}{2L^3} \,\langle \mathcal{O}_{\phi}\rangle && =
  -2\phi_2+\left(\frac{1}{3}-4\lambda_4\right)\phi_0^3.
\end{alignedat}
\label{ec:boundary_VEVs}
\end{equation}
For an $SU(\nc)$ gauge theory the prefactor $\kappa_5^2/2L^3$ in these equations typically scales as $\nc^{-2}$, whereas the stress tensor scales as $\nc^2$. The rescaled quantities are therefore finite in the large-$N$ limit. The stress tensor and the expectation of the scalar operator are related through the Ward identity
\be
\label{ward}
\langle T^\mu_\mu \rangle = -\Lambda \langle \mathcal{O} \rangle \,.
\ee


\subsection{Implementation}
\label{sec:code}

As already mentioned, we have implemented the algorithm of
\Sec{sec:evol-alg} in a new numerical code called
\texttt{Jecco}~\cite{jecco-2022}, written in Julia~\cite{Julia-2017}. Julia is a
dynamically-typed language with good support for interactive use and with
runtime performance approaching that of statically-typed languages such as C or
Fortran. Even though a relative newcomer to the field of scientific computing,
its popularity has been steadily growing in the last few years. It boasts a
friendly community of users and developers and a rapidly growing package
ecosystem.

\texttt{Jecco} was developed as a Julia module and is freely available at
\url{https://github.com/mzilhao/Jecco.jl}. This code is a generalization of the
2+1 C code introduced in~\cite{Attems:2017zam}, and completely written from
scratch.
The codebase is neatly divided into generic infrastructure, such as general
derivative operators, filters, and input/output routines (which are defined in the main \texttt{Jecco} module) and physics, such as initial data, evolution equations, and diagnostic routines (which are defined in submodules).

In \texttt{Jecco} we have implemented finite-difference operators of arbitrary
order through the Fornberg algorithm~\cite{fornberg1998classroom} as well as
Chebyshev and Fourier differentiation matrices. These methods are completely
general and can be used with any Julia multidimensional array. We have also
implemented output methods that roughly follow the openPMD
standard~\cite{openPMD-2018} for writing data.

\subsubsection{Discretization}

For our numerical implementation of the algorithm in \Sec{sec:evol-alg}
we have discretized the $x$ and $y$ directions on uniform grids where periodic
boundary conditions are imposed, while along the $u$~direction we break the
computational domain into several (touching) subdomains with $N_u$ points. In
each subdomain a \textit{Lobatto-Chebyshev} grid is used where the collocation
points, given by
\begin{equation}
 X_{i+1} =  -\cos\left(\frac{\pi\,i}{N_u}\right) \qquad (i=0,1,\ldots,N_u-1)\,,  \end{equation}
are defined in the range $[-1:+1]$, and can be mapped to the physical grid by
\begin{equation}
  u_i = \frac{u_R + u_L}{2} + \frac{u_R - u_L}{2} X^{}_i \qquad (i=1,\ldots,N_u)\,,
\end{equation}
where $u_L$ and $u_R$ are the limits of each of subdomain. For the subdomain
that includes the AdS boundary ($u=0$), the inner grid variables of
\Sec{sec:redef} are used; all remaining subdomains use the outer grid
variables.

Derivatives along the $x$ and $y$ directions are approximated by (central)
finite differences. Although in \texttt{Jecco} operators of arbitrary order are
available, we have mostly made use of fourth-order accurate ones for our
applications.
In the radial direction $u$, the use of the Chebyshev-Lobatto grid allow us to use pseudo-spectral collocation methods~\cite{Boyd2001}.
These methods are based in approximating solutions in a basis of Chebyshev polynomials $T_n(X)$ but, in addition to the spectral basis, we have an additional \textit{physical} representation -- the values that functions take on each grid point -- and therefore we can perform operations in one basis or the other depending on our needs.
Discretization using the pseudo-spectral method consists in the exact imposition
of our equations at the collocation points of the Chebyshev-Lobatto grid.


The radial equations that determine our grid functions have the schematic form
of equation~\eqref{eq:ODEs},
where $f$ represents the metric coefficients and scalar field $\phi$. Once our coordinate $u$ is discretized, the differential operator becomes an algebraic one acting over the values of the functions in the collocation points taking the form (at every point in the transverse directions $x,y$)
\begin{equation}
  \sum_{j=1}^{N_u}
\left[
  A_f^i(t,x,y) \mathcal D_{uu}^{ij} + B_f^i(t,x,y) \mathcal D_u^{ij} + C_f^i(t,x,y) \mathbb{I}^{ij}
\right] f^j(t,x,y) = -S_f^i(t,x,y)
\label{eq:linsystem-discrete}
\end{equation}
(no sum in $i$),
where $\mathcal D_{uu}$, $\mathcal D_{u}$ represent the derivative operators for
a Chebyshev-Lobatto grid in the physical representation (see for instance~\cite{trefethen2000spectral} for the explicit expression) and $i$, $j$
indices in the $u$ coordinate.
Boundary conditions are imposed by replacing full rows in this operator by the
values we need to fix:
at the inner grid $g1$, we impose the boundary conditions in (\ref{ec:bc_inner}); at the outer grids these are read off from the obtained values in the previous subdomain.

The resulting operators 
are then factorized through an LU decomposition and the linear
systems~\eqref{eq:linsystem-discrete} are subsequently solved using Julia's left
division (\texttt{ldiv!}) operation. Recall that we need to solve one such
radial equation per grid point in the $x,y$ transverse directions. Since these
equations are independent of each other, we can trivially parallelize the
procedure using Julia's \texttt{Threads.@threads} macro.

Equation~(\ref{ec:xi_evol}) for $\partial_t\xi$ is a linear PDE in $x,y$. To solve it, after discretizing in a $N_x \times N_y$ grid, we flatten the solution vector using lexicographic ordering
\[
  \bm{g} \equiv
  \begin{pmatrix}
    \partial_t\xi(t,x_1, y_1 )    \\
    \partial_t\xi(t,x_2, y_1 )    \\
    \vdots                        \\
    \partial_t\xi(t,x_{N_x}, y_1 ) \\
    \partial_t\xi(t,x_{1}, y_2 )   \\
    \vdots                         \\
    \partial_t\xi(t,x_{N_x}, y_{N_y} )
  \end{pmatrix}
\]
and introduce enlarged differentiation matrices, which can be conveniently built as Kronecker products
\begin{equation}
  \begin{aligned}
  \hat{\mathcal D}_x & = \mathbb{I}_{N_y\times N_y} \otimes \mathcal D_x, &
  \qquad
  \hat{\mathcal D}_y & = \mathcal D_y \otimes \mathbb{I}_{N_x\times N_x}, \\
  \hat{\mathcal D}_{xx} & = \mathbb{I}_{N_y\times N_y} \otimes \mathcal D_{xx}, &
  \qquad
  \hat{\mathcal D}_{yy} & = \mathcal D_{yy} \otimes \mathbb{I}_{N_x\times N_x}, 
\end{aligned}
  \label{eq:2DD}
\end{equation}
where $\mathcal D_x$, $\mathcal D_y$, $\mathcal D_{xx}$, $\mathcal D_{yy}$ are the first and second derivative finite-difference
operators. The cross derivative operator is built as a matrix product, $\hat{\mathcal D}_{xy} =  \hat{\mathcal D}_x   \hat{\mathcal D}_y$.
The PDE~(\ref{ec:xi_evol}) then takes the algebraic form
\begin{equation}
  \sum_{J=1}^{N_x \times N_y}
\left[
  A_{xx}^I \hat{\mathcal D}_{xx}^{IJ}
  + A_{xy}^I \hat{\mathcal D}_{xy}^{IJ}
  + A_{yy}^I \hat{\mathcal D}_{yy}^{IJ}
  + B_{x}^I  \hat{\mathcal D}_x^{IJ}
  + B_{y}^I \hat{\mathcal D}_y^{IJ}
  + C^I \mathbb{I}^{IJ}
\right] \bm{g}^J = -S_{\bm{g}}^I
\label{eq:PDE-discrete}
\end{equation}
(no sum in $I$), where $I,J = 1, \ldots,N_x \times N_y$. The
$x$ and $y$ directions are periodic, so no boundary conditions need to be
imposed. See for example \cite{Krikun2018} for a pedagogical overview of these
techniques.

As before, the operator defined inside the square brackets is factorized
through an LU decomposition and the linear system~\eqref{eq:PDE-discrete} is
then solved with the left division operation. Since all the matrices are
sparse, we store them in the Compressed Sparse Column format using the type
\texttt{SparseMatrixCSC}.

\subsubsection{Time evolution}

For the time evolution we use a method of lines procedure, where we find it
convenient to pack all evolved variables (across all subdomains) into one single
state vector. This state vector is then marched forwarded in time with the
procedure of \Sec{sec:evol-alg} using the \texttt{ODEProblem} interface
from the DifferentialEquations.jl Julia
package~\cite{DifferentialEquations.jl-2017}. This package provides a very long
and complete list of integration methods. For our applications, since evaluating
the time derivative of our state vector is an expensive operation, we find it
convenient for reasons of speed and accuracy to use the Adams-Bashforth and
Adams-Moulton family of multistep methods. Depending on the application, we find that the (third order) fixed step method \texttt{AB3} and the adaptive step size
ones \texttt{VCAB3} and \texttt{VCABM3} seem to work particularly well.
The integration package automatically takes care of the starting values by using a lower-order method initially.

We use Kreiss-Oliger dissipation~\cite{Kreiss1973} to remove spurious
high-frequency noise common to finite-difference schemes. In particular, when
using finite-difference operators of order $p-1$, we add Kreiss-Oliger
dissipation of order $p$ to all evolved quantities $f$ as
\begin{equation}
\label{eq:KO-op}
f \leftarrow f +  \sigma \frac{(-1)^{(p+3)/2}}{2^{p+1}} \left(
  h_x^{p+1}\frac{\partial^{(p+1)}}{\partial x^{(p+1)}}
  + h_y^{p+1} \frac{\partial^{(p+1)}}{\partial y^{(p+1)}}
\right) f
\end{equation}
after each time step,
where $h_x$ and $h_y$ are the grid spacings and $\sigma$ is a
tuneable dissipation parameter 
which we typically set to 0.2 unless explicitly stated otherwise.
This procedure effectively works as a low-pass filter.

Along the $u$-direction
we can damp high order modes directly in the spectral representation. After each time step, we apply an exponential filter to the spectral coefficients of our $u$-dependent evolved quantities $f$ (see for instance~\cite{KANEVSKY200641}). 
The complete scheme is
\begin{equation}
\left \{  f_i\   \right\}                          \stackrel{\rm FFT}{\longrightarrow}
\left \{  \hat f_k \right\}                               \stackrel{}{\longrightarrow}
\left \{ \hat f_k\; e^{ -\alpha (k/M)^{\gamma M} } \right\}  \stackrel{\rm FFT}{\longrightarrow}
\left \{f_i \right\}  \label{eq:u-filter}
\end{equation}
where $M\equiv N_u - 1$, $k=0,\ldots,M$,
$\alpha = \log \epsilon$ where $\epsilon$ is the machine epsilon
(for the standard choice of $\epsilon = 2^{-52}$, $\alpha = 36.0437$)
and $\gamma$ is a tuneable parameter which we typically fix to $\gamma = 8$. This effectively dampens the coefficients of the higher-order Chebyshev polynomials.

We performed a thorough set of tests on this implementation, which is detailed in Appendix~\ref{sec:results}.

\section{Bubble dynamics}
\label{bubbles}
The \texttt{Jecco} code described in the previous section was first applied to the study of gravitational waves produced by the spinodal instability in a cosmological first-order phase transition \cite{Bea:2021zol}. We now turn to a new application, namely the dynamics of bubbles in a strongly-coupled, four-dimensional gauge theory. For this purpose we will focus on a holographic model of the type described by equations~(\ref{eqn:Einstein-scalar_model}) and \eqn{eq:potential} with the same value of the parameters \eqn{param} as in \cite{Bea:2021zsu}, namely
\be
\label{paramchoice}
\phi_M=0.85 \sac \phi_Q = 10 \,.
\iff  \lambda_4= -0.346021 \sac \lambda_6=0.1 \,.
\ee
The motivation for the general class of models under consideration is that they provide  simple examples of non-conformal theories with first-order phase transitions (for appropriate values of $\phi_M$ and $\phi_Q$) whose dual gravity solutions are completely regular even at zero temperature. The motivation for the choice \eqn{paramchoice} is that it leads to a sizeable bubble wall velocity, as we will see in \Sec{expsec}.

\subsection{Thermodynamics}
\begin{figure}[t]
\begin{center}
  \includegraphics[width=0.9\textwidth]{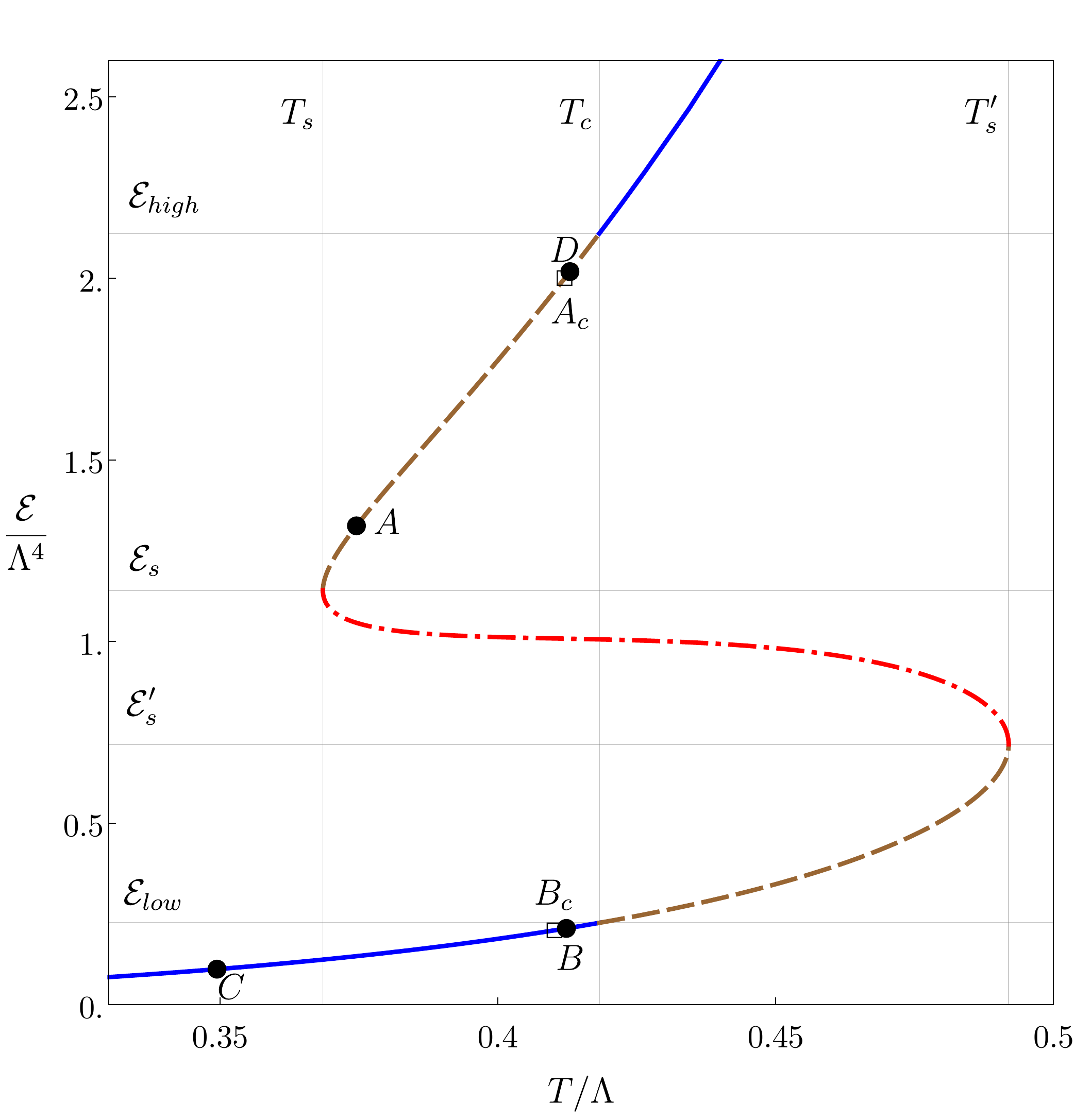}
  \caption{Energy density as a function of temperature for the gauge theory dual to the holographic model (\ref{eqn:Einstein-scalar_model})-\eqn{eq:potential} with parameters \eqn{paramchoice}. The squares 
$B_c$ and $A_c$ correspond, respectively, to the  states inside and outside of the closest-to-critical bubble studied in \Sec{critsec}. 
The dots $B$ and $C$ correspond to the initial states inside and outside the expanding bubble studied in \Sec{expsec}, respectively. At late times, the state $B$ inside the bubble evolves into $C$, and a heated region is created in front of the bubble that can be characterized in terms of the point $D$ in the phase diagram.
    \label{Fig:phasediagram}
}
   \end{center}
\end{figure}

The thermodynamics of the gauge theory can be extracted from the homogeneous black brane solutions on the gravity side (see e.g.~\cite{Gubser:2008ny}).
Figure~\ref{Fig:phasediagram} shows the result for the energy density as a function of  temperature, where we  see the usual multivaluedness associated to a first-order phase transition.
 At high and low temperatures there is only one phase available to the system. Each of these phases is represented by a solid, blue curve. At the critical temperature 
\[
T_c = 0.418\Lambda 
\]
the state that minimizes the free energy  moves from one branch to the other. The first-order nature of the transition is encoded in the non-zero latent heat, namely in the discontinuous jump in the energy density given by 
\be
\label{lat}
\mathcal{E}_{\mathrm{latent}}=\Ehigh-\Elow \sac 
\Elow = 0.225\Lambda^4 \sac
\Ehigh= 2.123 \Lambda^4 \,.
\ee
Note that the phase transition is a transition between two deconfined plasma phases, since both phases have energy densities of order $N^2$ and they are both represented by a black brane geometry with a horizon on the gravity side. 

In a region 
\be
\label{ts}
\Ts = 0.3879\Lambda < T < \Ts'=0.4057\Lambda
\ee
 around the critical temperature  there are three different states available to the system for a given temperature. The thermodynamically preferred one is the state that minimizes the free energy, namely a state on one of the blue curves. The states on the dashed, brown curves are not globally preferred but they are locally thermodynamically stable, i.e.~they are metastable. This follows from the fact that specific heat
\be
\label{heat}
c_v \equiv \frac{d \mathcal{E}}{dT} 
\ee
is positive on these branches.  At the temperatures $\Ts$ and $\Ts'$ the metastable curves meet the dotted-dashed, red curve, known as the ``spinodal branch''. States on this branch are locally unstable since their specific heat is negative and have energies comprised between 
\be
\mathcal{E}'_{s}=0.717\Lambda^4 \sac
\mathcal{E}_{s} =1.141\Lambda^4 \,.
\ee
Note that the characteristic scale for all the quantities above is set by the microscopic scale in the gauge theory, $\Lambda$, given holographically by $\Lambda=\phi_0$ in terms of the leading term in the near-boundary fall-off of the scalar field in \eqn{eq:asympt-phi}.

\subsection{Initial data}
\label{sec:indata}
As any other thermal system with a first order phase transition, the  gauge theory can be overcooled past the critical temperature $T_c$. The homogeneous, overcooled state, represented by a point on the upper, brown branch in \fig{Fig:phasediagram}, is stable against small fluctuations, including thermal ones, but not against sufficiently large fluctuations.  A particular class of large fluctuations are bubbles, namely inhomogeneous configurations in which  the energy density of a certain region of space within the overcooled homogeneous phase is reduced. For sufficiently large bubbles, the energy density in the centre of this region lies in the stable branch of the phase diagram, represented by the lower, blue curve in \fig{Fig:phasediagram}, and the bubble  smoothly interpolates between the stable and the metastable phases. 

In a homogeneous and isotropic thermal system it is expected that the nucleated bubbles are spherical. However, given our symmetry restrictions we will study cylindrical bubbles. This is enough to bring about two new physical aspects compared to our  previous  work  \cite{Bea:2021zsu} for planar configurations. The first one is that the surface tension now plays a role. In particular, we will be able to identify a critical bubble in which the inward-pointing force due to the surface tension exactly balances the outward-pointing force coming from the pressure difference between the inside and the outside of the bubble. The second one is that the asymptotic profile of an expanding bubble possesses more structure than in the planar case.

Our first task is to construct initial data corresponding to a bubble. By definition, this is a  configuration consisting of a cylindrical region filled with the stable phase (the inside of the bubble) connected to an asymptotic region filled with the metastable phase (the outside of the bubble) through an appropriate interface. The stable and metastable  phases correspond to the points labelled $B$ and $A$  in \fig{Fig:phasediagram}, respectively, and both have $T<T_c$. As we will now explain, our strategy to construct these bubbles will be to start with a phase-separated state, which has $T=T_c$,  and to rescale it appropriately. 

\begin{figure}[thp]
 \centering
  \includegraphics[width=0.53\textwidth]{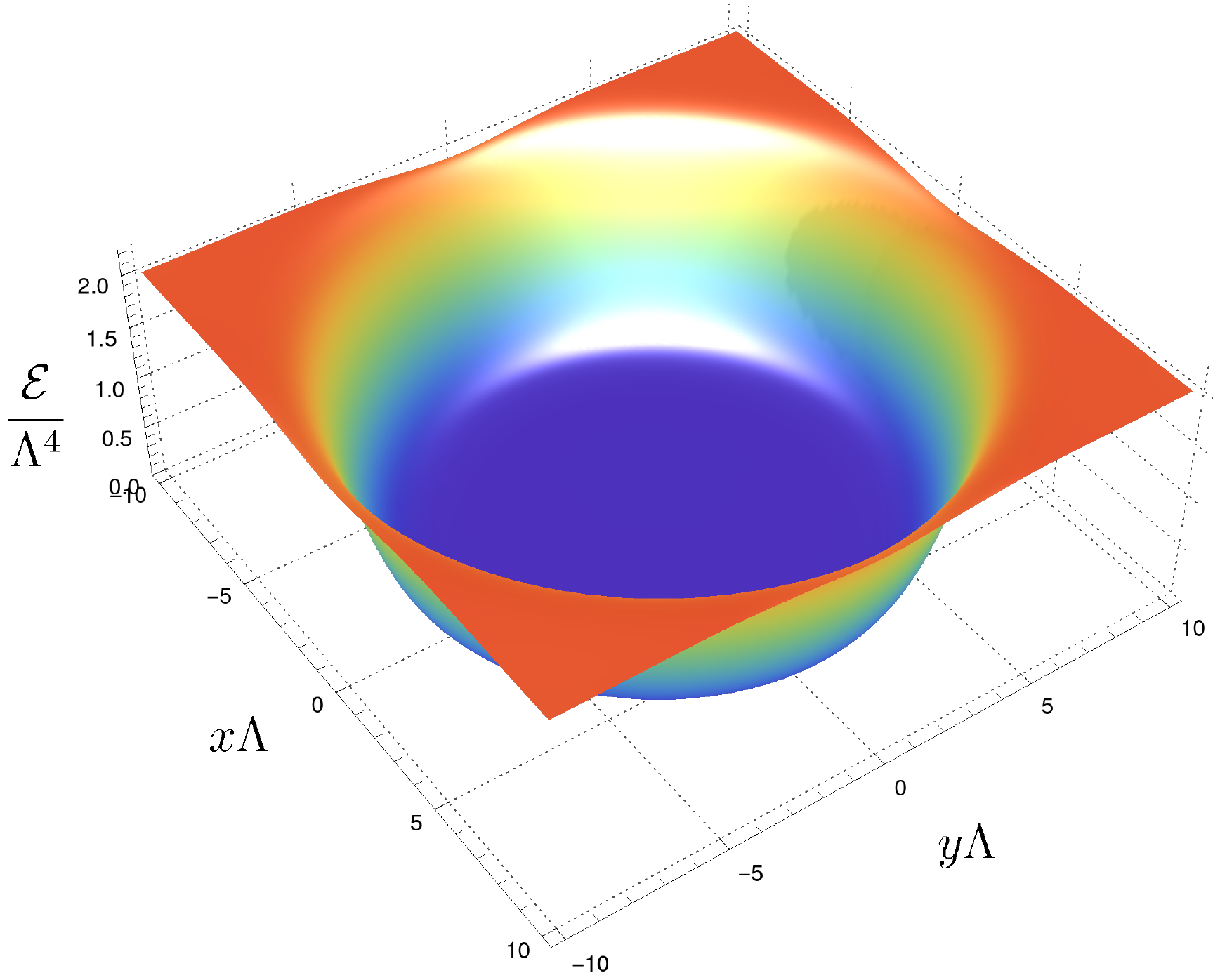} \\
  \includegraphics[width=0.53\textwidth]{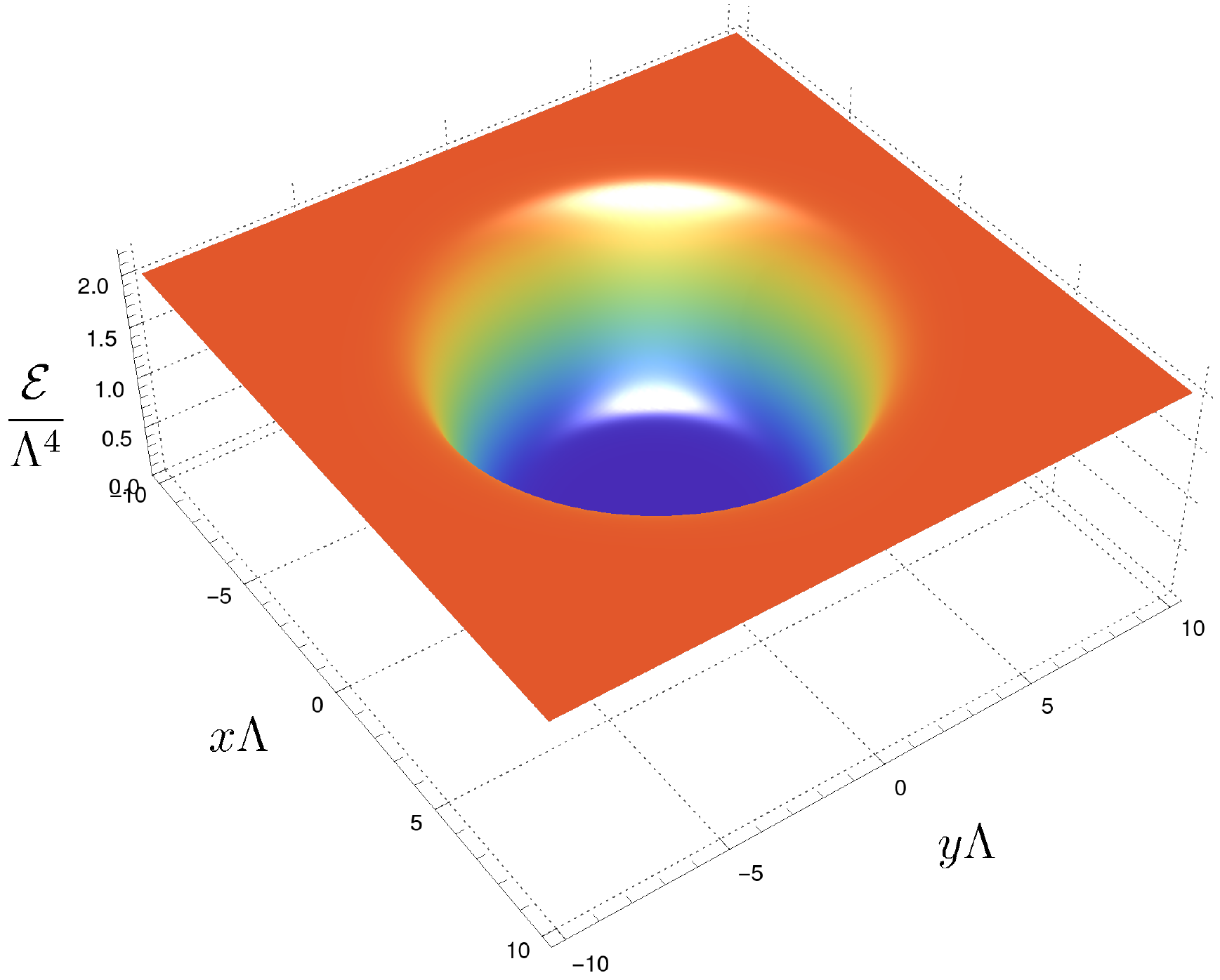} \\
  \includegraphics[width=0.53\textwidth]{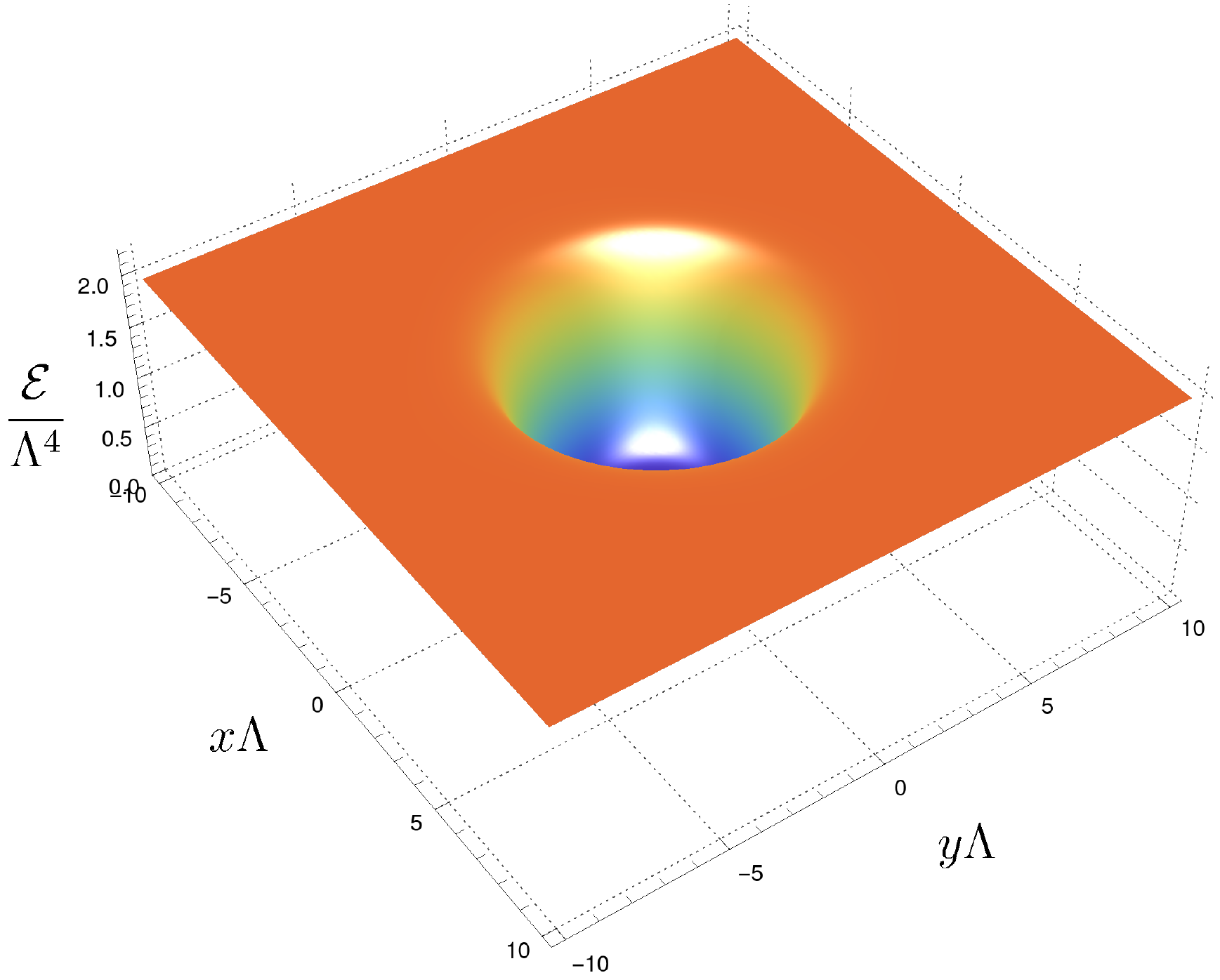}
  \caption{Phase-separated configurations in a box of size $L_x\Lambda=L_y\Lambda=20$ with average energy densities $\bar{\mathcal{E}}/\Lambda^4 = 1.0$ (top), $\bar{\mathcal{E}}/\Lambda^4=1.6$ (middle) and $\bar{\mathcal{E}}/\Lambda^4=1.8$ (bottom). 
    \label{Fig:phase_separated}
    }
\end{figure}
Phase-separated states are configurations in which the two homogeneous phases with energy densities $\Ehigh$ and $\Elow$  coexist in equilibrium at $T=T_c$. This is possible because at this temperature the free energy densities, and hence the pressures, are equal in the two phases. Three examples of such configurations in a box of constant size are shown in \fig{Fig:phase_separated}.
The difference between the three cases is the relative fraction of the total volume occupied by each phase. For a box of fixed size, changing this relative fraction is equivalent to changing the average energy density in the box, $\bar{\mathcal{E}}$. The larger the average energy density, the larger the size of the high-energy region, and vice-versa. We will use this fact to our advantage when we search for the critical bubble below. 

Strictly speaking, phase-separated states only exist in infinite volume, since only in that case the two coexisting phases become arbitrarily close to being homogeneous sufficiently far away from the interface. The middle and bottom panels of \fig{Fig:phase_separated} correspond to states that are fairly close to this limit, but  deviations can still be seen with the naked eye. For example, the energy density in the region outside the bubbles is slightly below $2\Lambda^4$, whereas the energy density in the high-energy phase at $T=T_c$ has $\Ehigh$ above $2\Lambda^4$, as given in~\eqn{lat}. The state in \fig{Fig:phase_separated}(top) is even more affected by finite box-size effects because the size of the low-energy region is comparable to the size of the box. In any case, these deviations will have no implications for our purposes, since we are not interested in phase-separated states per se but only in using them to construct initial data for bubble configurations.  

The value of $\bar{\mathcal{E}}$ in a box of fixed size is conserved upon time evolution. Therefore,  phase-separated states with an average energy density in the region 
 \mbox{$\Es' \leq \Ebar \leq \Es$} can be generated by starting with a homogeneous state in the spinodal region of \fig{Fig:phasediagram}, perturbing it slightly, and letting evolve until it settles down to a phase-separated configuration \cite{Attems:2019yqn,Bea:2021zol}. To initialize the code we specify some $\phi_2$ that is not too far away from the value of the thermal state and generate a simple bulk profile for the scalar, $\phi(t=0,u)$, given by the truncated series in~(\ref{eq:asympt-phi}) to third order. This is not the geometry associated to the black brane of such energy density, but it would relax fast to the true static solution. The value for $a_4$ is obtained by using the energy expression in~\eqref{ec:boundary_VEVs}. On top of it we add a sinusoidal perturbation, so the final $a_4$ reads
\begin{equation}
  a_4(t=0,x,y)=\bar{a}_4\left[1+\delta a_4 \left(\cos\left(\frac{2\pi}{L_x}(x-x_{\rm mid})\right)+\cos\left(\frac{2\pi}{L_y}(y-y_{\rm mid})\right)\right)\right],
  \label{eq:a4_pert}
\end{equation}
where $\bar{a}_4$ is the value we obtained above, $L_x$ and $L_y$ are the lengths of the box, $x_{\rm mid}$ and $y_{\rm mid}$ correspond to the central point and $\delta a_4$ represents the amplitude of the perturbation, equal for both $x$- and $y$-directions. The fastest way to arrive at a phase-separated configuration is to assign the largest possible value to $\delta a_4$ compatible with keeping the apparent horizon within our grid. We have found that $\delta a_4\sim 10^{-3}$ is a convenient choice. The state in \fig{Fig:phase_separated}(top) was generated following this method with $\phi_2=0.3\Lambda^3$. After a time $t\Lambda = 300$ the system has settled down to the configuration shown in the figure.

Phase-separated configurations with average energy densities in the regions 
\be
\Es \leq \Ebar < \Ehigh \qquad \mbox{and}\qquad \Elow < \Ebar \leq \Es' 
\ee
also exist, but they cannot be found directly via time evolution of an initial state in the spinodal region. Instead, to obtain them we follow \cite{Bea:2020ees}. We take initial data corresponding to a phase-separated state with $\Ebar$ in the spinodal region, and we modify it by increasing or decreasing the value of $\bar{a}_4$ so that the new $\Ebar$ takes the desired value. We then let the system evolve. In a time around 
 $t=100/\Lambda$ the system relaxes to a new inhomogeneous, static configuration. The phase-separated configurations in the middle and bottom panels of \fig{Fig:phase_separated} have \mbox{$\Es \leq \Ebar \leq \Ehigh$} and were obtained with this procedure.

\begin{figure}[thb]
 \centering
  \includegraphics[width=0.7\textwidth]{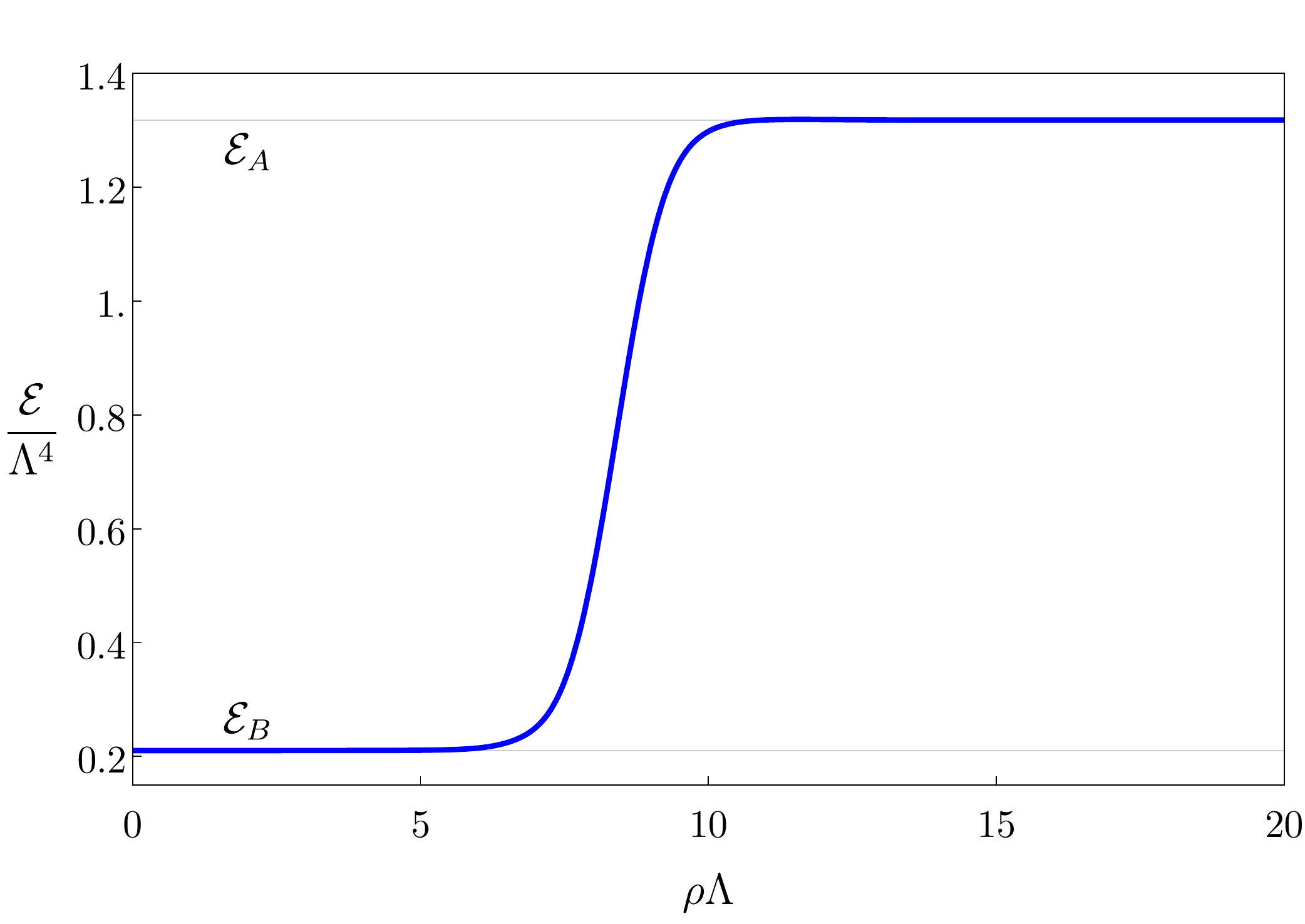}
  \caption{Initial energy density profile of a bubble. 
    \label{Fig:initial_E_generic_bubble}
    }
\end{figure}

The phase-separated states interpolate between the energy densities $\Elow$ and $\Ehigh$. To construct  initial data for bubble configurations that interpolate between two energy densities $\mathcal{E}_B$ and $\mathcal{E}_A$ we proceed as follows. Let $f_{\textrm{PS}}$ be any of the functions specifying the initial data of a phase-separated state. This could be one of the metric components in the bulk or the scalar field, in which case $f_{\textrm{PS}}=f_{\textrm{PS}}(u,x,y)$, or one of the boundary functions such as $a_4$, in which case 
$f_{\textrm{PS}}=f_{\textrm{PS}}(x,y)$. We assume that the centre of the region with energy density $\Elow$ is at $x=y=0$, and that the point at the edge of the box 
$x=y=L/2$ lies in the region with energy density $\Ehigh$. Let $f_A$ and $f_B$ be the corresponding functions for the states $A$ and $B$. Since these states are homogeneous, $f_A$ and $f_B$ depend on $u$ for a bulk function and are just constants for a boundary function. We then define the corresponding initial data for a bubble through the rescaling
 \begin{equation}
f_{\textrm{bubble}}(u,x,y) = f_B(u)+\Big( f_A(u)-f_B(u)\Big) 
\left[ \frac{f_{\textrm{PS}}(u,x,y)-f_{\textrm{PS}}(u,0,0)}
{f_{\textrm{PS}}(u,L/2,L/2)-f_{\textrm{PS}}(u,0,0)} \right] \,.
\end{equation} 
If $f$ is a boundary function then there the dependence on $u$ is absent. At any fixed value of $u$, the term in square brackets interpolates smoothly between 0 at the centre of the low-energy region and 1 at the edge of the box. As a consequence, 
$f_{\textrm{bubble}}(u,x,y)$ interpolates smoothly between $f_B$ and $f_A$, as desired. A state generated with this procedure is shown in \fig{Fig:initial_E_generic_bubble}.
If the subsequent time evolution leads to an expansion of the bubble, it is convenient to further enlarge the size of the box before starting the evolution, in order to prevent the bubble from reaching the boundary of the box before it has reached an asymptotic state. This can be done simply by ``adding'' more metastable bath outside the initial box.

Variations of an initial bubble state can be obtained in a simple way. For example, we can choose different states $B$ for a fixed $A$. As in \cite{Bea:2021zsu}, we expect that the subsequent time evolution will quickly select a dynamically preferred state $C \neq B$ inside the bubble. We could also multiply the bulk metric functions $B_1$ and $B_2$ in \eqn{eq:metric} by some factor, thus changing the pressure distribution (the anisotropy) along the wall but not the energy profile. We could further consider initial bubbles whose cross sections are not perfectly circularly symmetric by starting with an initial phase-separated state whose low-energy region is comparable to the size of the box, as in \fig{Fig:phase_separated}(top).

\subsection{Critical bubbles}
\label{critsec}
Consider a cylindrical bubble of radius $\rho$ such that the states inside and outside the bubble correspond to the points marked as $B_c$ and $A_c$ in \fig{Fig:phasediagram}, respectively. The pressure difference between these states  generates an outward-pointing force on the bubble wall. In turn, the surface tension of the bubble wall results in an inward-pointing force on the wall. A critical bubble is one for which these two forces exactly balance each other. Since these bubbles are static, they correspond to equilibrium states. As a consequence, the temperature must be constant across the entire system and, in particular, it must be equal to $T_{A_c}$. It follows that the state $B_c$ is determined by $A_c$. If the radius of the bubble is large compared to the width of the interface between $A_c$ and $B_c$, then the radius of the critical bubble takes the form 
\begin{equation}
\label{eq:rhoc}
\rho_c= \frac{\gamma}{\P_{B_c}-\P_{A_c}} \,.
\end{equation}
This follows from approximating the interface by a zero-width surface with free energy density $\gamma$, assigning a well defined pressure $\P_{B_c}$, and hence a free energy density $-\P_{B_c}$, to the interior of the bubble, and requiring that the critical bubble locally extremizes the free energy. The fact that this extremum is a maximum means that the critical bubble is in unstable equilibrium. This expression for the critical radius is only valid for large critical bubbles, which are realized when $T_{A_c}$ is close to the phase transition temperature $T_c$, namely for $T_{A_c} \lesssim T_c$. This is the reason for our choice of the point 
${A_c}$ in \fig{Fig:phasediagram}. If the bubble is not large enough then the phase inside  the bubble is not approximately homogeneous and it cannot be clearly separated from the interface. In this case one cannot assign a meaningful surface tension to the interface or a well defined pressure to the interior of the bubble. This situation is realized when $T_{A_c}$ is sufficiently close to the turning point at $T=T_s$, namely when $T_s \lesssim T_{A_c}$. In this paper we will only discuss large critical bubbles; small bubbles will be analysed elsewhere.

The fact that critical bubbles are unstable means that supercritical bubbles expand, whereas undercritical bubbles collapse. Critical bubbles are therefore the static configurations that separate these two sets of large, inhomogeneous, cylindrically-symmetric fluctuations of the plasma. This is precisely the feature that will allow us to identify the critical bubbles with \texttt{Jecco}. 

\begin{figure}[th]
\centering
  \includegraphics[width=0.95\textwidth]{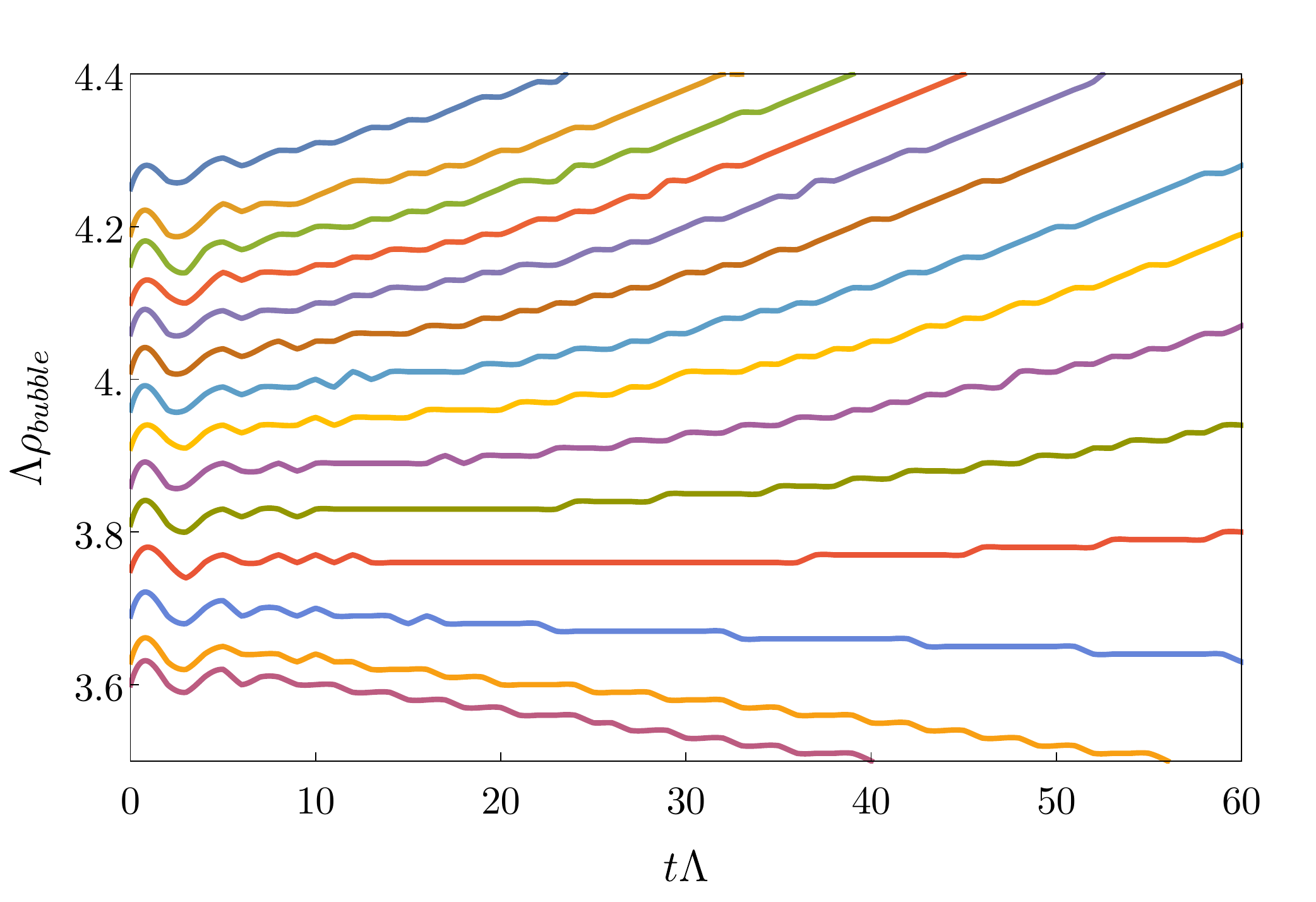}
  \caption{Time evolution of the wall position for several different initial bubbles. The critical bubble radius has to be $3.69 < \Lambda\rho_c <3.75$.
  \label{lots} }
\end{figure}
Following the procedure outlined in Sec.~\ref{sec:indata}, we generate a family of initial cylindrical bubbles with different radii and we numerically evolve them with  \texttt{Jecco}. As expected from the discussion above, large bubbles expand and small bubbles collapse. This is illustrated in \fig{lots}, where we plot the radius of each bubble, defined as the position of the inflection point of the energy density profile, as a function of time.
Each simulation presented in this section were performed in MareNostrum 4 using 1 node with 48 cores. The typical runtime was around 250h.
We see that bubbles with initial radius $\Lambda \rho_c \geq 3.75$ eventually expand, whereas bubbles with radius $\Lambda \rho_c \leq 3.69$ eventually collapse. This means that the critical radius must be in between these two values. Substituting into \eqn{eq:rhoc} we then obtain an estimate for the surface tension $\gamma$. Thus,
\[
3.69 < \Lambda \rho_c < 3.75 \sac 
0.116 < \frac{\gamma}{\Lambda^3} < 0.118 \,.
\]

\begin{figure}[t!p]
\includegraphics[width=0.475\textwidth]{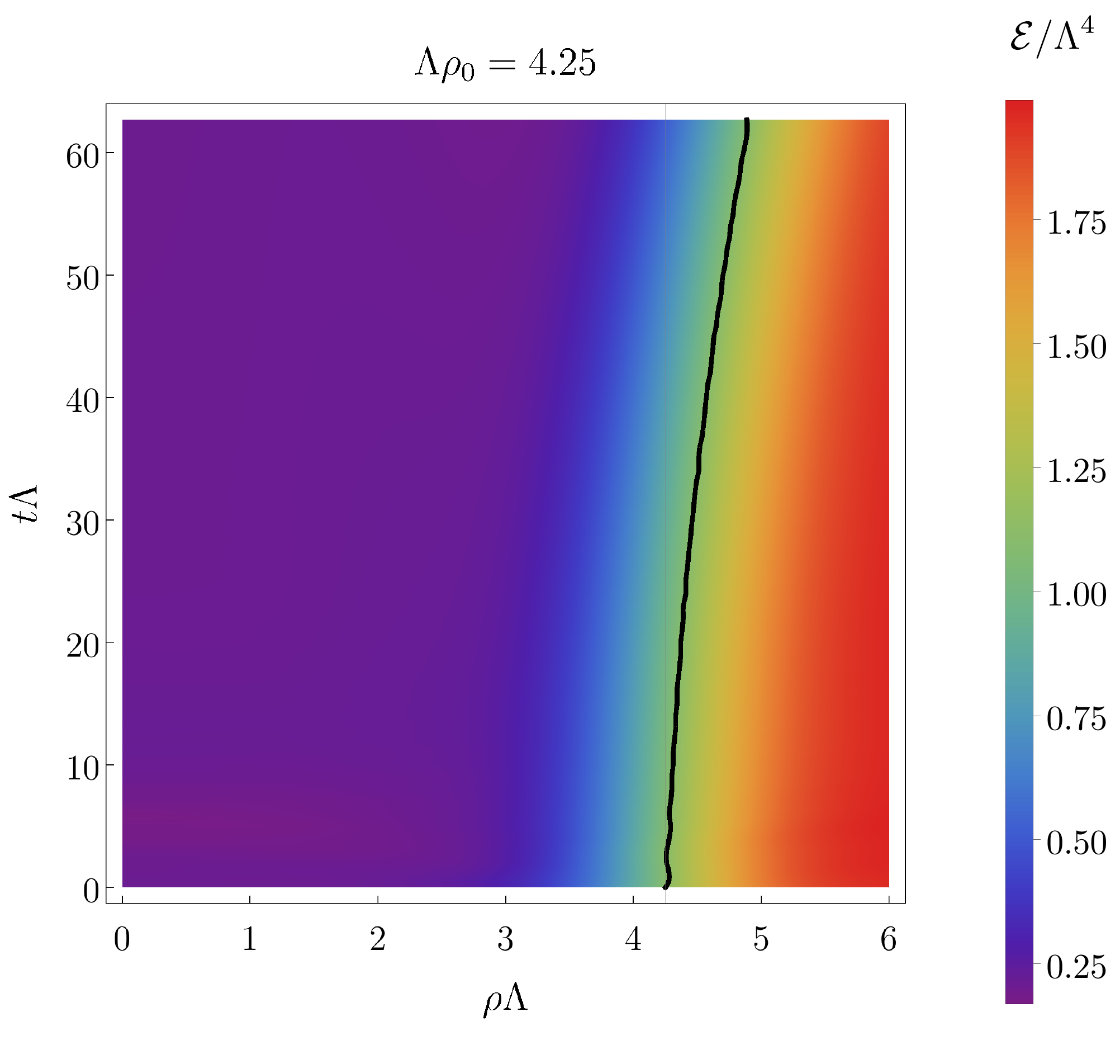} \hfill
\includegraphics[width=0.475\textwidth]{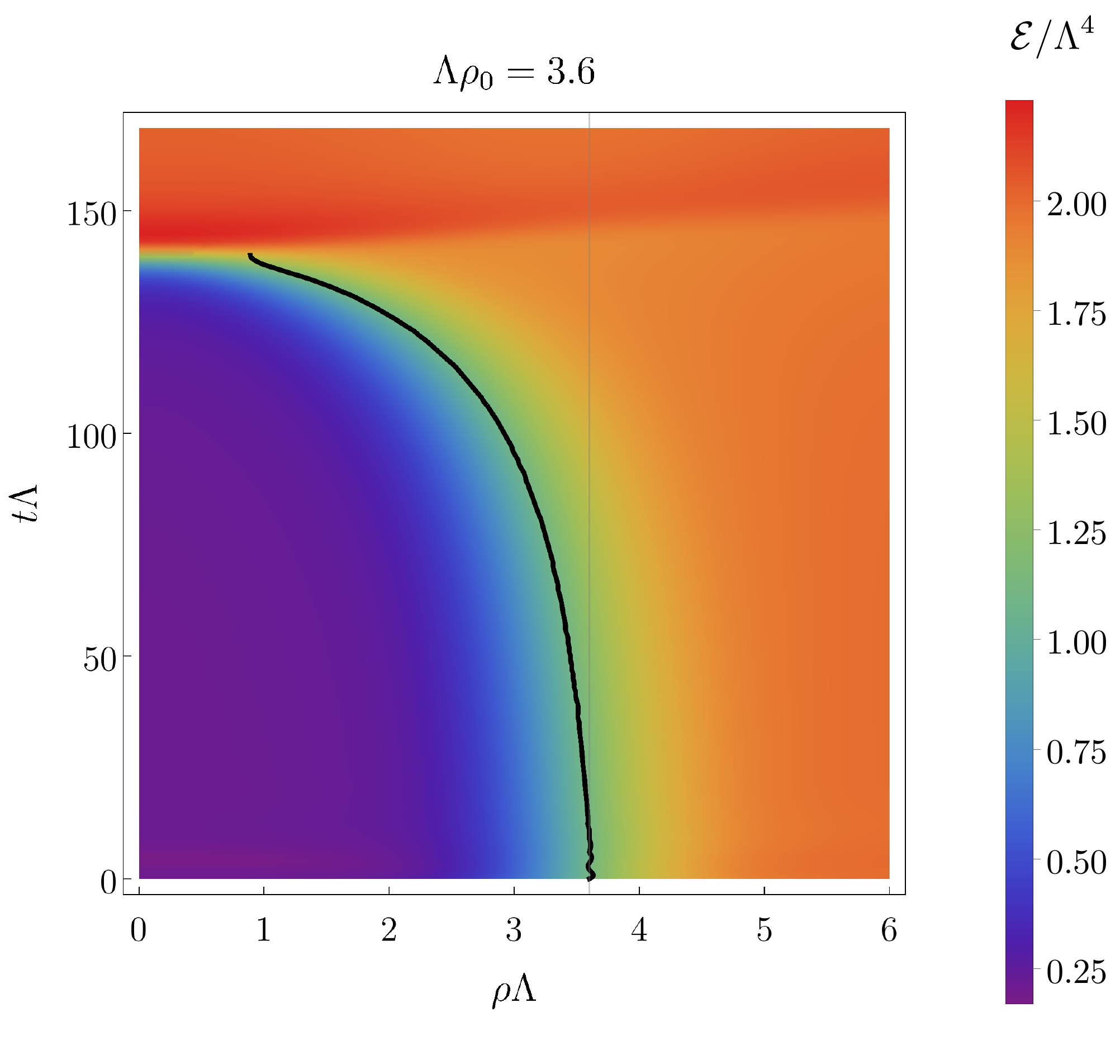} \\
\includegraphics[width=0.475\textwidth]{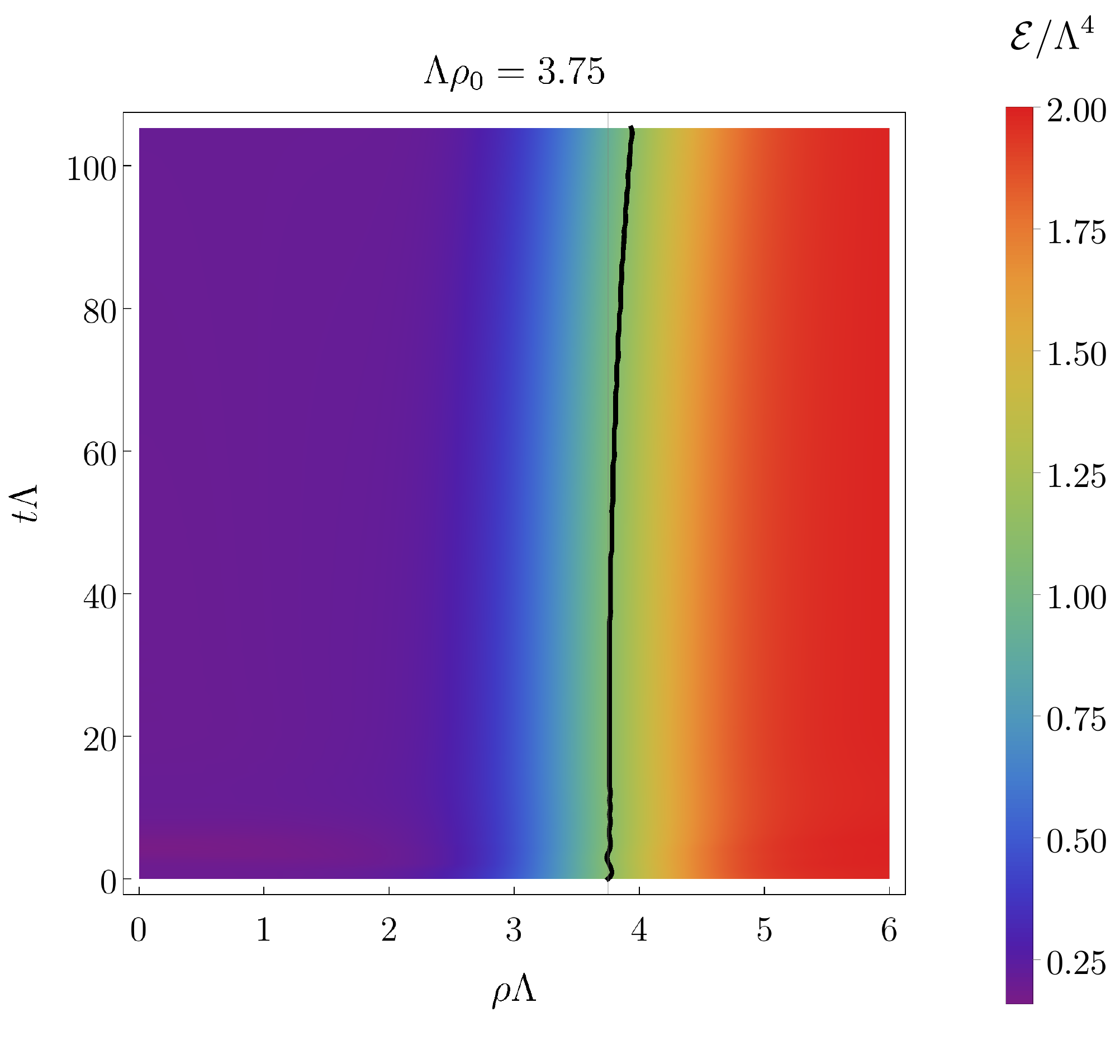} \hfill
\includegraphics[width=0.475\textwidth]{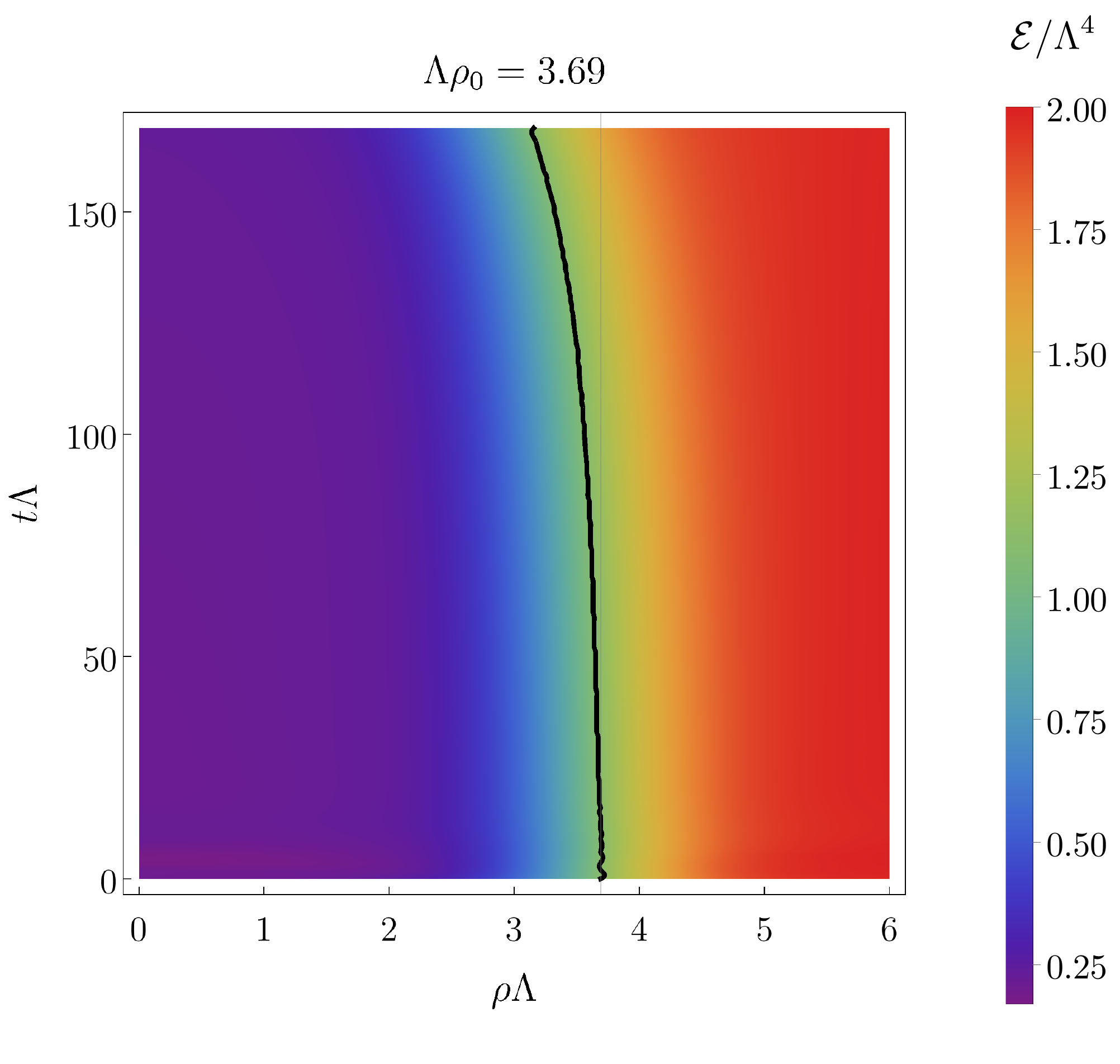}
\caption{Time evolution of bubbles with different initial radii. The black curves represent the position of the wall, defined as the inflection point in the energy density profile. The radius of the critical bubble lies in the interval $3.69 < \Lambda \rho_c<3.75$. The bubbles on the left column are supercritical and they expand. The bubbles on the right column are subcritical and they collapse. The bubbles in the bottom  row are closer to the critical bubble than those on the top row and  hence they evolve more slowly.  Videos of each of the evolutions can be found at \href{https://youtube.com/playlist?list=PL6eUQq2UUQ4JJTD_pRfJt-ShPGKxGBRW9}{https://youtube.com/playlist?list=PL6eUQq2UUQ4JJTD_pRfJt-ShPGKxGBRW9} \label{cont} }
\end{figure}

\begin{figure}[thp]
\includegraphics[width=0.48\textwidth]{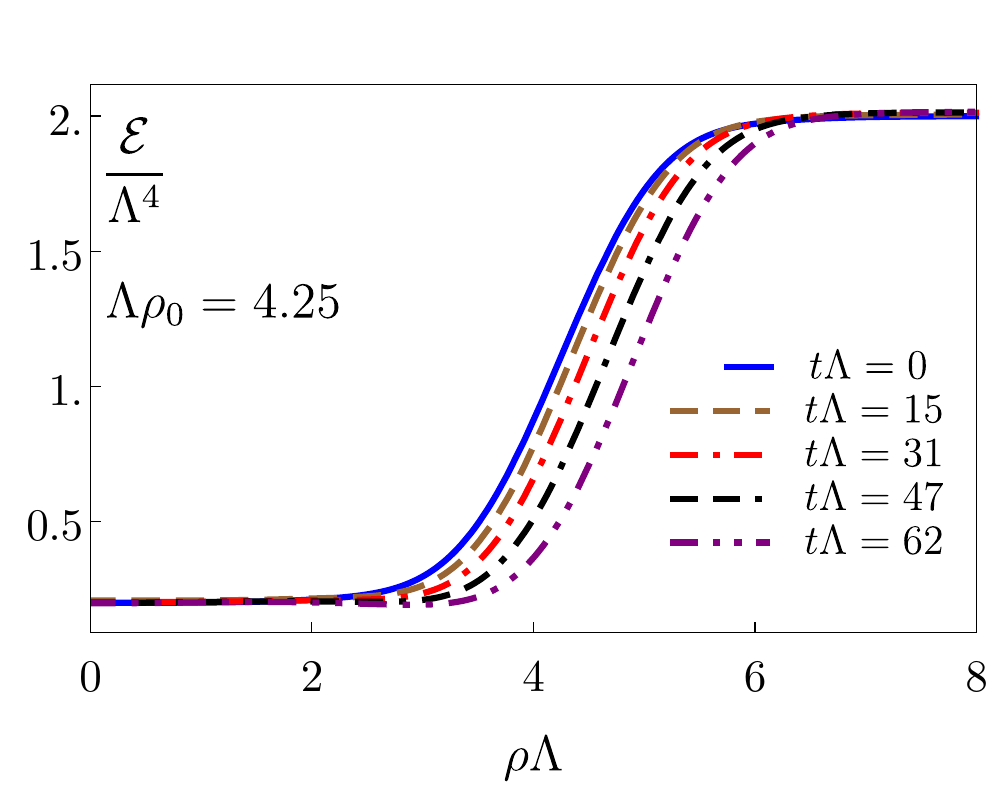} \hfill
\includegraphics[width=0.48\textwidth]{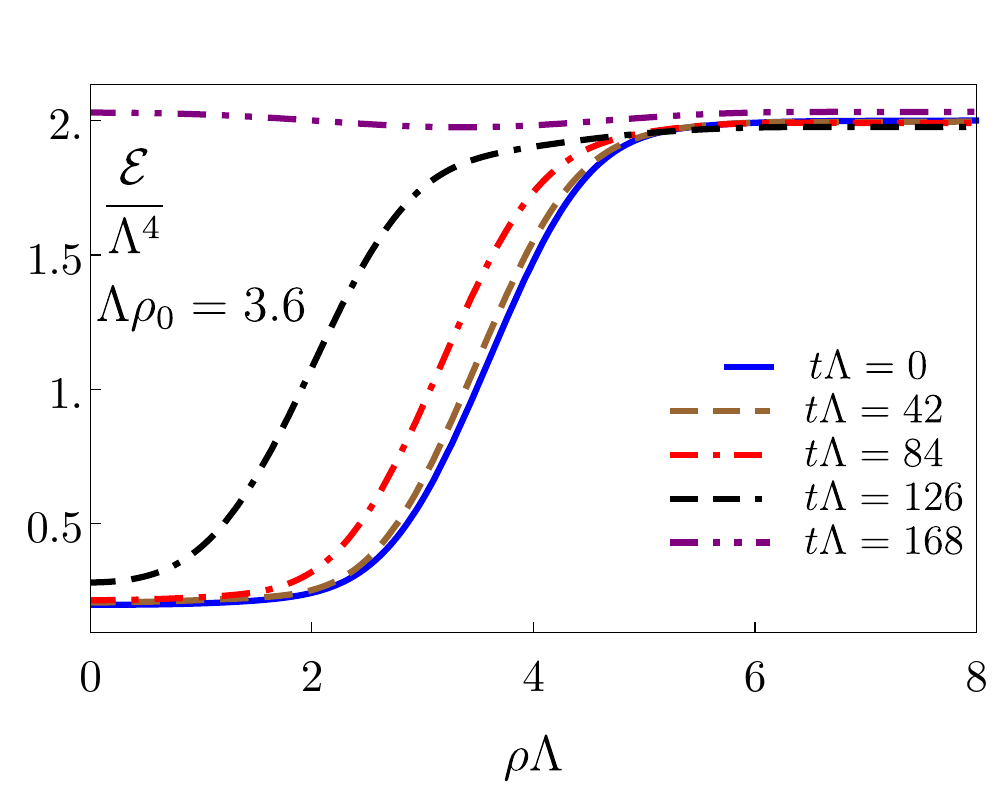} \\
\includegraphics[width=0.48\textwidth]{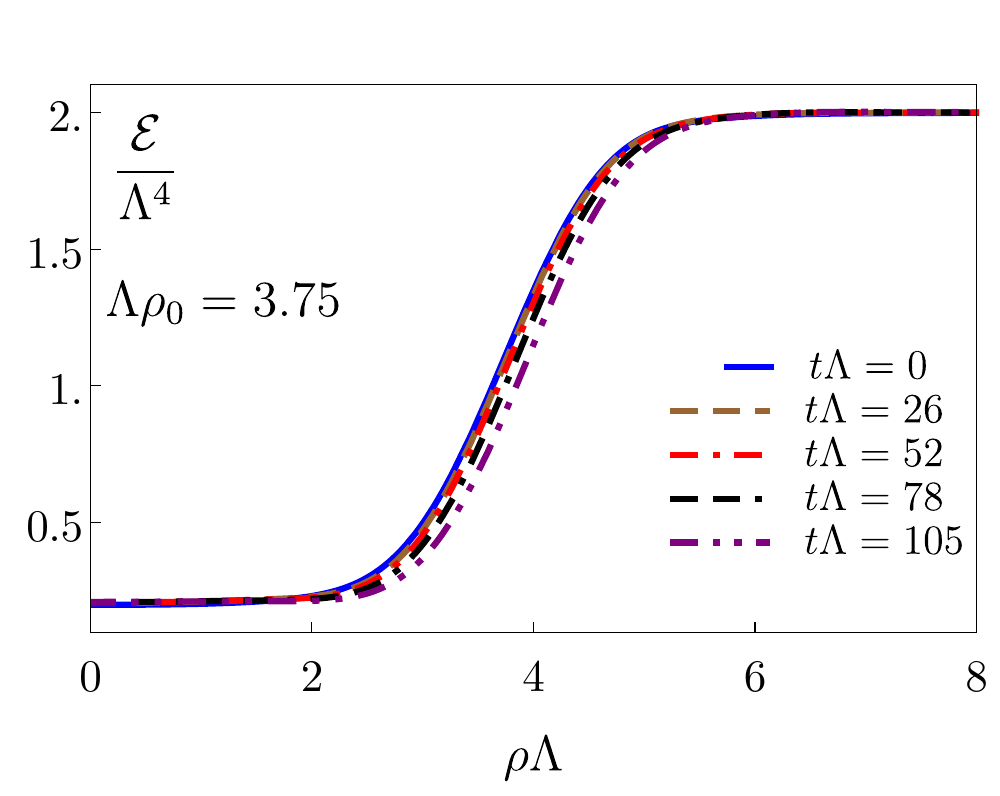} \hfill
\includegraphics[width=0.48\textwidth]{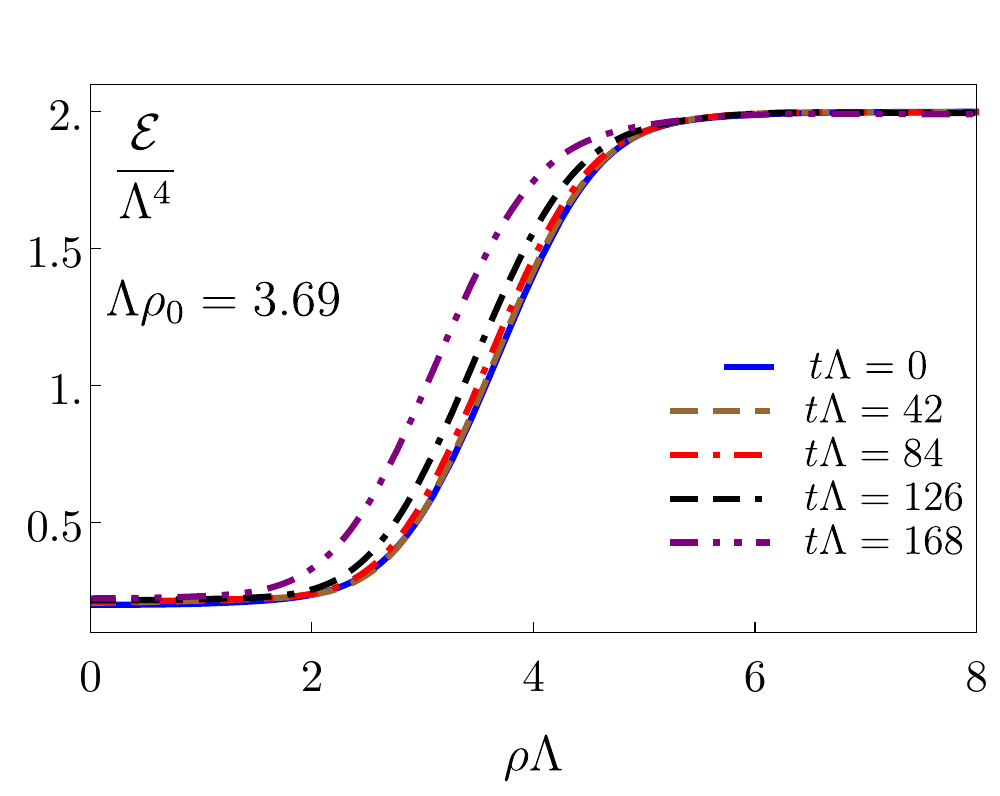}
  \caption{Snapshots of the energy density profile of bubbles with different initial radii.  The radius of the critical bubble lies in the interval $3.69 < \Lambda\rho_c <3.75$. The bubbles on the left column are supercritical and they expand. The bubbles on the right column are subcritical and they collapse. The bubbles in the bottom  row are closer to the critical bubble than those on the top row and  hence they evolve more slowly. \label{snap} 
    }
\end{figure}

\begin{figure}[thb]
\centering
  \includegraphics[width=0.95\textwidth]{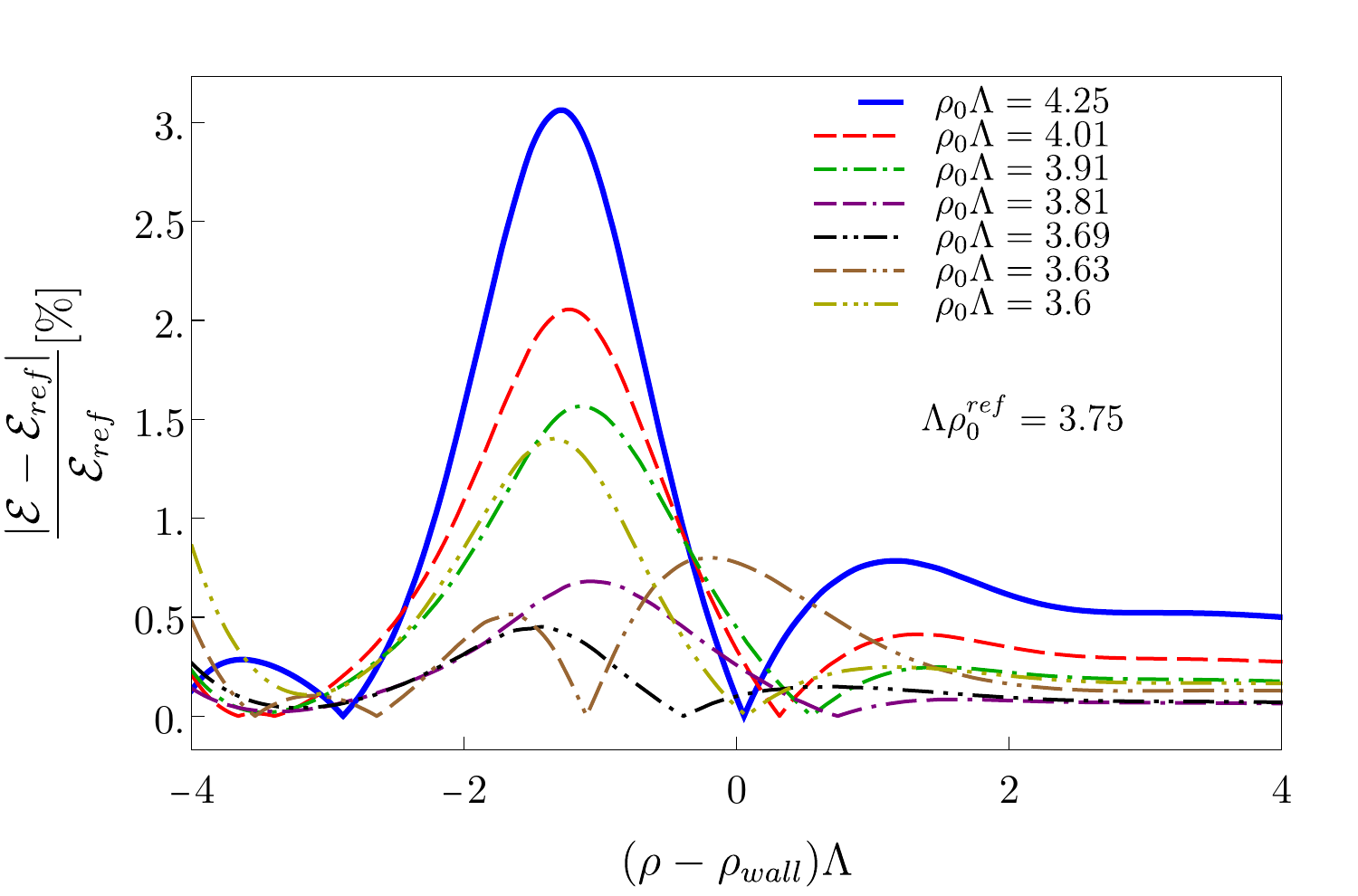}
  \caption{Relative difference between the energy density profiles at $t\Lambda\sim 20$ of bubbles with different initial radii. We take as a reference the profile for a bubble with initial radius $\Lambda\rho=3.75$, which is close to the critical radius. We see that, as this value is approached both from above and from below, the profiles converge to a single profile. \label{compa} }
\end{figure}

As we approach the critical bubble, the dynamics becomes slower and slower. 
This feature can be seen in the 
 contour plots of \fig{cont} and in the  energy density snapshots of \fig{snap}.
In these figures the bubbles in the bottom row evolve more slowly than those in the top row because their initial radii are closer to $\rho_c$. 
By fine-tuning the radius of the initial bubble we can get closer and closer to the critical bubble.  \fig{compa} shows that, as we approach this limit both from above and from below, the bubble profile converges to a single profile. 
In this figure we evaluate the profiles at $\Lambda t = 20$ so that the result is not contaminated by the fast-decaying, transient oscillations present around $\Lambda t=0$ in \fig{lots}. 

The fact that we can approach the critical bubble by fine-tuning a single parameter is consistent with the fact that the critical bubble should possess 
a single unstable mode (see e.g.~\cite{Laine:2016hma,Gould:2021ccf}). Indeed, the latter property means that, in the infinite-dimensional space of configurations around the critical bubble, the hypersurface of stable perturbations has codimension one. As we change a single parameter in our initial data, we trace a curve in the space of configurations that will generically intersect this hypersurface. If we were to start the time evolution exactly on this hypersurface, we would remain within it and we would be attracted to the exact, static critical bubble solution. By tuning the radius of the bubble in our initial data we come close to this situation and therefore the dynamics becomes slower and slower. 
   
Since the critical bubble is a static solution, an alternative method to determine it would be to solve an elliptic problem in two dimensions in AdS, along the lines of \cite{Bea:2020ees}.   
   
\subsection{Expanding bubbles}
\label{expsec}
We now turn to the analysis of expanding  bubbles, which play an important role in the dynamics of first order phase transitions. 
At sufficiently late times, the wall of these bubbles is expected to move with a constant velocity, which results from the balance between the friction that the plasma exerts on the  wall and the pressure difference between the inside and the outside of the bubble. Moreover, the energy density profile should approach a characteristic and time-independent shape when plotted as a function of $\rho/t$. In this section we will use holography to determine both the bubble wall velocity and the asymptotic profile. 

The simulation presented in this section was performed in MareNostrum 4 using 1 node with 48 cores. The typical runtime was around 800h.

\subsubsection{Wall profile, wall velocity and hydrodynamics} 
For computational reasons, it is easier to identify the late-time limit for bubbles that expand at high velocity, since for these configurations the evolution is faster and we need to run our code for a shorter time to reach the late-time,  asymptotic limit. Based on the mechanical picture we described above we expect that, as the pressure difference between the inside and the outside of the bubble  grows, the wall velocity will grow too. Therefore, we will focus on bubbles formed in the large overcooling limit, when the metastable phase is close to the limit of local stability and the pressure difference between the inside and outside of the bubble is the largest. For this reason we will choose the state $A$ outside the bubble as indicated in \fig{Fig:phasediagram}, whereas for the state inside we choose the one indicated as $B$. Following \Sec{sec:indata}, we then construct a bubble that interpolates monotonically between the states $B$ inside and $A$ outside, as in \fig{Fig:initial_E_generic_bubble}. This is our initial state at $t=0$.

 \begin{figure}[thp]
 \begin{center}
  \includegraphics[width=0.7\textwidth]{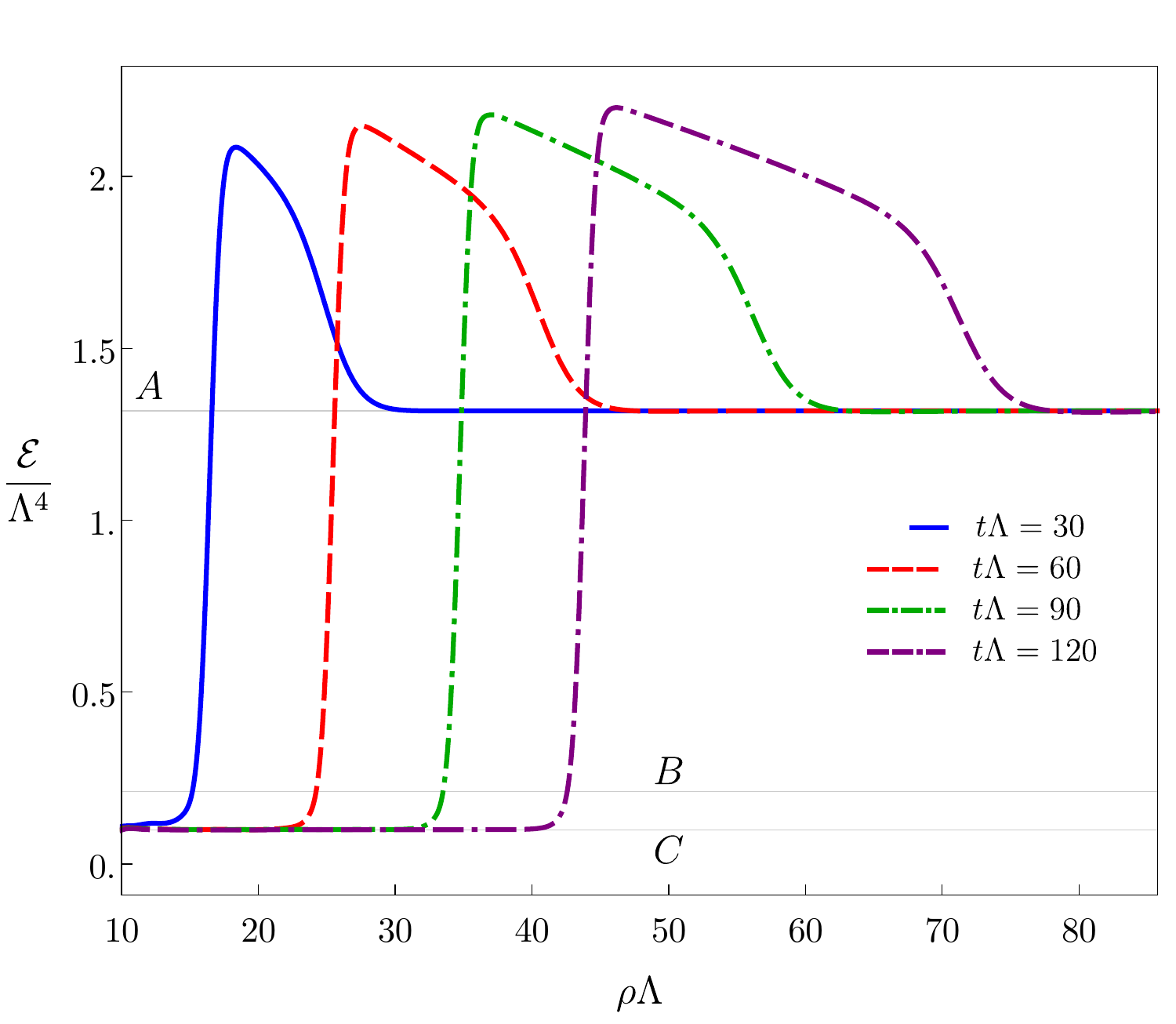}\\
    \includegraphics[width=0.7\textwidth]{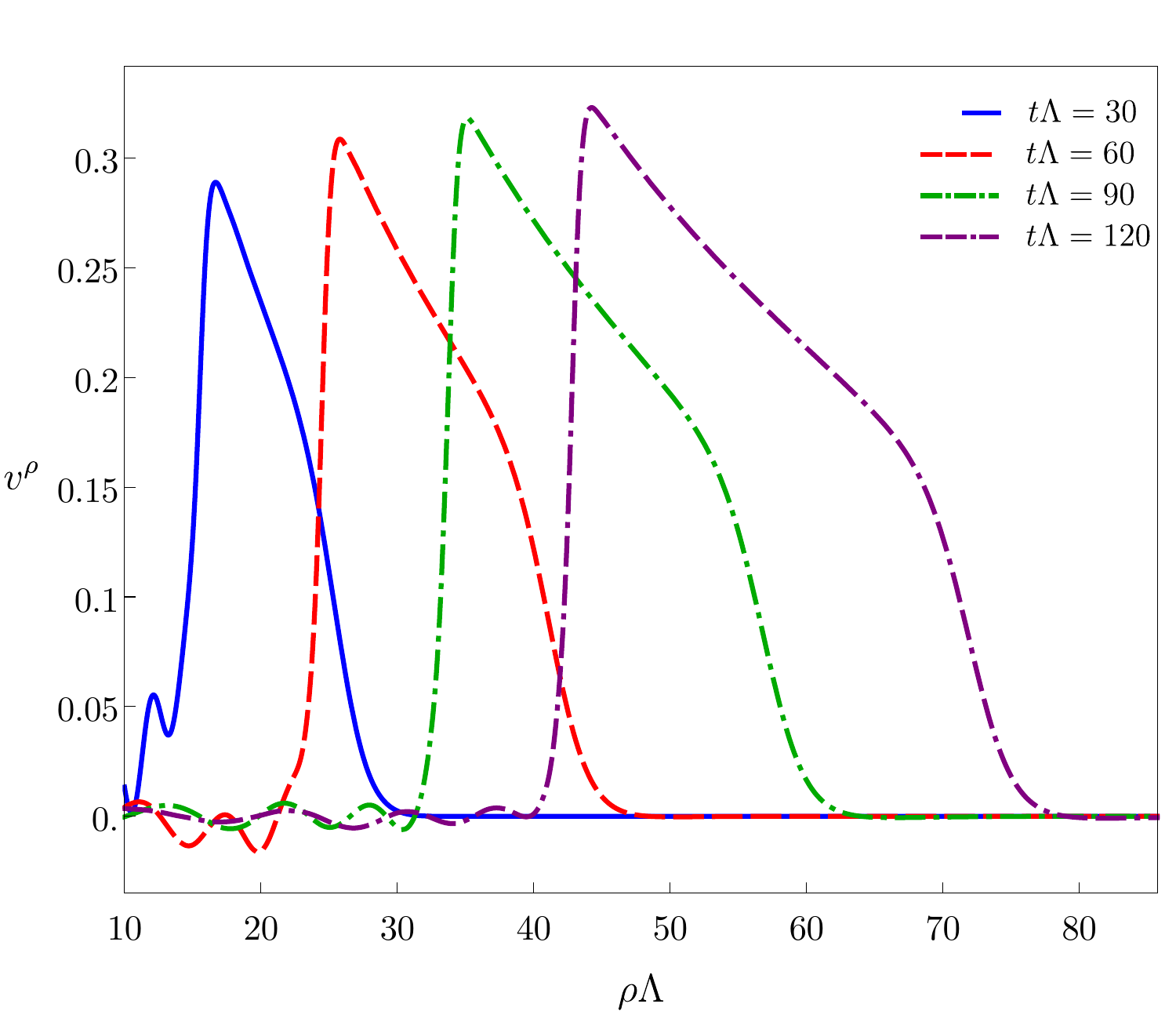}
  \caption{(Top) Snapshots of the energy density profile (top) and of the fluid velocity (bottom) for an expanding bubble. The bubble at $t=0$ interpolates monotonically between the states $B$ inside and $A$ outside, as in \fig{Fig:initial_E_generic_bubble}. At late times the state inside the bubble evolves dynamically to $C$. The states $A, B$ and $C$ are indicated by black dots in \fig{Fig:phasediagram}. A full time evolution video of the energy density can be found at \href{https://youtu.be/wFLp0FSeO8Q}{https://youtu.be/wFLp0FSeO8Q}}
    \label{Fig:expandingbubblesnapshots}
    \end{center}
   \end{figure}
\begin{figure}[htp]
\includegraphics[width=0.48\textwidth]{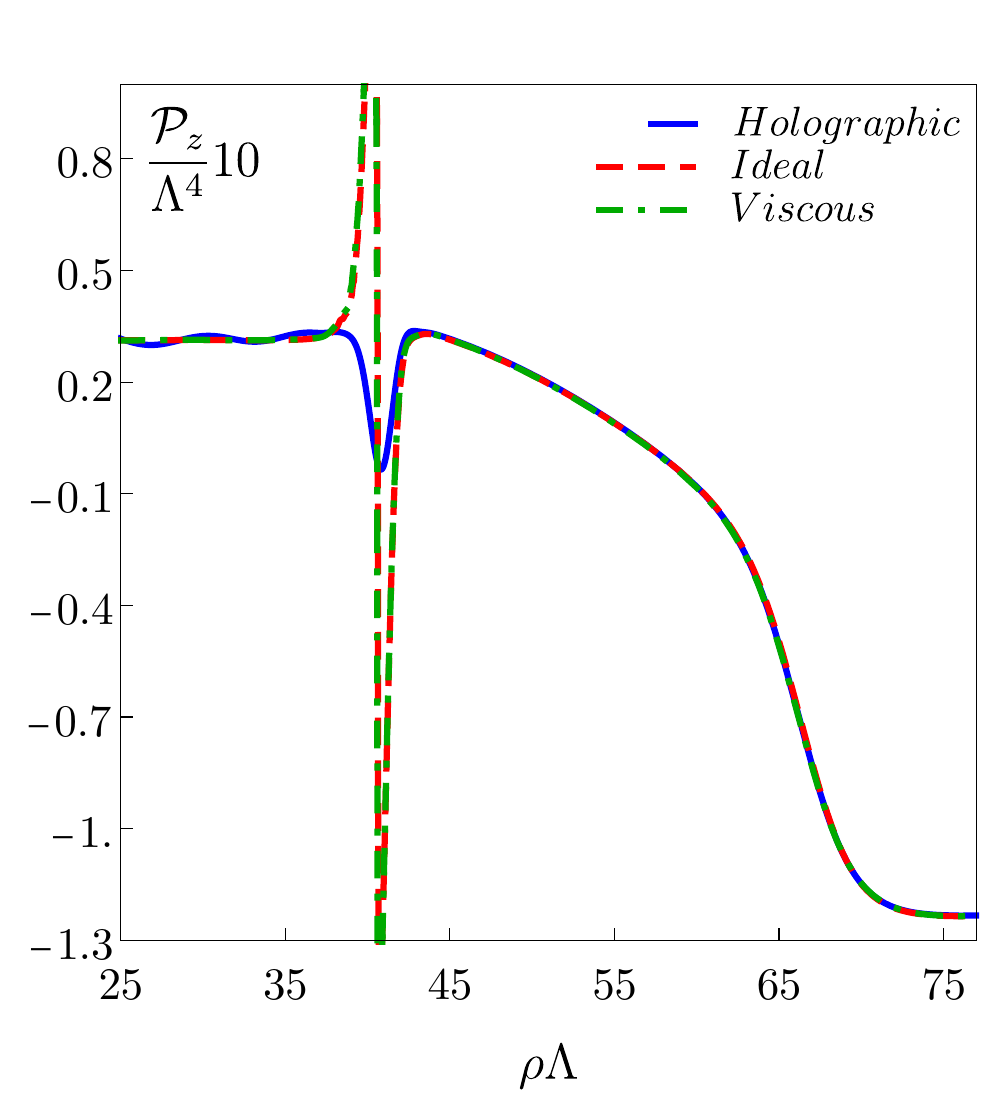} \hfill
\includegraphics[width=0.48\textwidth]{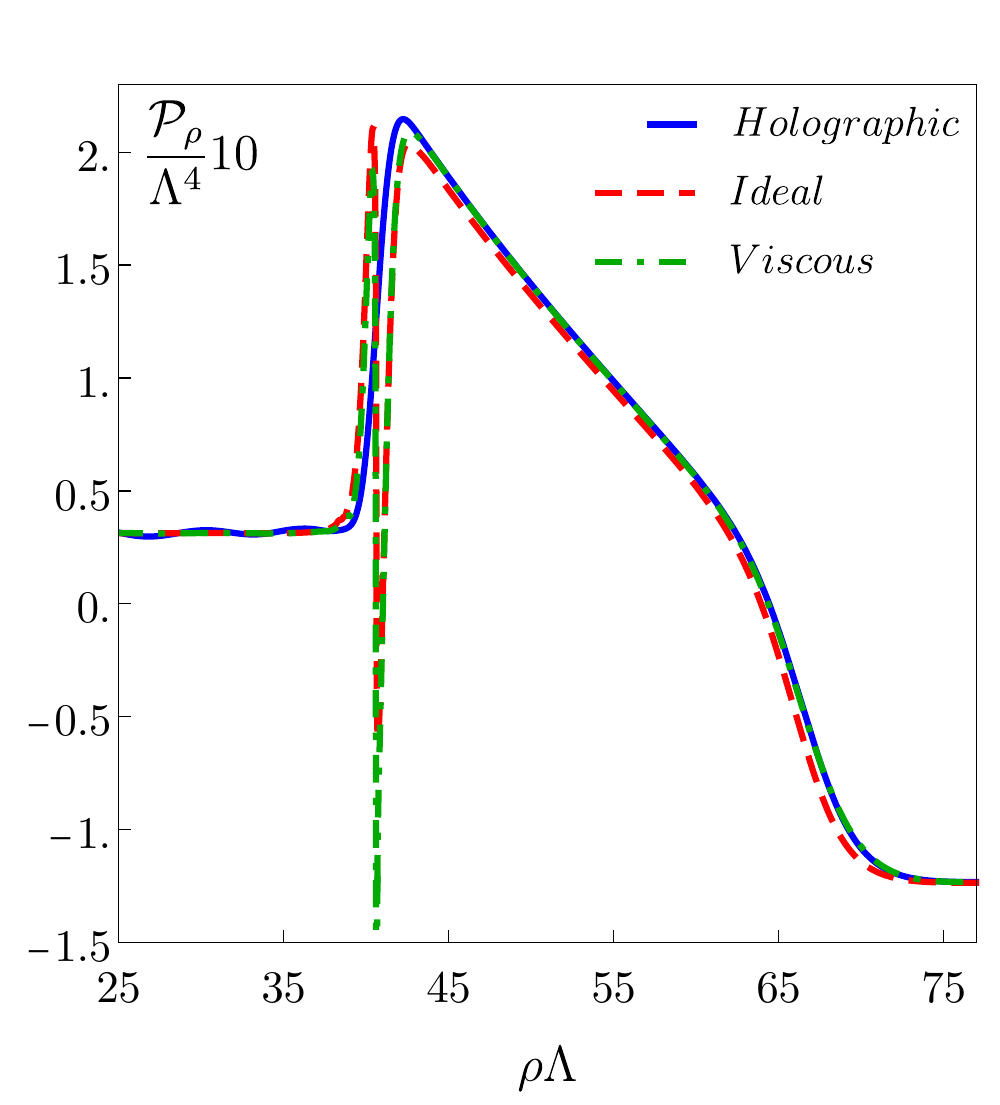} \\
\includegraphics[width=0.48\textwidth]{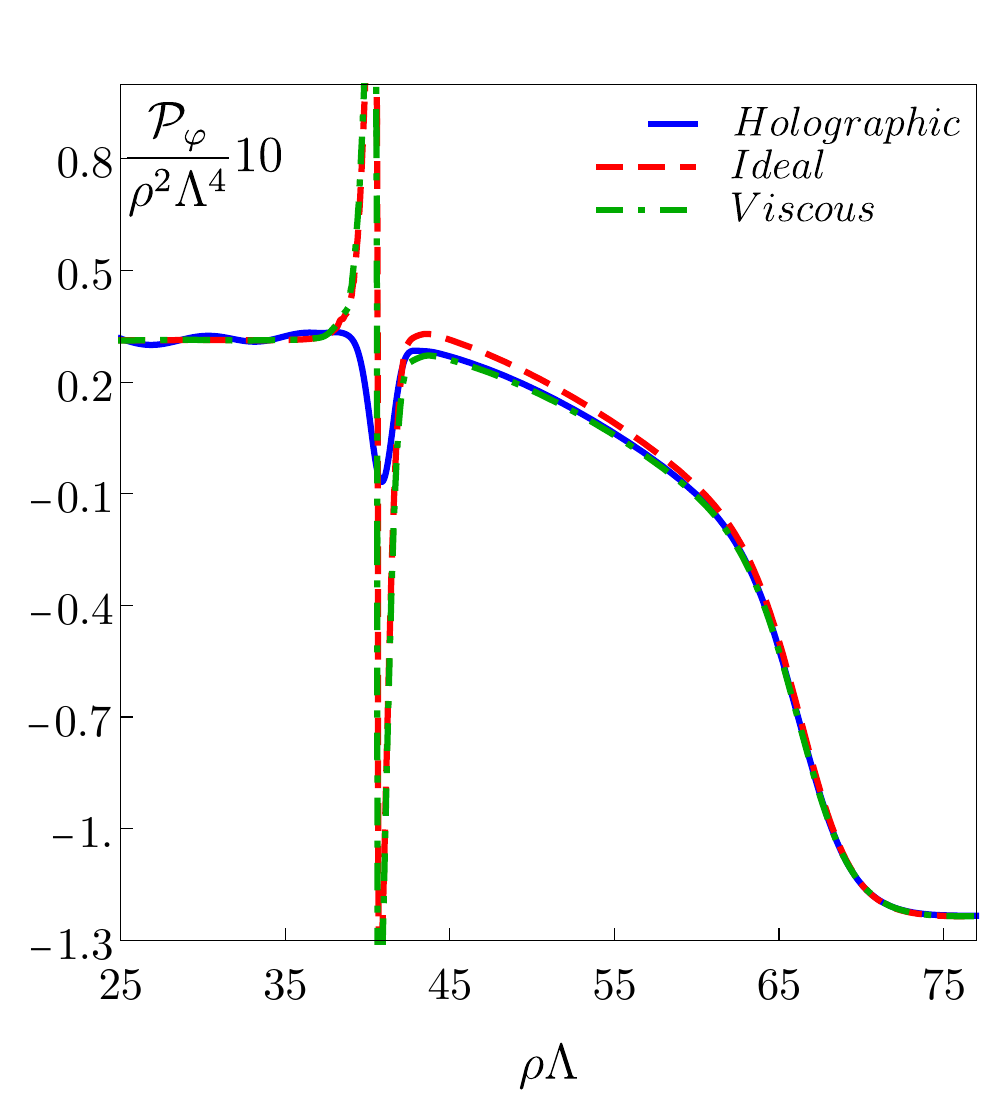} \hfill
\includegraphics[width=0.48\textwidth]{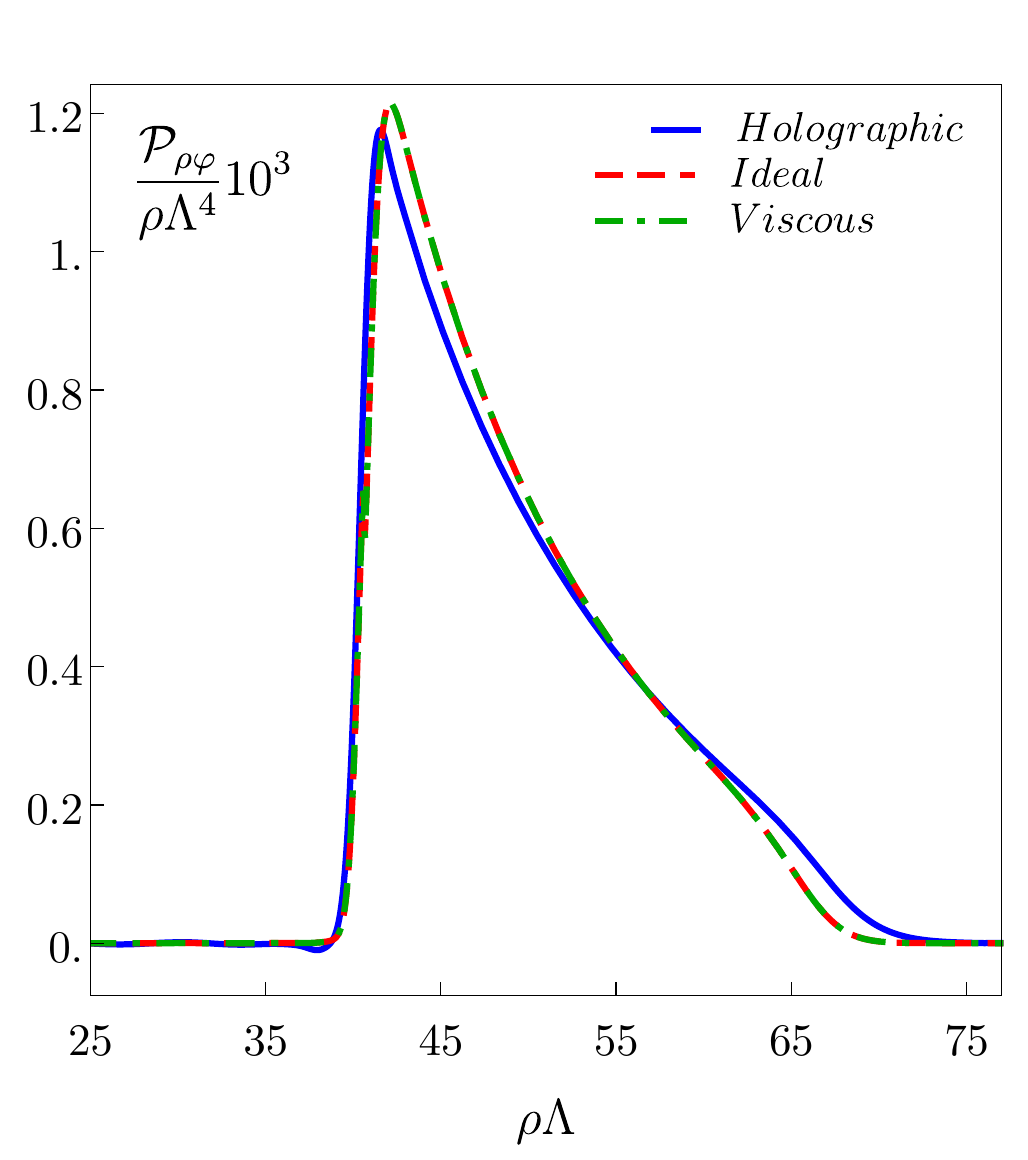}
\caption{Comparison of the holographic stress tensor with the ideal and viscous hydrodynamic approximations based on the constitutive relations at a time $\Lambda t=110$ at which the bubble wall is located at $\Lambda \rho=40.7$. $\mathcal{P}_z, \mathcal{P}_\rho,   \mathcal{P}_\varphi$ and $\mathcal{P}_{\rho\varphi}$ are the stress tensor components in the $zz-, \rho\rho-, \varphi\varphi-$ and $\rho\varphi-$directions, respectively.  \label{fig:hydrocomp}}
\end{figure}

In \fig{Fig:expandingbubblesnapshots}(top) we show snapshots of the subsequent evolution of the energy density of the bubble and in \href{https://youtu.be/wFLp0FSeO8Q}{https://youtu.be/wFLp0FSeO8Q} we show a video of the full time evolution. As time progresses, the energy density in the interior of the bubble evolves until it reaches the value corresponding to the state $C$ in \fig{Fig:phasediagram}. This means that, as in \cite{Bea:2021zsu}, this state is dynamically determined. 
While the initial configuration at \mbox{$\Lambda t=0$}  interpolates monotonically between the stable and meta-stable branches of the phase diagram, the expanding bubbles quickly develop a non-monotonic energy density profile. As illustrated in \fig{Fig:expandingbubblesnapshots}, the propagation of the bubble leads to an overheating of the region in front of the bubble that gradually decreases back to $\mathcal{E}_A$ sufficiently far away from the bubble front. This overheated region  possesses non-vanishing energy and momentum fluxes, which allows us to define a flow velocity via the Landau matching condition, 
\be
T^{\nu \mu} u_\mu=-\mathcal{E}_{\mathrm{loc}} u^\nu \,,
\ee
with $\mathcal{E}_{\mathrm{loc}}$ the energy density of the fluid in the local rest frame. The flow velocity \mbox{$v=u^\rho/u^0$}, with $u^\rho$ the radial component of the flow field,  for this configuration is shown in \fig{Fig:expandingbubblesnapshots}(bottom). As we can see in these figures, the region between the bubble wall and the asymptotic metastable state grows linearly with time as the bubble expands. As a consequence, we expect that, at late times, the gradients of the bubble profile decrease and most of the dynamics is captured by hydrodynamics. 
We can test this expectation by checking the validity of the hydrodynamic constitutive relations for the stress tensor in the Landau frame.
After extracting the rest frame energy density and the fluid velocity from the holographic stress tensor, we can predict the rest of the components of the stress tensor via the constitutive relations with or without viscous corrections. The result of this comparison at $\Lambda t=110$ is shown in \fig{fig:hydrocomp}. 
We see that hydrodynamics becomes a very good approximation for the dynamics of the entire system except for the bubble wall, where the failure of hydrodynamics is expected on general grounds. 
%

\begin{figure}[htp]
\centering
\includegraphics[width=0.95\textwidth]{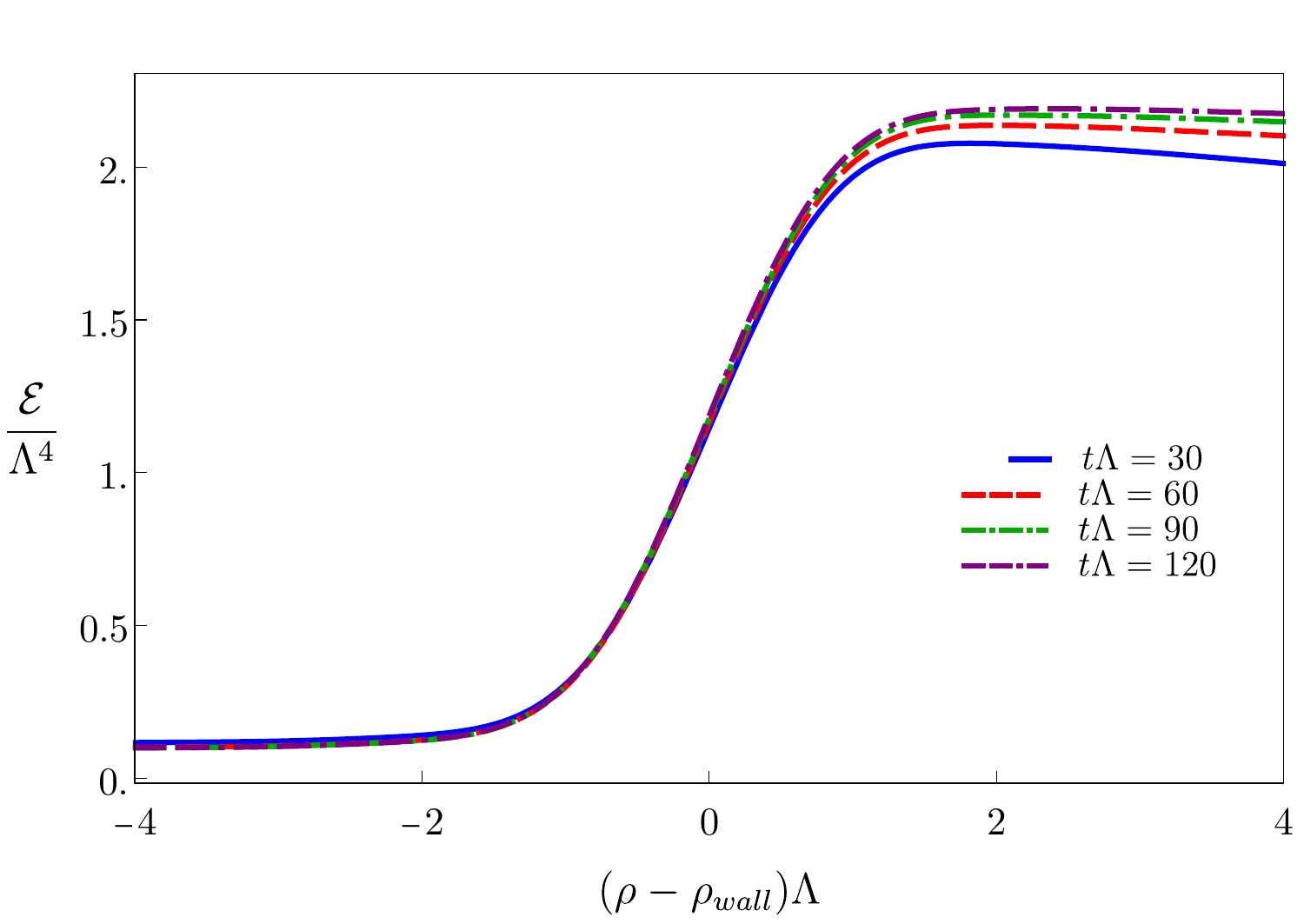}
  \caption{Comparison of the bubble wall profiles at several different times for the expanding bubble of \fig{Fig:expandingbubblesnapshots}. To facilitate the comparison, we shift the position of each curve such that the inflexion point of the different walls at different times coincide with one another.
    \label{fig:rigidwall}
    }
\end{figure}
Despite its non-hydrodynamic nature, the dynamics of the bubble wall becomes remarkably simple at sufficiently late times: it moves almost rigidly at constant velocity. The velocity $v \simeq 0.31$ can be extracted from \fig{Fig:expandingbubblesnapshots} via a linear fit to the wall position of the form 
\be
\label{vwallfit}
\rho_{\mathrm{wall}} (t)=\rho_{\mathrm{wall},0} + v_{\mathrm{wall}} t \,.
\ee
To illustrate the rigidity, in \fig{fig:rigidwall} we compare the bubble wall profiles  at several different times. 
To facilitate the comparison, we shift the position of each curve such that the inflexion point of the different walls at different times coincide with one another. We see that the way that the wall deviates from the inner region $C$ is identical for all sufficiently late times. In contrast, the maximum value of the energy density at the end of the wall grows slowly with time. As we will explain in the next section, this growth indicates that, in the times covered by our simulation, the bubble has not yet reached  the asymptotic late-time form. Despite this, \fig{fig:rigidwall}  shows that the wall has a fixed size set by the microscopic scale of the theory, $\Lambda$. In particular, the size of the wall does not grow with time, in contrast with the overheated  region in front of the bubble wall.  

\begin{figure}[thp]
\centering
  \includegraphics[width=0.95\textwidth]{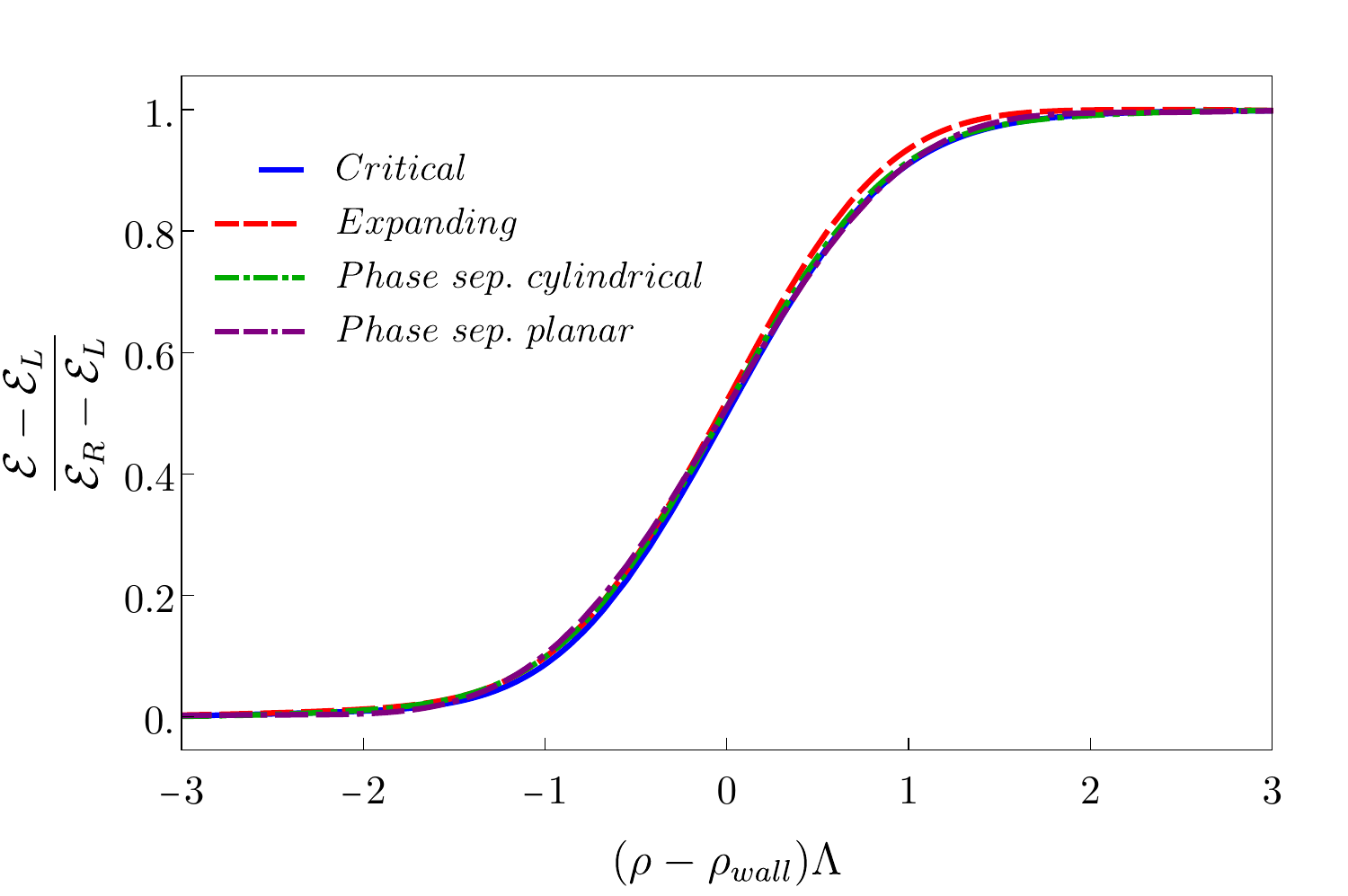}
  \caption{Comparison of wall profiles for several configurations. ``Critical''   refers to the bubble of \fig{cont} with $\Lambda \rho_0=3.75$. 
  ``Expanding'' refers to the bubble of 
  \fig{Fig:expandingbubblesnapshots}. ``Phase sep.'' refers to phase-separated configurations, be they planar or cylindrical. Each profile has been shifted and rescaled so that it interpolates between 0 on the left of the wall and 1 on the right. In the case of the expanding bubble, we define $\mathcal{E}_R$ as the value of the energy density at the maximum located right in front of the wall. 
    \label{Fig:comparison_wall}
    }
\end{figure}
In the case of planar bubbles, Ref.~\cite{Bea:2021zsu} showed that the late-time wall profile only depends on the asymptotic metastable state $A$. In other words, the profile is independent of the initial conditions used to generate the bubble in the first place, as long as they lead to an expanding bubble. We expect the same conclusion to hold for the cylindrical bubbles considered here, but it would be interesting to verify it explicitly. Assuming  this, it is interesting to check how the wall profile of an expanding bubble compares to those of (almost) static walls. For this purpose, in \fig{Fig:comparison_wall} we compare the  profile of the expanding wall of \fig{Fig:expandingbubblesnapshots} with that of the critical bubble of \Sec{critsec} and with the walls of phase-separated planar and cylindrical configurations.  
Following \cite{Bea:2021zsu}, to facilitate the comparison we shift and rescale each profile appropriately so that it interpolates between 0 on the left of the wall and 1 on the right. We achieve this by plotting not just the energy density $\mathcal{E} (\rho)$ but the combination 
\mbox{$(\mathcal{E} (\rho) - \mathcal{E}_L)/(\mathcal{E}_R- \mathcal{E}_L)$}, with $\mathcal{E}_L$ and $\mathcal{E}_R$ the values of the energy density on the left and on the right of the wall, respectively. In the case of the expanding bubble, we define $\mathcal{E}_R$ as the value of the energy density at the maximum located right in front of the wall. We see from the figure that, while  all profiles are fairly similar,  differences can be seen with the naked eye. These are more pronounced in the regions where the second derivative is larger, where they are of the order of 9\%.

\subsubsection{Late-time self-similar solution}
As we have seen, for sufficiently late time the bubble wall becomes rigid and moves at a constant velocity $\vwall$. This implies that the radius of the region inside the bubble grows linearly with time. Since the energy density in this region is lower than that in the asymptotic, metastable phase, this linear growth of the bubble radius must be compensated by a linear growth in the size of the overheated region in front of the bubble. At very late times, when all the microscopic scales become irrelevant, this behaviour leads to a self-similar solution for the bubble that only depends on the ratio $\rho/t$, as described in e.g.~\cite{Espinosa:2010hh}. In this section we study how our numerical solutions approach this late-time self-similar solution. For this purpose, we shift the time and radial coordinates by appropriate amounts 
$t_{\rm{shift}}$ and $\rho_{\rm{shift}}$ that we will define below. In other words, we define
\be
\label{scalingvariable}
\xi = \frac{\rho-\rho_{\rm{shift}}}{t-t_{\rm{shift}}} \,.
\ee 
These shifts are motivated by the fact that our initial configuration has a finite size, and that it takes a certain amount of time for the configuration to become sufficiently close to the late-time asymptotic solution. While at asymptotic times these shifts become irrelevant, we find that this procedure accelerates the convergence to the self-similar regime in our finite-time simulations.  

\begin{figure}[t]
\centering
  \includegraphics[width=0.75\textwidth]{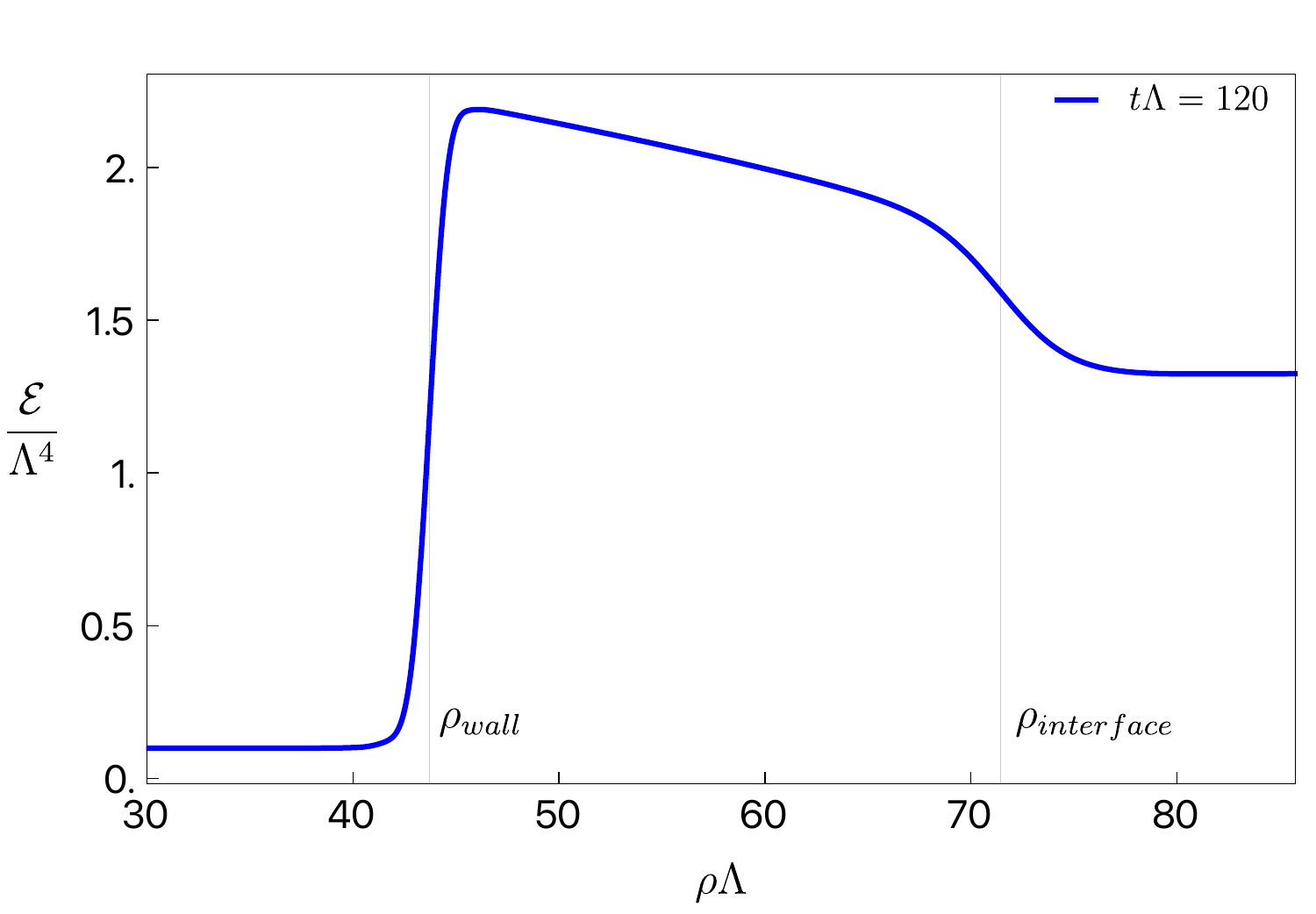}
  \caption{Expanding bubble profile at late times and inflection points, marked with vertical grey lines, used to define the size of the overheated region. 
    \label{inflection}
    }
\end{figure}
The shifts in question are defined as follows. Consider the overheated region in front of the bubble wall. This region is connected with the asymptotic region $A$ by an interface. We begin by locating the inflection point on this interface, indicated by a vertical line at $\rho=\rho_{\mathrm{interface}}$ in \fig{inflection}. We then consider sufficiently late times such that both the wall and the interface positions move with constant velocity. In this regime $\rho_{\mathrm{wall}}(t)$ is given by \eqn{vwallfit} and 
\be
\rho_{\mathrm{interface}}(t) = \rho_{\mathrm{interface},0} + 
v_{\mathrm{interface}} t \,.
 \ee
We then impose that, as soon as this regime starts, the values of $\xi$ at the positions of the wall and of the interface  immediately  agree their late-time limits. In other words, we adjust  the two parameters $t_{\rm{shift}}$ and $\rho_{\rm{shift}}$ so that the following two conditions are satisfied: 
\be
\xi_{\rm{wall}} = v_{\rm{wall}} \sac 
\xi_{\rm{interface}} = v_{\rm{interface}} \,.
\ee


\begin{figure}[tp]
 \begin{center}
  \includegraphics[width=0.9\textwidth]{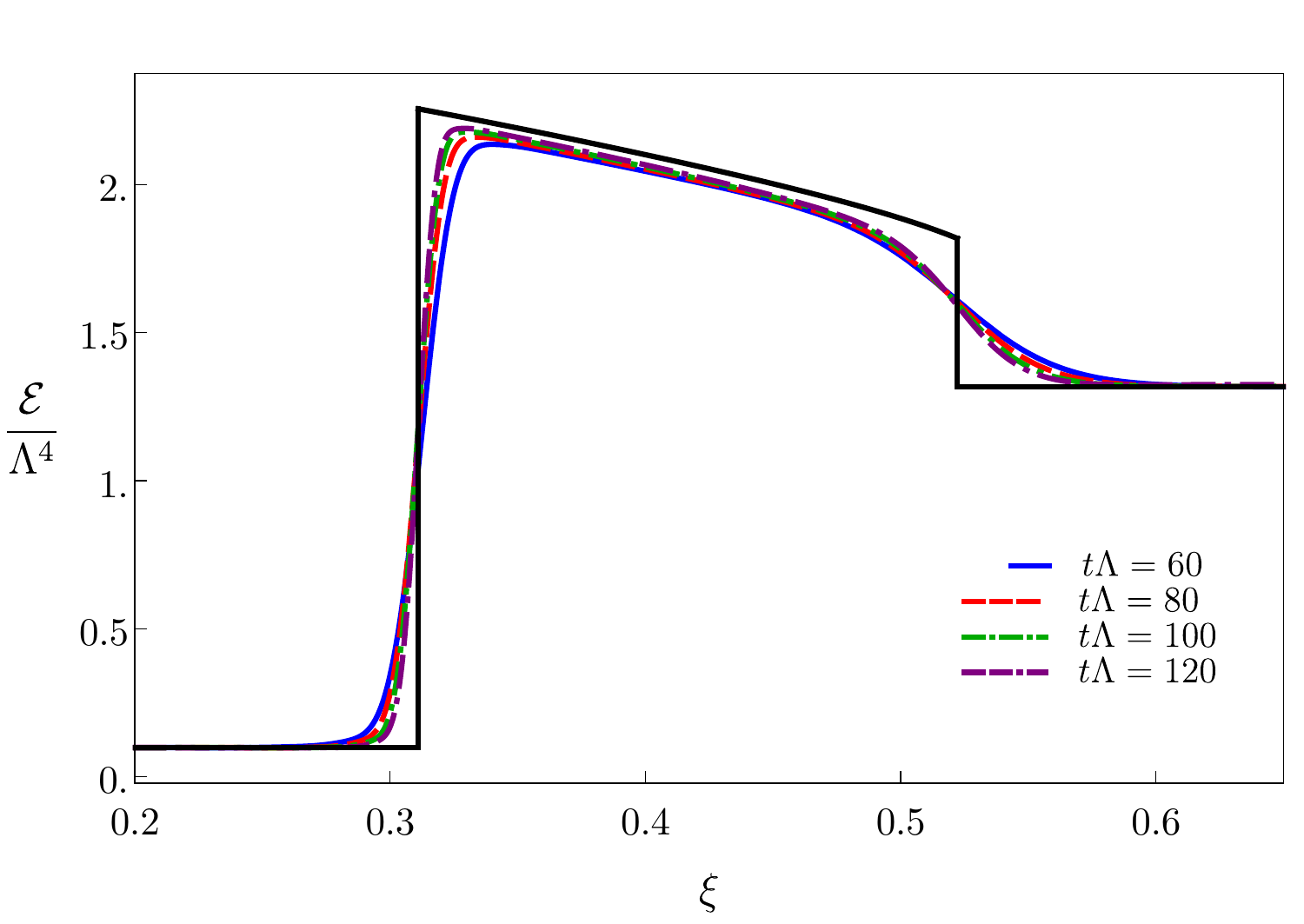}\\ 
    \includegraphics[width=0.9\textwidth]{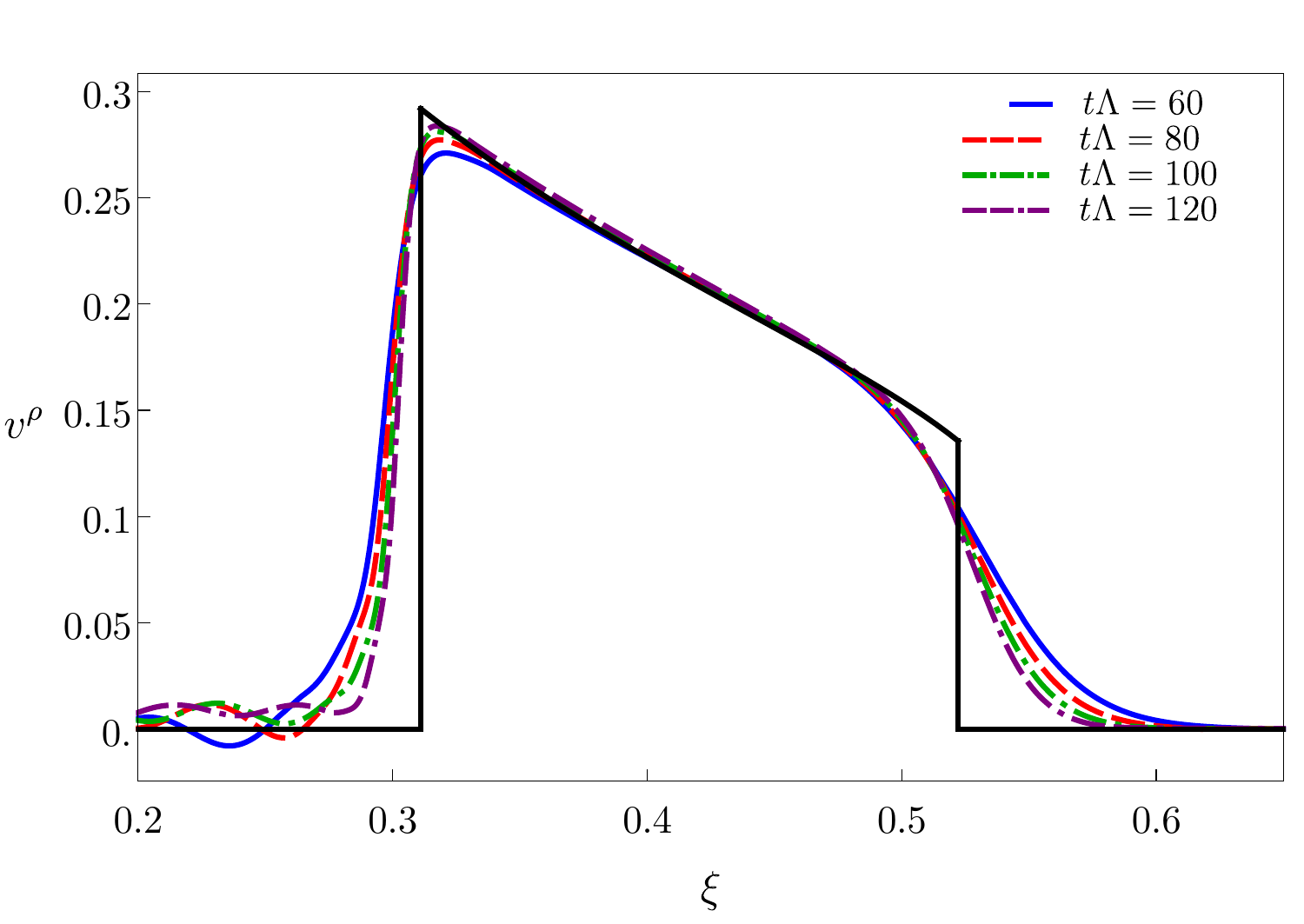}
  \caption{Energy density (top) and fluid velocity (bottom) profiles for different simulation times as a function of the scaling variable \eqn{scalingvariable}. The black solid curves correspond to the ideal hydrodynamic prediction. 
    \label{Fig:selfsimilarbubbles}
    }
  \end{center}
\end{figure}
In \fig{Fig:selfsimilarbubbles} we show the energy density and fluid  velocity profiles for different simulation times as a function of $\xi$. 
In both plots we see two regions of fast change that separate three smooth regions. The first region of fast change occurs around $\xi=\vwall$ and connects the state $C$ in the interior of the bubble, at rest and with a fixed energy density, with the overheated boosted region in front of the bubble. This abrupt behaviour is associated to the presence of the bubble wall. Since the size of the wall remains approximately constant in time, its width in the $\xi$-coordinate decreases with time. As a consequence,  the wall becomes a discontinuity at asymptotically late times.  The shape of the overheated  region in front of the wall is not constant in time. In particular, its slope in the $\rho$-coordinate decreases with time. However, going to the $\xi$-coordinate enhances this slope, since at late times 
$d\mathcal{E}/d \xi \sim t \, d\mathcal{E}/d \rho$. The curves in \fig{Fig:selfsimilarbubbles} indicate that these two effects exactly cancel each other at asymptotically late times, resulting in a constant, non-zero value of the slope in the $\xi$-coordinate in this limit. The second abrupt region occurs at $\xi \simeq 0.52$ and  corresponds to the interface between the overheated region and the asymptotic metastable region $A$. 
In the times covered by our simulations, the width of this interface grows with time, but this growth is slower than linear. However, it is possible that, at sufficiently late times, the width of this interface approaches a constant value. It would be interesting to verify this in the future through longer simulations. In any case,  
 this interface  also approaches a discontinuity in the $\xi$-coordinate at late times. Despite this, both the interface and the overheated region are well described by hydrodynamics at late times, as we saw in \fig{fig:hydrocomp}. 

This discussion suggests that, at asymptotically late times, the bubble profile should consist of a static inner region $C$ and an outer static region $A$ connected through discontinuities with an intermediate overheated  region with non-zero fluid velocity. This behaviour agrees with hydrodynamic analysis of large bubbles, as performed for example in~\cite{Espinosa:2010hh}. At very late times, when the bubble profile depends only on the scaling variable $\xi$, the ideal hydrodynamic equations lead to the following equation for the energy density and the velocity field of a cylindrical bubble
\begin{align}
\gamma ^2 \Big[ 1-\xi  \, v(\xi ) \Big] \Big[ c_s^2 \mu ^2-1\Big] v'(\xi )
-\frac{v(\xi )}{\xi }&=0 \,,
\label{eqxi} \\
\frac{c_s^2 }{\mathcal{W}}\, 
\Big[1-\xi  v(\xi) \Big] \mathcal{E}_{\rm loc}' (\xi)
-\gamma ^2 \Big[ \xi -v(\xi )\Big] v'(\xi ) &= 0\,,
\end{align}
where $\gamma=1/\sqrt{1-v^2}$ is the Lorentz factor, $c_s$ is the speed of sound, 
$\mathcal{E}_{\rm loc}$ is the energy density in the local rest frame of the fluid, 
\be
\mathcal{W} = \mathcal{E}_{\rm loc}+ 
\mathcal{P}_{\rm eq} (\mathcal{E}_{\rm loc}) 
\ee
is the enthalpy density, and 
\be
\mu=\frac{\xi-v}{1-\xi v} \,. 
\ee
It is well known that the ideal hydrodynamic equation \eqn{eqxi} for the fluid velocity does not posses non-trivial continuous solutions with zero velocity in the interior and exterior of the bubble. Therefore, in this approximation the description of an expanding bubble requires the introduction of discontinuities in the hydrodynamic fields. These discontinuities are constrained by energy-momentum conservation: although the local energy density or the fluid velocity may be discontinuous, the energy-momentum flux across the discontinuity must be continuous. For each value of the wall velocity, these ``junction conditions'' at the discontinuities, together with the hydrodynamic equations elsewhere, determine the entire bubble profile in terms of the energy density in $A$. This is the reason why a microscopic model is needed in order to determine the wall velocity. In our case, this model is provided by holography. Using the holographic prediction for $\vwall$ as an input, we have solved the hydrodynamic equations plus the junction conditions and we have determined the profiles represented by the black solid lines in \fig{Fig:selfsimilarbubbles}. The result is consistent with the holographic profiles at late times in the sense that the holographic curves approach the black curves more and more as time progresses. 

Incidentally, these results allow us to define an analogue of ``the state in front of the bubble wall'' for planar bubbles. In the planar case the entire overheated region in front of the bubble has constant energy density and moves with constant fluid velocity 
$v_D$ \cite{Bea:2021zsu}. Using this velocity one can boost the overheated region to its rest frame and thus define a state in the phase diagram of \fig{Fig:phasediagram}. This state was dubbed $D$ in \cite{Bea:2021zsu}, and the state in the overheated region was dubbed $D_{\mathrm{boosted}}$. The difference between $A$ and $D$ gives an intuitive idea of the intensity of the overheating in front of the wall, since in the absence of it we would have $A=D$. In the cylindrical case we can obtain a similar idea by defining the state $D_{\mathrm{boosted}}$ in terms of the maximum values of the black solid curves in \fig{Fig:selfsimilarbubbles} as we approach the bubble wall discontinuity from the right. The values we obtain are 
\be
\E_{D_{\mathrm{boosted}}}= 2.26 \Lambda^4 \sac 
v_D = 0.292 \sac \E_D = 2.06 \Lambda^4 \,.
\ee
The state $D$ is represented by a black dot in \fig{Fig:phasediagram}.

\section{Final remarks}
\label{sec:final}

We have presented a new code called~\texttt{Jecco} (Julia Einstein
Characteristic Code), which is able to evolve Einstein's equations coupled to a
scalar field in asymptotically AdS spacetimes using a characteristic
formulation.
This implementation generalises the one presented in~\cite{Attems:2017zam} to
3+1~dimensional settings and further allows, for instance, the usage of
other choices for the scalar potential $V(\phi)$.
The code is written in the Julia programming language~\cite{Julia-2017} and is
freely available at github~\url{https://github.com/mzilhao/Jecco.jl} and
Zenodo~\cite{jecco-2022}.

\texttt{Jecco} is written in a modular way,
making it 
an interesting tool to attack other physical setups.
Different problems can be implemented as separate Julia modules (containing,
for example, evolution equations, initial data, and diagnostic tools)
which could be tackled by taking advantage of the general infrastructure in \texttt{Jecco} (such as finite-difference and pseudo-spectral derivative operators, filtering tools, and input/output routines).

In the main body of this paper we have presented the formulation, equations of motion, numerical methods, and the corresponding implementation currently present in the code.
%
%
Moreover, in Appendix~\ref{sec:results} we show several tests of this implementation in various setups, including convergence tests, comparisons with analytical solutions and an independent
numerical implementation, recovering thermodynamical and quasi-normal mode
properties of known solutions, and checking the constitutive relations of
hydrodynamics through the fluid/gravity prescription. We obtained very good
results in all the tests performed, which reassures us that the implementation
is working as intended.



The first new physical application of \texttt{Jecco} was the calculation of the gravitational wave spectrum produced by a first-order phase transition that takes place via the  instability of the spinodal branch of the phase diagram of \fig{Fig:phasediagram} \cite{Bea:2021zol}. In this paper we have presented a second application to the dynamics of bubbles in a strongly-coupled four-dimensional gauge theory. This extends  our previous work on planar bubbles \cite{Bea:2021zsu} to cylindrical bubbles and brings about two  new physical aspects. The first one is that the surface tension now plays a role, and therefore a critical bubble exists in which the inward-pointing force due to the surface tension exactly balances the outward-pointing force coming from the pressure difference between the inside and the outside of the bubble. We have shown that our numerical code allows us to construct configurations that are arbitrarily close to this critical bubble. The fact that we can do this with a time evolution code by fine-tuning a single parameter (which we chose to be the radius of the bubble) is compatible with the fact that the space of perturbations of a critical bubble has only one unstable direction. Nevertheless, since the critical bubble is static, it would be interesting to find it by solving an elliptic 2D problem in AdS along the lines of \cite{Bea:2020ees}. This would allow for an efficient exploration of the bubble properties for the entire range of temperatures on the metastable branch. 

The second new physical aspect brought about by cylindrical bubbles is that the asymptotic, self-similar profile of an expanding bubble possesses a richer structure than in the planar case. We have verified this by plotting our holographic result for the gauge theory stress tensor at late times as a function of the appropriate scaling variable. We have also compared the holographic result with the hydrodynamic approximation. As expected, we have found that hydrodynamics  provides a good approximation everywhere except at the bubble wall. 

An immediate extension of this work is to consider multiple expanding bubbles \cite{bubbles}. This is an extremely interesting problem because the resulting bubble collisions will generate gravitational waves. As  in previous applications of holography to the quark-gluon plasma \cite{Casalderrey-Solana:2011dxg,Busza:2018rrf} or to condensed matter systems \cite{Zaanen:2015oix,Hartnoll:2016apf,Nastase:2017cxp}, we expect that the first-principle nature of the holographic approach will shed new light on this problem too.

\acknowledgments
It is a pleasure to thank Bartomeu Fiol, Oscar Henriksson, David~Hilditch, Mark Hindmarsh, Carlos Hoyos, Christiana~Pantelidou and Oriol Pujol\`as for discussions.  YB acknowledges support from the European Research Council Grant No.\ ERC-2014-StG 639022-NewNGR and the Academy of Finland grant no. 333609. TG acknowledges financial support from FCT/Portugal Grant No.\ PD/BD/135425/2017 in the framework of the Doctoral Programme IDPASC-Portugal. 
AJ acknowledges support from the European Research Council Grant No.\ ERC-2016-AvG 692951-GravBHs. MSG acknowledges financial support from the APIF program, fellowship APIF\_18\_19/226. JCS, DM and MSG are also supported by grants SGR-2017-754, PID2019-105614GB-C21, PID2019-105614GB-C22 and the ``Unit of Excellence MdM 2020-2023'' award to the Institute of Cosmos Sciences (CEX2019-000918-M). MZ acknowledges financial support provided by FCT/Portugal through the IF programme grant IF/00729/2015 and CERN project CERN/FIS-PAR/0023/2019, as well as the support by the Center for Research and Development in Mathematics and Applications (CIDMA) through FCT/Portugal, references UIDB/04106/2020, UIDP/04106/2020 and the projects  PTDC/FIS-AST/3041/2020 and CERN/FIS-PAR/0024/2021. We further acknowledge support from the European Union's Horizon 2020 research and innovation (RISE) program H2020-MSCA-RISE-2017 Grant No.\ FunFiCO-777740.
The authors thankfully acknowledge the computer resources, technical expertise and assistance provided by CENTRA/IST. Computations were performed in part at the cluster ``Baltasar-Sete-S\'ois'' and supported by the H2020 ERC Consolidator Grant ``Matter and strong field gravity: New frontiers in Einstein's theory'' grant agreement No.\ MaGRaTh-646597. We also thank the MareNostrum supercomputer at the BSC (activity Id FI-2020-3-0010, FI-2021-1-0008 and FI-2021-3-0010) for significant computational resources.


\begin{appendix}

\section{Tests of \texttt{Jecco}}
\label{sec:results}

To gauge the performance, accuracy and reliability of \texttt{Jecco} we conduct
a number of tests. These tests include comparing the data from numerical
simulations against known analytical results, as well those from the 2+1
\texttt{SWEC} code introduced in~\cite{Attems:2017zam}. We also perform
convergence tests and contrast obtained results against expected physical
quantities and properties of our model systems, such as the black brane entropy
density and the frequencies of its quasi-normal modes.
Unless specifically mentioned, results will be presented in ``code units'', where $G=c=\hbar=L=1$.


We note that we solve the equations of motion of our Einstein-scalar
model~\eqref{eqn:Einstein-scalar_model} using the ingoing Eddington-Finkelstein
gauge (equation~(\ref{eq:metric})), which is a Bondi-like gauge, and the
resulting PDE system is expected to be only weakly
hyperbolic~\cite{Giannakopoulos:2020dih}.
%
We thus restrict our tests to smooth data, 
where the effect of weak hyperbolicity is not
expected to be manifested~\cite{Giannakopoulos:2020dih}.

As mentioned in the main text, for the moment we have only implemented
shared-memory parallelism using Julia's \texttt{Threads.@threads} macro. We have
performed some simple scaling tests with an AMD Ryzen 9 5950X 16-Core Processor
and we see a speedup factor of 2.7 when running with 4 threads, 3.5 with 8 threads, and
4.5 with 16 threads. The bottleneck comes from an operation within the DifferentialEquations.jl package which does not seem to be parallelized. We plan to investigate this further in the near future.

\subsection{Analytical black brane}
\label{subsubsec:conf_homo_BB_test}

In these tests the code is initiated in a homogeneous black brane
configuration, which is a static exact solution of the equations of
motion with $\phi_0=0$ (conformal case). The functions specified in the initial data vanish and the
only non-vanishing boundary data are~$a_4 = -4/3$. For most of these
tests, we do not perform a time evolution but instead we just solve
the whole nested system at $t=0$ and compare the last bulk function to be
computed, that is~$A$, against its analytic form:
\begin{align}
  A = \frac{1}{u^2} + \frac{2 \xi}{u} + \xi^2
  + \frac{a_4 \, u^2}{1 + 2 \, \xi \, u + \xi^2 \, u^2 }
  \,,
  \label{eqn:A_exact_homo_BB}
\end{align}
using the field redefinitions of \Sec{sec:redef}
appropriately. From~\eqref{eqn:A_exact_homo_BB} we see that the gauge
fixing can be performed via
\begin{align}
  \xi = \left(-a_4\right)^{-1/4} - 1/u_{H}
  \,,
  \label{eqn:horizon_fix_homo_BB}
\end{align}
with~$u_{H}=1$ the gauge fixed position of the apparent horizon for
the tested configuration. Since~\texttt{Jecco} provides us with the
possibility of multiple outer spectral domains, we wish to understand
to what extent faster configurations compromise the accuracy of the
numerical solution. We vary the number of nodes in the $u$-domains, as
well as the number of outer $u$-domains, to examine the accuracy of the
code for different configurations of the spectral grid. The 
inner~$u$-domain discretizes the region~$[0,0.1]$ and the outer one the
region~$[0.1, 1.0]$. The domain of both the transverse directions~$x$
and~$y$ is~$[-5,5)$ and is discretized uniformly with 128 nodes in
each case.

\begin{figure}[t]
  \includegraphics[width=1.\textwidth]{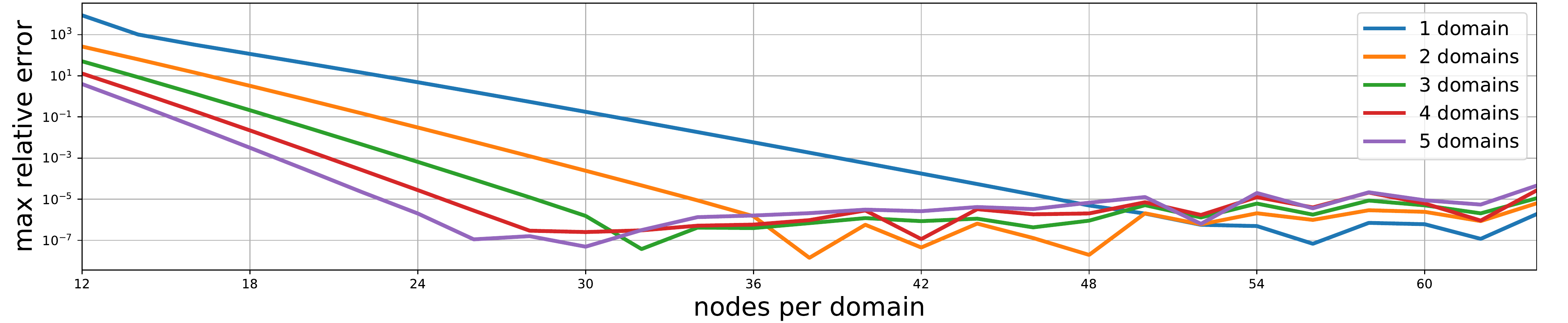}
  \caption{The maximum relative errors for the bulk function~$A$,
    in the outer radial domains,
    for different configurations of the test against the analytical
    homogeneous black brane static solution. The same accuracy for
    this test is achieved e.g.~by three outer radial domains with 32
    nodes per domain, and a single domain with 56 nodes. The former
    configuration is faster.
    \label{Fig:max_rel_errors_Aout}
  }
\end{figure}

The maximum relative error of~$A$ for the inner spectral domain
remains below~$O(10^{-10})$ for a range of nodes between 12 and
36. The respective error for different configurations of outer
spectral domains is shown in \fig{Fig:max_rel_errors_Aout}. A
maximum relative error below~$O(10^{-5})$ in the outer region can be
achieved with one or multiple domains, where the latter typically
provides faster configurations.
%
The difference in orders of magnitude between the maximum
  relative error of the inner and outer domains is due to the near
  boundary field redefinition. This redefinition factors out the near
  boundary radial dependence of the field and allows for a more
  accurate numerical solution.
For completeness, we perform a time evolution for one of the
aforementioned configurations, even if the evolution is expected to be
trivial since we are investigating a static setup. For a configuration
with 12 nodes in the inner domain and 28 nodes on each of the three
outer domains we have verified  that the maximum error maintains its
expected value even after 550 timesteps, which corresponds to~$t_f=2$
in code units. For the time integration the third order Adams-Moulton
method with adaptive step is used.

For a generic physical setup we find that some experimentation may be required to find the optimal numerical parameters, like the number of outer domains and nodes per domain, the choice of time integrator, etc.
For instance, if accuracy of temporal derivatives of the solution is important
one might consider chosing a fixed timestep integrator with a small timestep instead of an adaptive one. If the main focus is the late-time behaviour of the
solution, perhaps an adaptive step integrator is preferable.



\subsection{Comparison with \texttt{SWEC}}
\label{subsubsec:pert_BB_test}

For this test the code is initialized with an $x$-dependent perturbation on top
of a homogeneous black brane configuration. The initial data are
\begin{equation}
  \begin{aligned}
  B_1(0,u,x,y) & = 0.01 u^4 \,,\\
  a_4(0,x,y) & = -\frac{3}{4} \left[1
  +  \delta a_{4}
  \cos
  \left(
  2 \pi k_x \frac{x-x_{\textrm{mid}}}{x_{\textrm{max}}-x_{\textrm{min}}}
  \right)
\right] \,,\\
  \xi(0,x,y) & = \left(\frac{4}{3}\right)^{1/4} - 1\,,
\end{aligned}
\end{equation}
where~$\delta a_4 = 5 \cdot 10^{-4}$, and
the remaining free data functions ($B_2$, $G$, $\phi$, $f_{x2}$, $f_{y2}$)
are set to zero.
We compare the error of the numerical
solution provided by~\texttt{Jecco} against that of the \texttt{SWEC}
code used in~\cite{Attems:2017zam}, for the same setup.


We use one inner radial domain spanning the region~$ u \in [0,0.1]$
discretized with 12 grid points, and another (outer) domain spanning
the region~$u \in [0.1, 1.01]$ with 48 grid points. The transverse
direction $x$ spans $x \in [-10, 10)$, which is discretized with 128
grid points, while the $y$ has trivial dynamics for this setup (and 6
grid points are used so that the finite difference operator fits in
the domain).
The time evolution is performed using the fourth-order accurate
Adams-Bashforth method.
The evolution is performed for a total of 2000 time steps.  The choice
of a single outer radial domain in~\texttt{Jecco} is made for a more
explicit comparison against~\texttt{SWEC}, since the latter does not
offer the possibility of multiple outer radial domains. It is worth
noticing, however, that there are still differences between the setups in
the two codes. For instance, the inner and outer domains
of~\texttt{Jecco} share only one common radial point, whereas
in~\texttt{SWEC} there is an overlapping~$u$-region between them.

\begin{figure}[t]
  {\includegraphics[width=0.47\textwidth
    ]{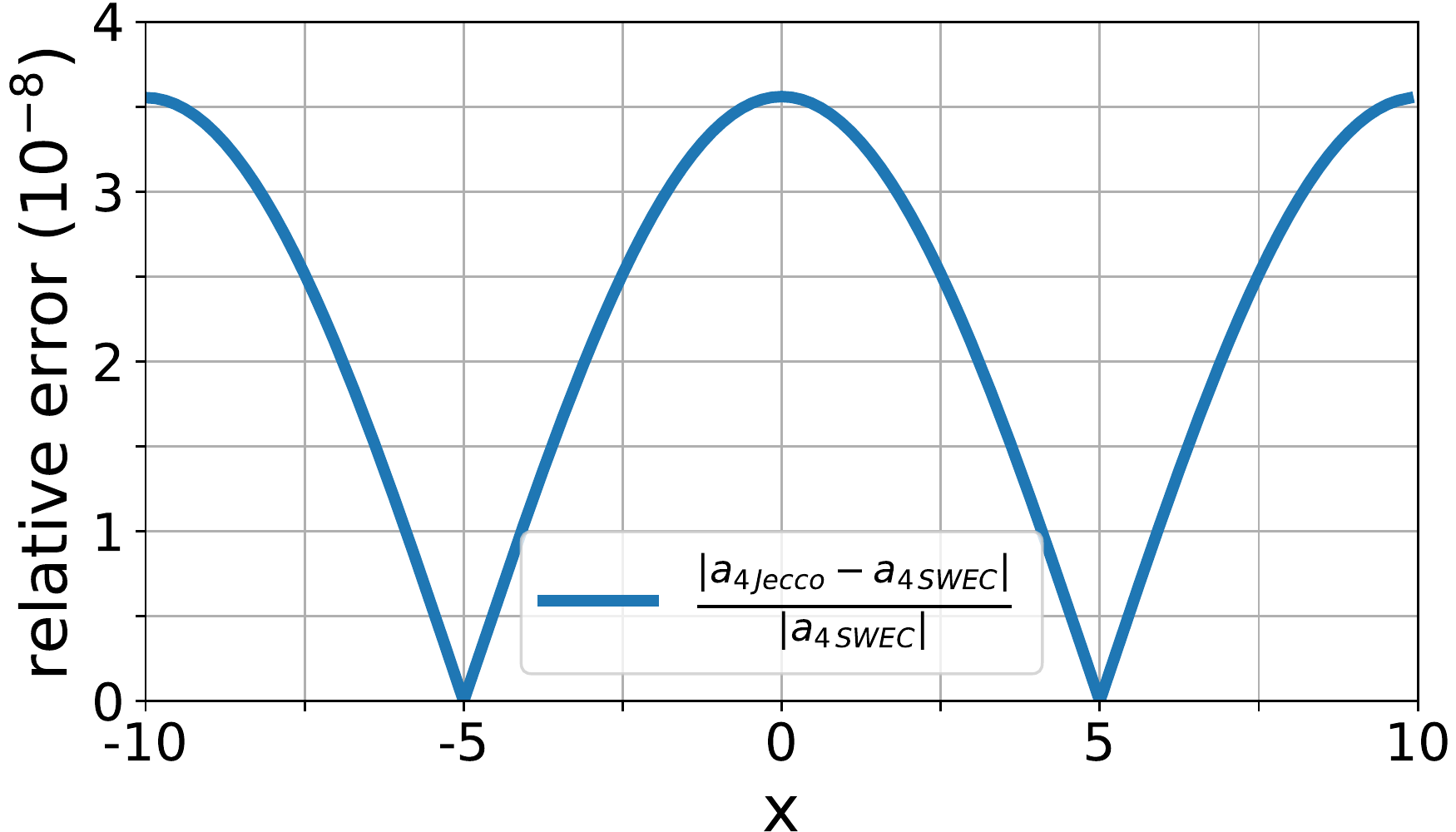}}
  \hfill
  {\includegraphics[width=0.47\textwidth
    ]{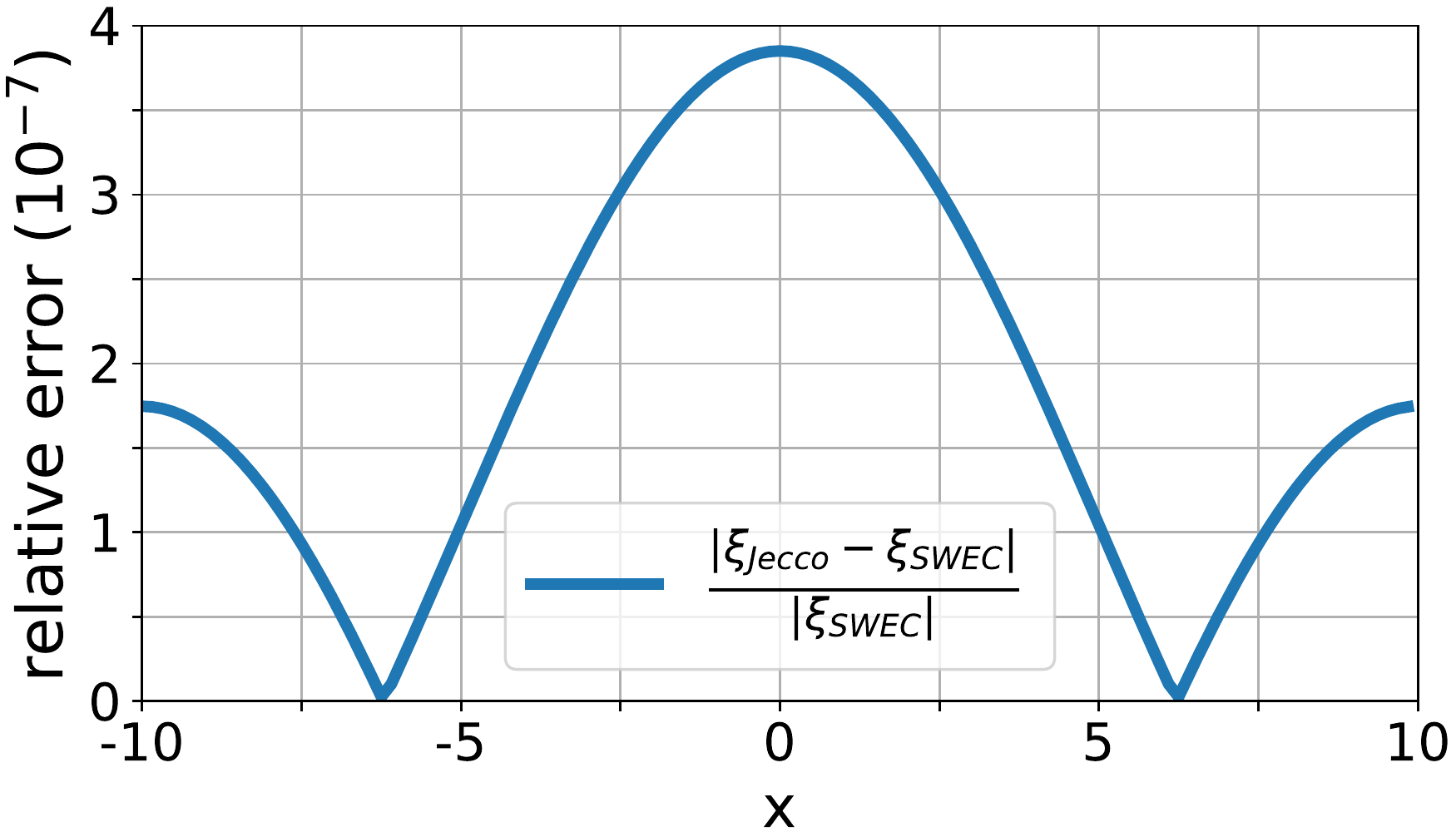}}
  \caption{Relative errors for the $a_4$ and $\xi$ functions
    at the end of the evolution. Results obtained with the \texttt{SWEC} code
    are used as benchmark.
    \label{Fig:benchmark_jecco_swec}
  }
\end{figure}

We show relative differences between the $a_4$ and $\xi$ functions
obtained in the two codes in
\fig{Fig:benchmark_jecco_swec}.
The pattern observed was similar for the metric function $B_1$.
To compare the output of the two codes exactly on the same grid points
we perform cubic spline interpolation on the data and use the values
of the interpolated functions for the comparison. It is reassuring
that the results from the two codes agree so well.


\subsection{Convergence tests}
\label{subsubsec:BB_advectxi}

We now show
\textit{convergence} tests using numerical solutions
obtained only from~\texttt{Jecco}. For this, we solve the same physical
setup with increasing resolution and inspect the rate at which the
numerical solution tends to the exact one.
The rate at which numerical error tends to zero with increasing
resolution is determined by the approximation~\textit{accuracy}. The
latter is the degree to which a discretized version of a PDE system
approximates the correct continuum PDE system, and such a discretized
version is called~\textit{consistent}. If its numerical solution is
bounded at some arbitrary finite time by the given data of the problem
in a discretized version of a suitable norm, it is furthermore
called~\textit{stable}. The Lax equivalence theorem states that
consistency of the finite difference scheme 
and stability with respect to a specific norm
guarantee convergence for linear problems (and the
converse)~\cite{LaxRic56}.


For our present case, since the spatial discretization is performed with a mixture of
finite-difference and pseudo-spectral techniques,
we fix the number of grid points along the spectral direction
and vary only the number of grid points in the uniform grid
along the transverse directions $x,y$.
The finite-difference operators dominate the numerical error, so
the expected convergence rate is
controlled by the rate at which we increase the resolution in the
uniform grid, as well as the approximation order of the operators.

Let us denote by~$f$ the solution to the
continuum PDE problem and by~$f_h$ its numerical approximation. We have
\begin{align}
  f = f_h + O(h^n)
  \,,
  \label{eqn:numerical_error}
\end{align}
where $h$ is the grid spacing and $n$ the accuracy of the finite-difference
operators.


Consider performing
numerical evolutions with coarse, medium and fine
resolutions $h_c$, $h_m$ and $h_f$ respectively. Then one can
construct the quantity
\begin{align}
  Q \equiv \frac{h_c^n - h_m^n}{h_m^n-h_f^n}
  = \frac{f_{h_c} - f_{h_m}}{f_{h_m}-f_{h_f}}
  \,,
  \label{eqn:convegence_factor}
\end{align}
often called the~\textit{convergence factor}, which informs us about the
rate at which the numerical error induced by the finite-difference
scheme converges to zero.
%
Comparison of grid functions corresponding to different
resolutions is to be understood by the use of the common grid
points among the different resolutions.


Using a physical setup with
known exact solution provides a clear benchmark to compare with,
and we can prepare such a setup by evolving a homogeneous black brane
with only gauge dynamics.
This can be achieved by using a different choice for the evolution of the gauge function~$\xi$ than the one specified in \Sec{sec:gauge-fixing}. In particular, we impose the advection equation
\begin{align}
  \partial_t \xi(t,x,y) = - v_x \, \partial_x \xi(t,x,y)
  \,, \label{eq:advect-xi}
\end{align}
which introduces non-trivial dynamics to the numerical evolution.


The only non-vanishing initial data for this setup is the boundary
function~$a_4$, which we set to $a_4(t,x,y) = -1$, and the gauge function~$\xi$, which we initialize to
\begin{align}
  \xi(0,x,y) = \xi_0 + A_x \sin\left( \frac{2 \pi \, n_x}{L_x}
  \left( x_{\textrm{max}} - x \right) \right)
  \,,
  \label{eqn:advect_xi_initial_xi}
\end{align}
where $L_x \equiv   x_{\textrm{max}} -   x_{\textrm{min}} $.
For such a configuration, the solution to equation~(\ref{eq:advect-xi}) is
\begin{align}
  \xi(t,x,y) = \xi_0 + A_x \sin\left( \frac{2 \pi \, n_x}{L_x}
  \left( x_{\textrm{max}} - x + v_x t \right) \right)
  \,,
  \label{eqn:advect_xi_sol}
\end{align}
and the exact solution of the metric function~$A$ is given
by~\eqref{eqn:A_exact_homo_BB}, where $\xi$ is now provided
by~\eqref{eqn:advect_xi_sol}.

For the tests presented herein we have fixed
\begin{align*}
  \xi_0 = 0
  \,, \quad A_x = 0.1
  \,, \quad n_x = 1
  \,, \quad
  x_{\textrm{max}} = 5
  \, \quad x_{\textrm{min}} = -5
  \,.
\end{align*}
For the numerical discretization we have employed one inner radial domain with
12 grid points (spanning the region~$u \in [0,0.1]$) and three equal-sized outer
domains for the region~$u \in [0.1, 1.2]$ with 28 grid points each.
%
%
For the transverse directions we use 16, 32, and 64 grid points for
coarse, medium and fine resolution respectively. The time integration
is done with the third-order accurate Adams-Moulton method, with
adaptive timestep. We have performed these tests with both
second- and fourth-order accurate (periodic)
finite difference operators, where
Kreiss-Oliger dissipation is used with the
prescription of equation~(\ref{eq:KO-op}) with~$\sigma =
0.01$. 
The tests were run on a laptop 
  with an Intel Core i7-10510U at 1.80GHz CPU. For the fourth-order
  accurate finite difference case, the coarse resolution ran
  with a single thread and was completed within 36 minutes. The
  corresponding medium and high resolution cases
  were performed with two threads 
  and were
  completed within 66 and 271 minutes, respectively.


\begin{figure}[t]
  {
    \includegraphics[width=0.235\textwidth
    ]
    {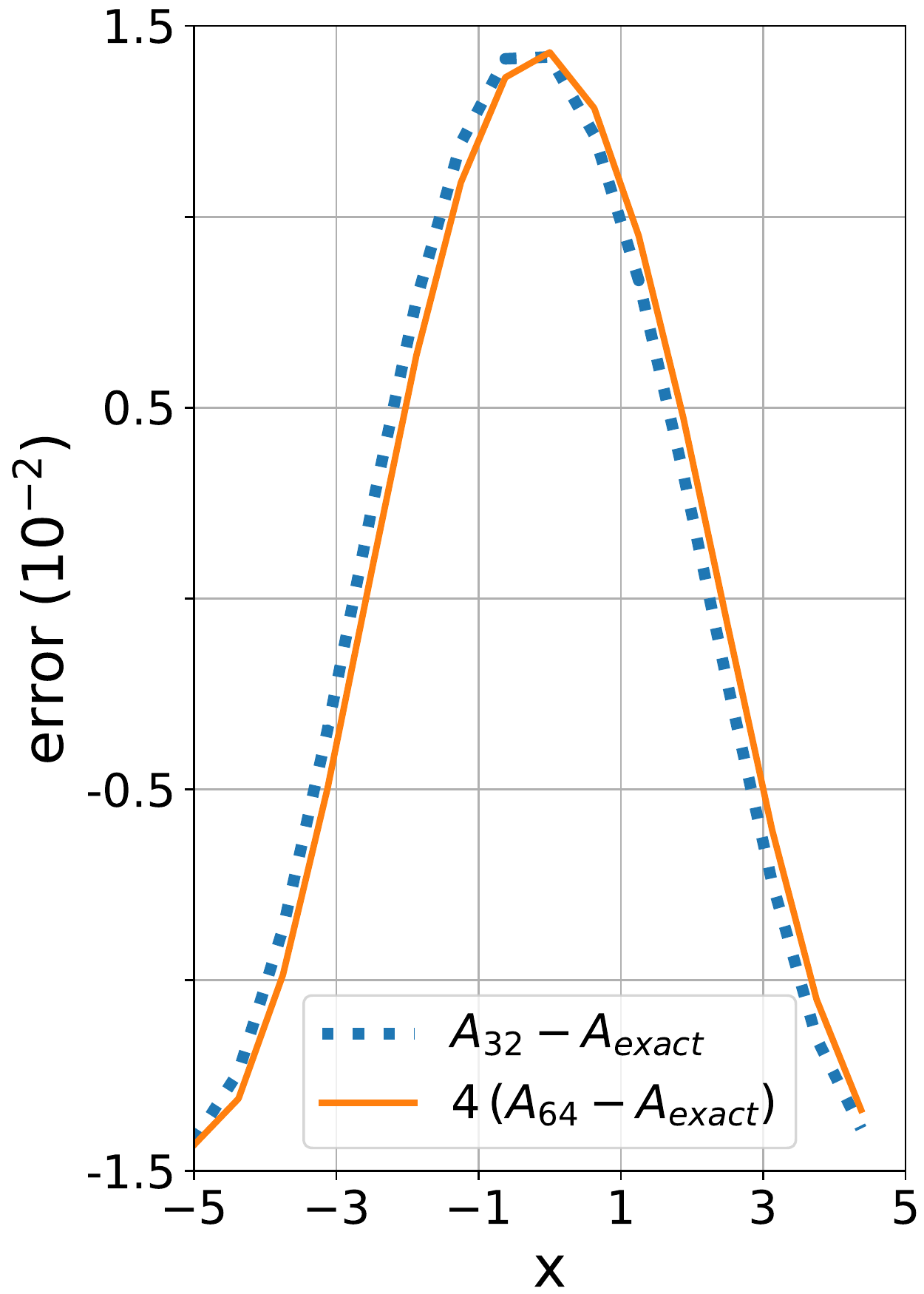}
  }
  {
    \includegraphics[width=0.76\textwidth
    ]
    {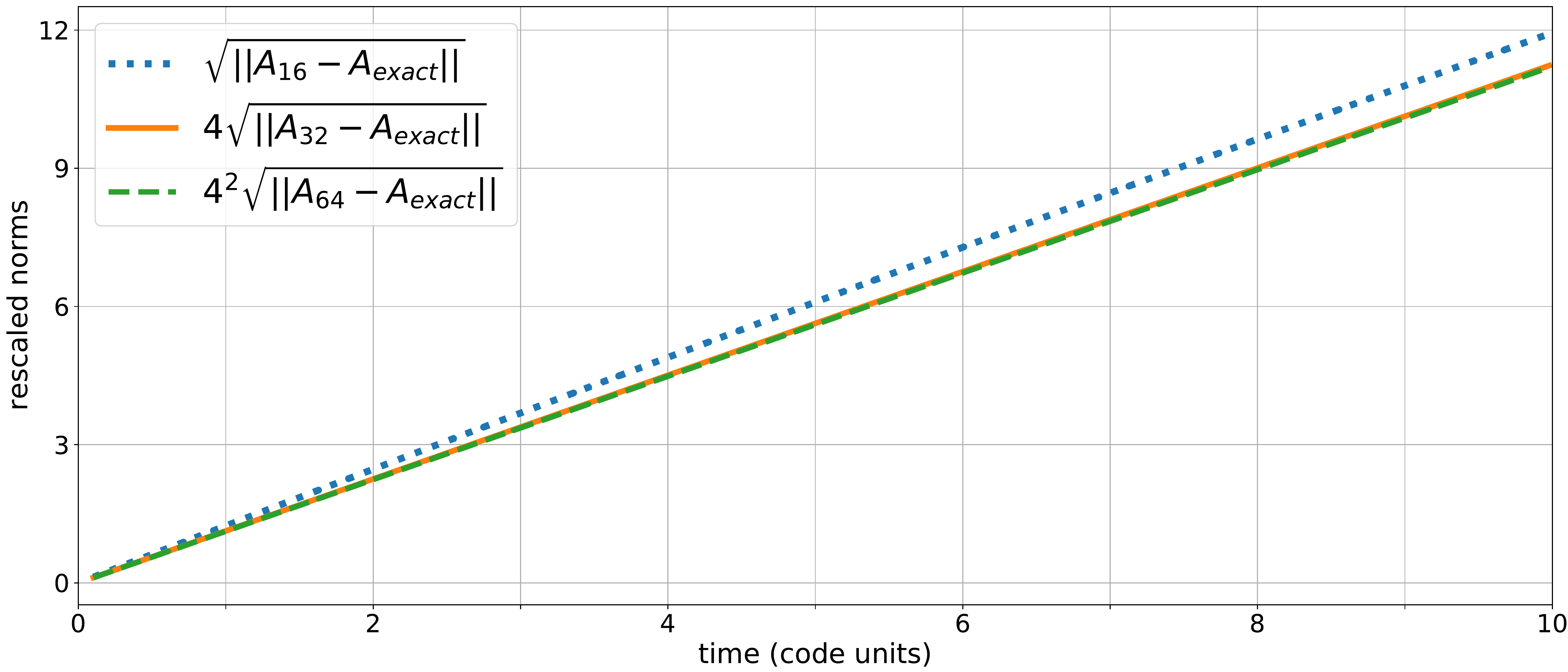}}
  \\
  {
    \includegraphics[width=0.235\textwidth
    ]
    {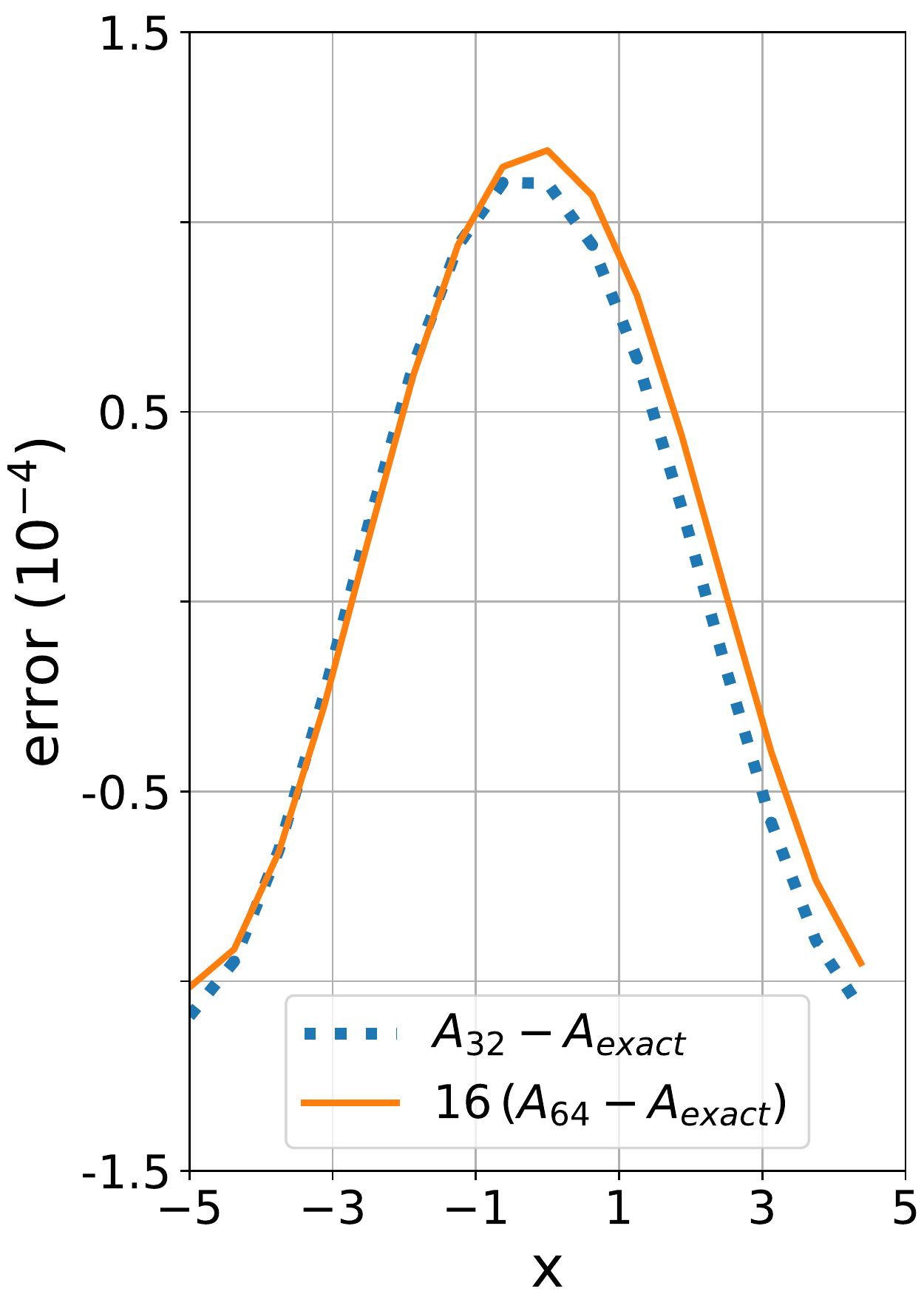}
  }
  {
    \includegraphics[width=0.76\textwidth
    ]
    {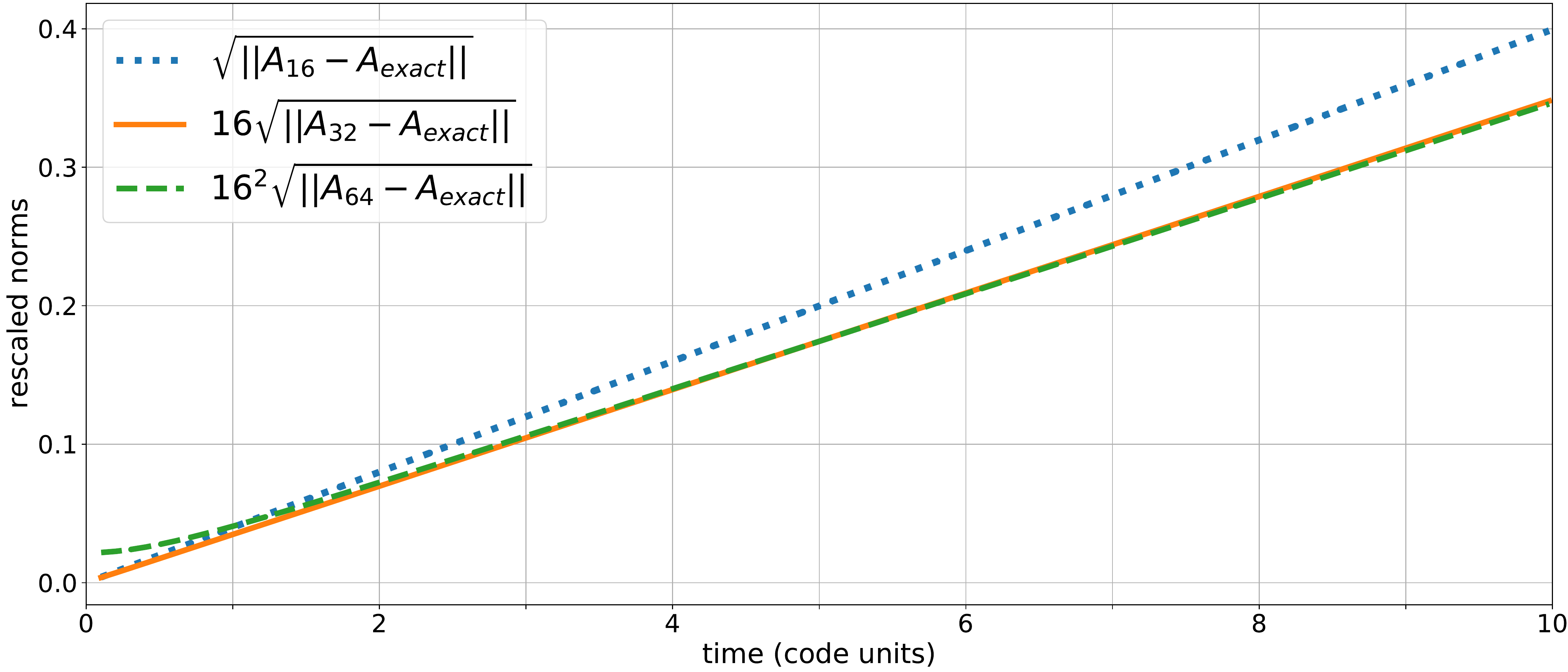}} 
  \caption{(Left) Pointwise convergence of the metric
      function~$A$ along the~$x$ direction, at~$t=9.98$ (code
      units), $u=0.83$ and~$y=0.625$, for the medium and fine
      resolutions.  
      (Right) Convergence rate 
      for the metric function~$A$ in terms of rescaled norms. Perfect
      overlap of curves should be understood as perfect
      convergence. Top rows show the second-order finite difference approximation
      case and bottom rows show the fourth-order one. The
      expected convergence factor for the former is~$Q=4$ and the
      latter~$Q=16$ for our specific setup.
\label{Fig:conv_plots_BB_advectxi}
}
\end{figure}

Convergence tests for the $A$ metric function can be seen in \fig{Fig:conv_plots_BB_advectxi}.
As mentioned above, the comparison of the grid functions against the exact solution is
performed only on grid points that are common to all three
resolutions.
The expected convergence factor for this setup is~$Q=4$ for second-order finite difference operators and~$Q=16$ for fourth-order ones, which is
indeed what we observe in the left column.
%
The same convergence rate is expected when we perform a norm
comparison. The discretized version of the~$L^2$-norm that we employ
here is simply the square root of the sum of the squared grid function
under consideration (over all domains).
%
In the right column of the figure 
we again see very good agreement for the norm convergence rate.



We also perform convergence tests for the setup that results in the
top phase-separated configuration of
Fig.~\ref{Fig:phase_separated}. In this case, the initial data
comprises of the sinusoidal perturbation~\eqref{eq:a4_pert}
with~$\bar{\alpha}_4 = 1$ and~$\delta \alpha = 10^{-3}$, as well
as~$\phi_0 = \Lambda = 1$,~$\phi_2=0.3$,
and~$\bar{\mathcal{E}}=1$. The size of the box
is~$L_x\Lambda = L_y \Lambda = 10$. The discretization of the
transverse and holographic domains, as well as the time integrator are
the same as for the previous convergence test, with the only
difference that the outer holographic domain here resolves the
region~$u \in[0.1,1.05]$. We use second order finite difference
operators and set~$\sigma=10^{-5}$. Since we do not have an exact
solution, we perform self convergence tests using only numerical
results. The comparisons are performed again using the common points
of the coarse grid.

\begin{figure}[t]
  {
    \includegraphics[width=0.235\textwidth
    ]
    {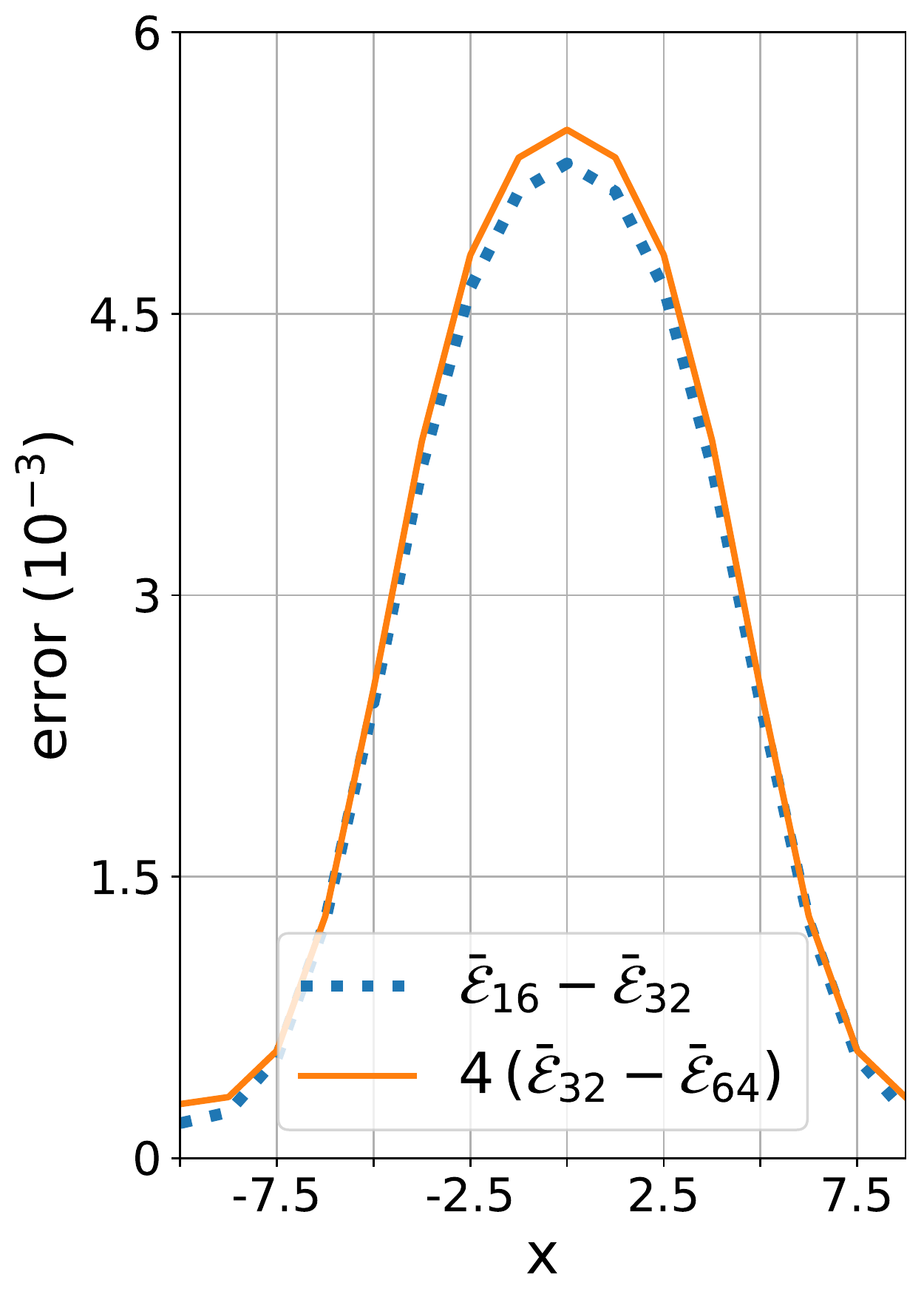}
  }
  {
    \includegraphics[width=0.76\textwidth
    ]
    {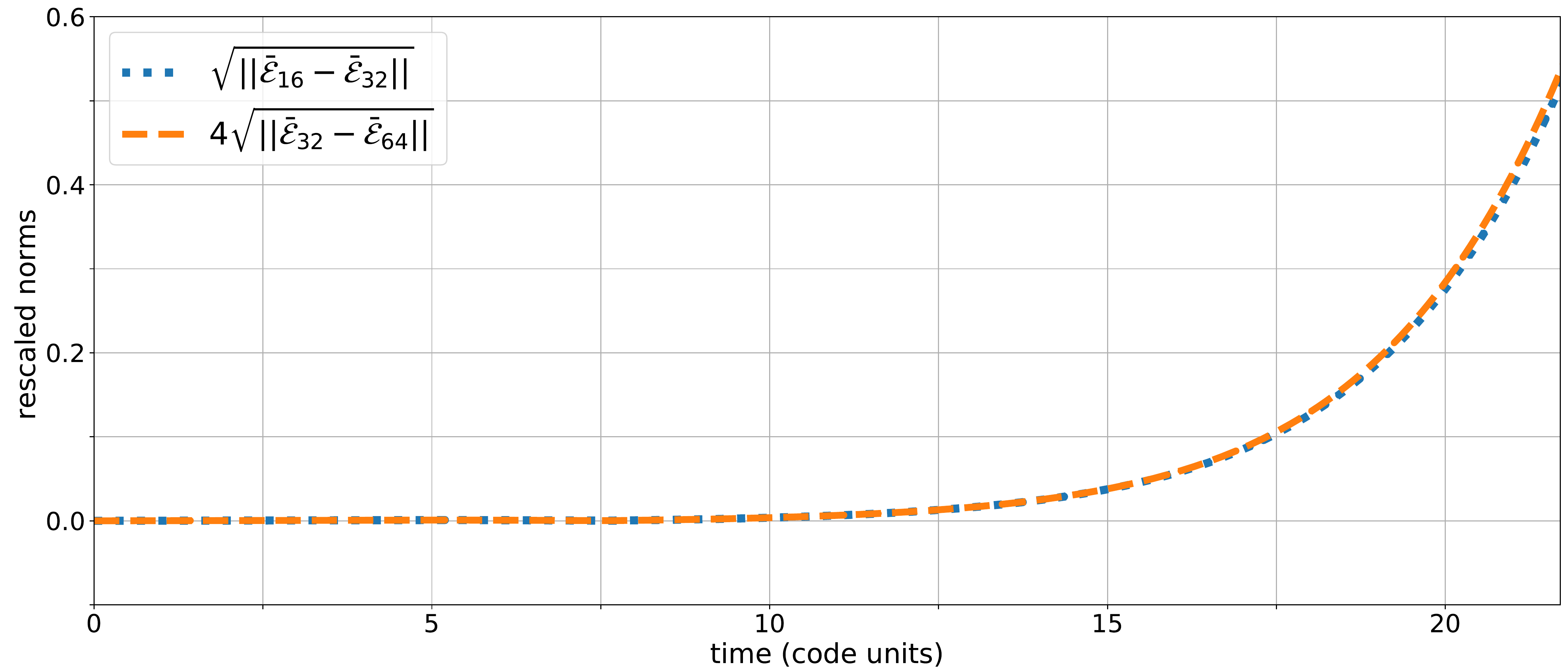}}
  \caption{
    (Left) Pointwise convergence of the boundary energy density~$\bar{\mathcal{E}}$ along
    the~$x$ direction for fixed~$y=-1.25$, at~$\Lambda t=21.69$ (code units).
    (Right) Convergence rate 
    for the boundary energy density in terms of rescaled norms, until~$\Lambda t=21.69$. Perfect
    overlap of curves should be understood as perfect convergence. We see
    good second order convergence throughout the evolution.
\label{Fig:bubble_from_spinodal_convergence}
}
\end{figure}

\begin{figure}[t]
  {
    \includegraphics[width=0.96\textwidth
    ]
    {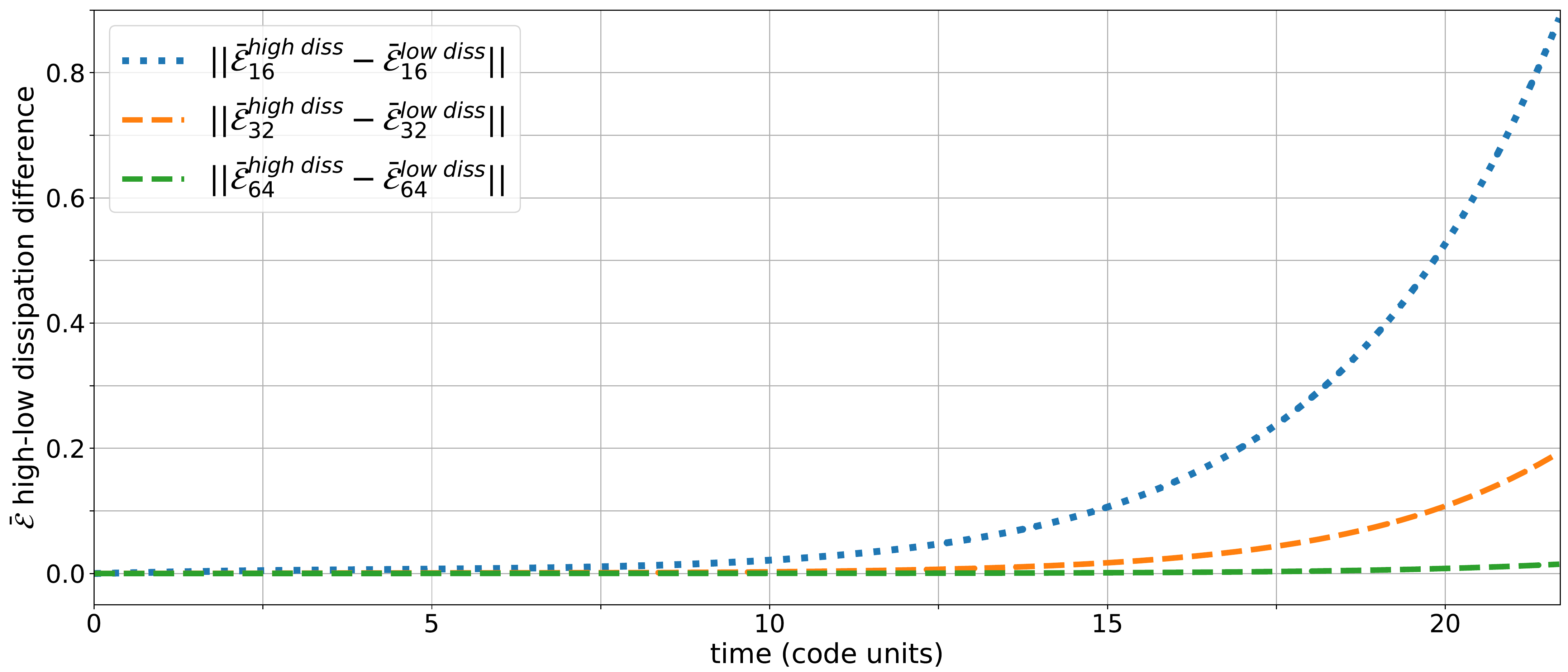}}
  \caption{
    The difference in~$\bar{\mathcal{E}}$ for runs with low and high
    Kreiss-Oliger dissipation, with~$\sigma=10^{-5}$ and~$\sigma=0.2$,
    respectively. The norm of this difference is illustrated
    until~$\Lambda t = 21.69$. We observe that this difference decreases
    with increasing resolution.
    \label{Fig:bubble_from_spinodal_low_high_diss_diff}
}
\end{figure}

In Fig.~\ref{Fig:bubble_from_spinodal_convergence} we present
pointwise and norm convergence tests for the boundary energy
density~$\bar{\mathcal{E}}$ of the above configuration. Notice that
the runs performed for these tests reach~$t \Lambda = 21.69$, whereas
the top phase-separated profile of Fig.~\ref{Fig:phase_separated}
corresponds to~$t \Lambda=300$ of the setup. Since we are using a low value
for the dissipation parameter ($\sigma = 10^{-5}$), it is not possible to perform such long
runs. The reason for this choice is that high values of~$\sigma$ seem to non-trivially affect the convergence properties of these configurations. However, we have checked that when
performing the same runs with~$\sigma=0.2$, which is sufficient for
long runs, the difference when comparing to the setups
with~$\sigma=10^{-5}$ drops fast with increasing resolution, as
illustrated in
Fig.~\ref{Fig:bubble_from_spinodal_low_high_diss_diff}.


\subsection{Thermodynamics tests}

Let us now explore how well the code can recover known properties of non-conformal homogeneous black branes. For concreteness we will focus on cases 
with $\lambda_4=-0.25$ and $\lambda_6=0.1$ for the model given by equation~\eqref{eq:potential}.

We initialize the code to some homogeneous (along $x$ and $y$) and isotropic state, setting $B_1=B_2=G=0$ and, as we are not interested in non-zero momenta, $f_{x2}=f_{y2}=0$. We set
$a_4$ and the (initial) gauge parameter $\xi$ as follows
\begin{equation}
a_4 = -\frac{4}{3}\left(\mathcal{E}+\phi_0\phi_2+\left(\lambda_4-\frac{7}{36}\right)\phi_0^4\right), \qquad \xi = \left(-a_4\right)^{-1/4}-\frac{1}{u_{H}},
\label{eq:a4_initial_NCBB}
\end{equation}
where $\mathcal{E}$ is the energy density of the black brane,
$u_{H}=1.0$ is the location at which the apparent horizon will be placed, and we choose $\phi_2=0.29819$ and $\phi_0=1.0(=\Lambda)$.\footnote{The value of $\phi_2$ should not be too far away from the equilibrium value of the non-conformal black brane with energy density $\mathcal{E}$. Otherwise, the initial choice for the gauge function $\xi$ may not ensure that the apparent horizon lies inside the numerical domain.}
Motivated by its near boundary behaviour, we initialize the scalar field to
%
\begin{equation}
\phi(u)=\phi_0 u-\xi_0\phi_0u^2+u^3\left(\xi^2\phi_0+\phi_2\right).
\label{eq:phi_initial_NCBB}
\end{equation}
We then let the code evolve. Since this scalar profile is not an equilibrium configuration, the system will relax in a few time units to the non-conformal uniform black brane with the given energy density~$\mathcal{E}$.


\begin{figure}[t]
  {\includegraphics[width=0.475\textwidth]{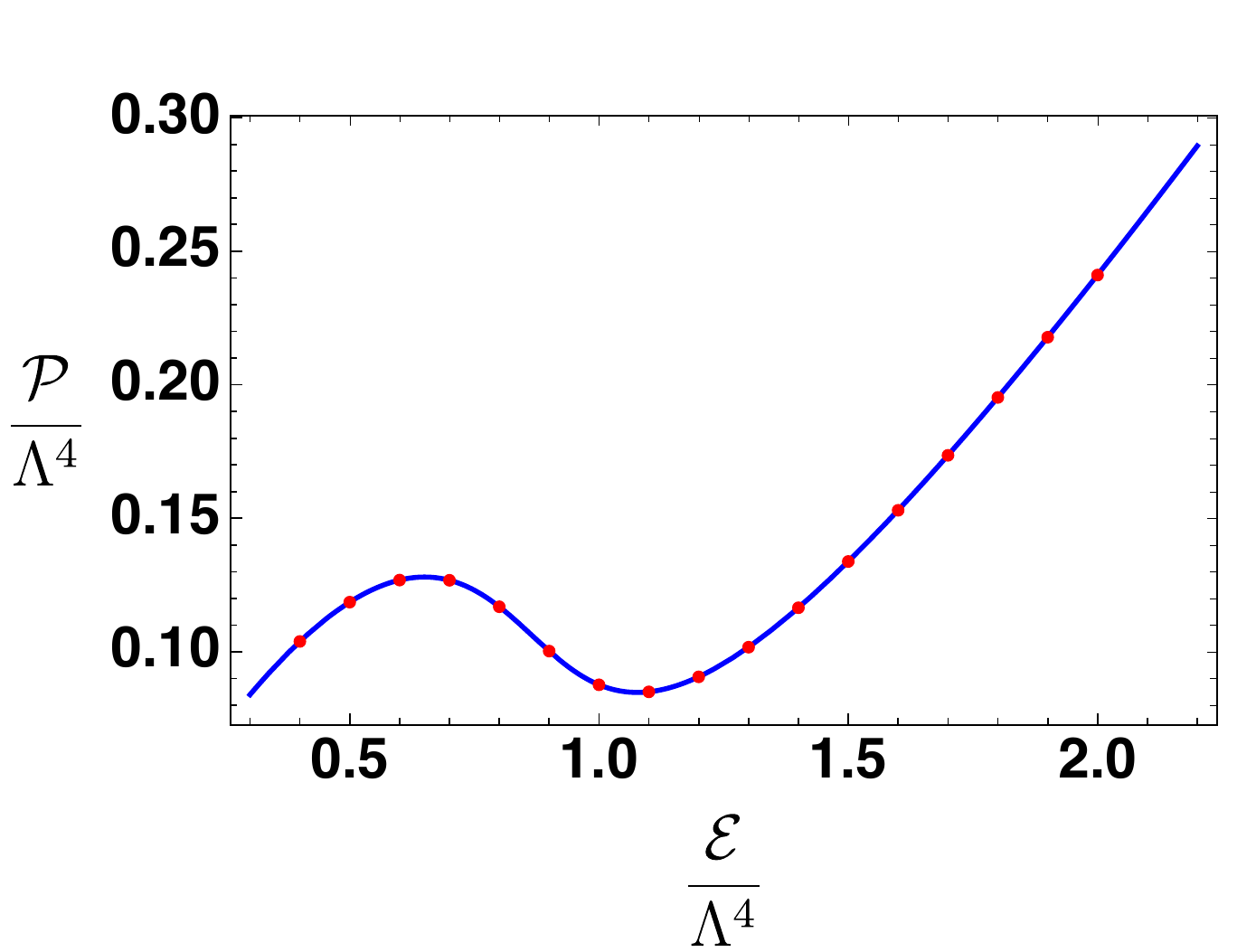}} \quad
  {\includegraphics[width=0.475\textwidth]{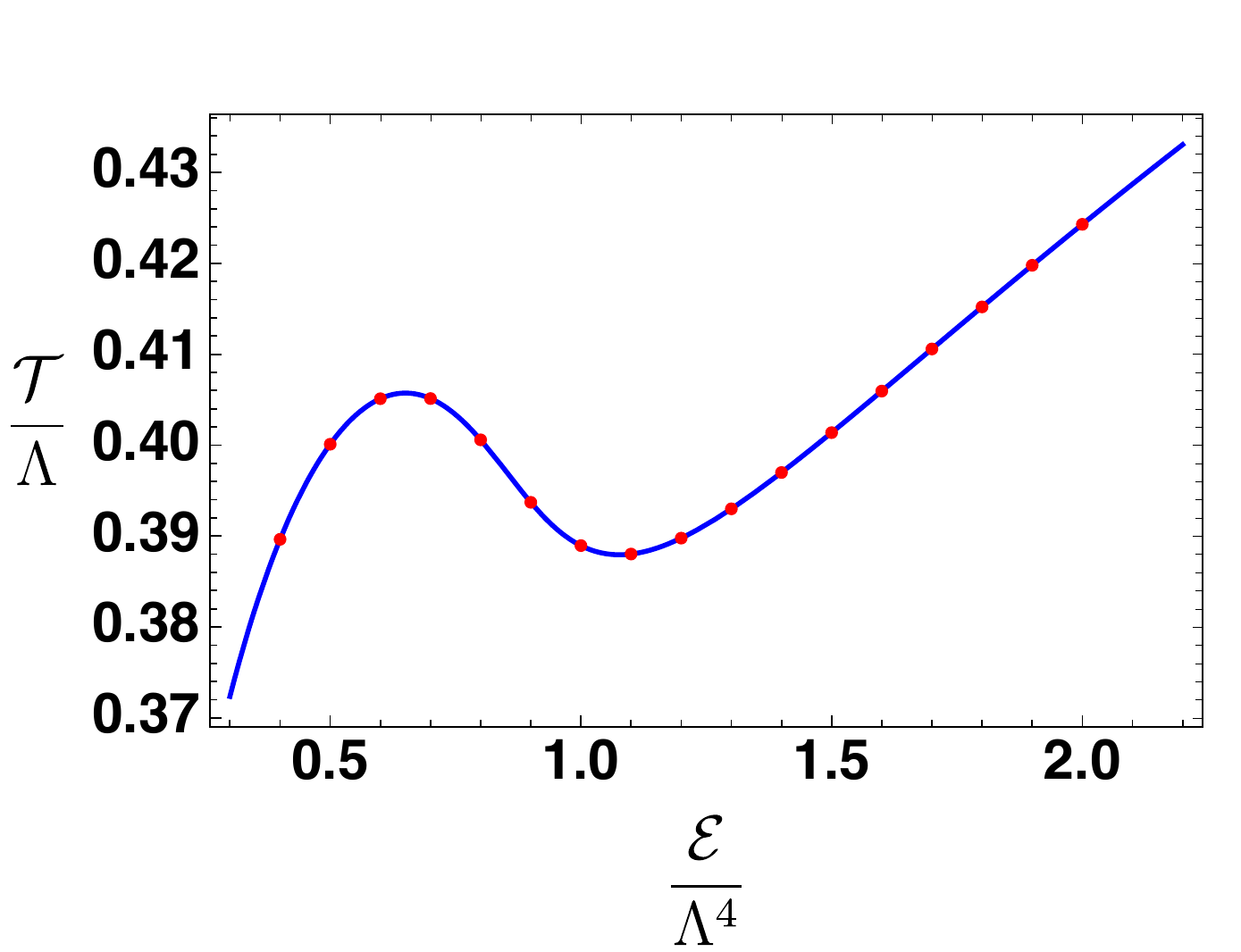}} \\
  {\includegraphics[width=0.475\textwidth]{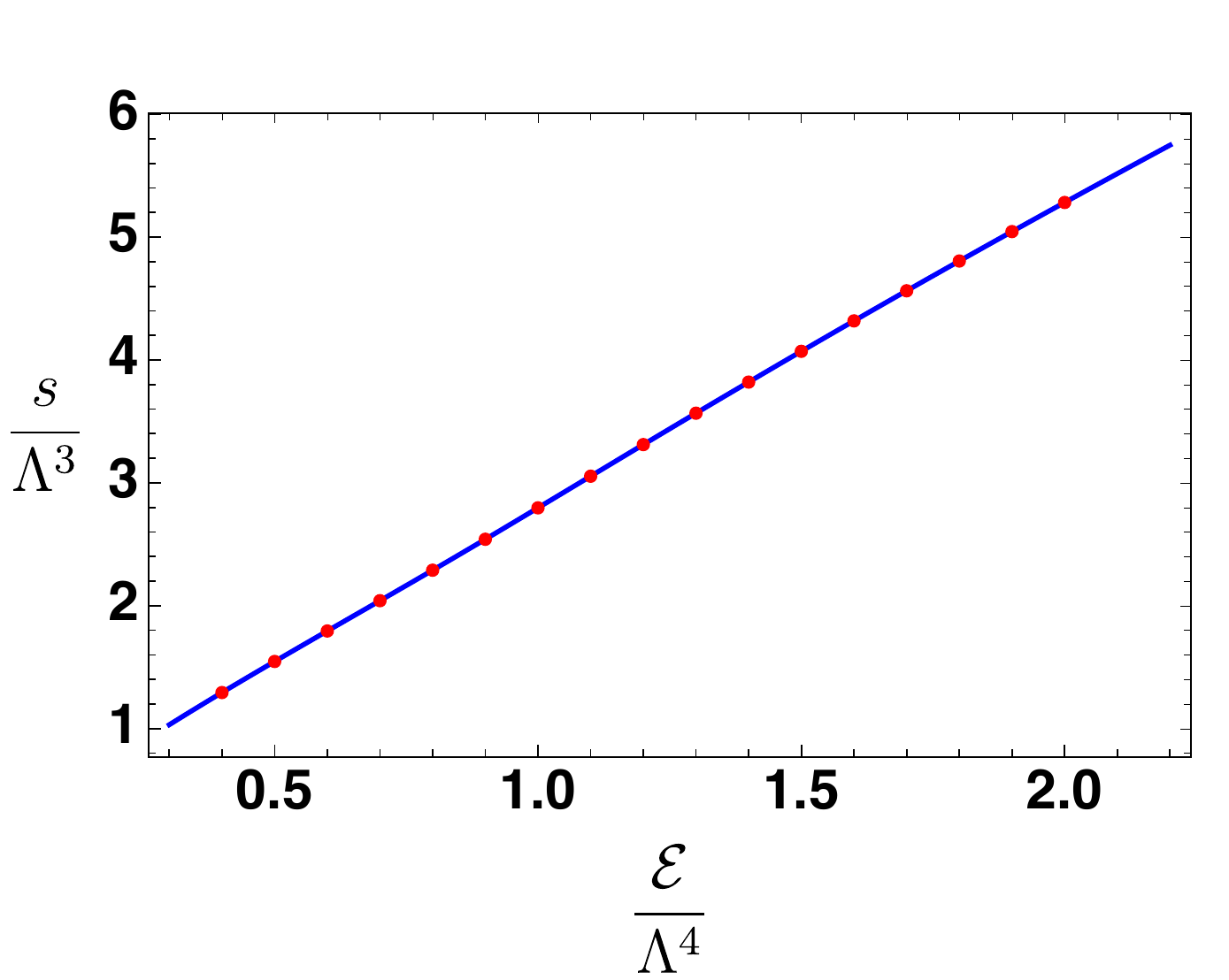}} \quad
  {\includegraphics[width=0.475\textwidth]{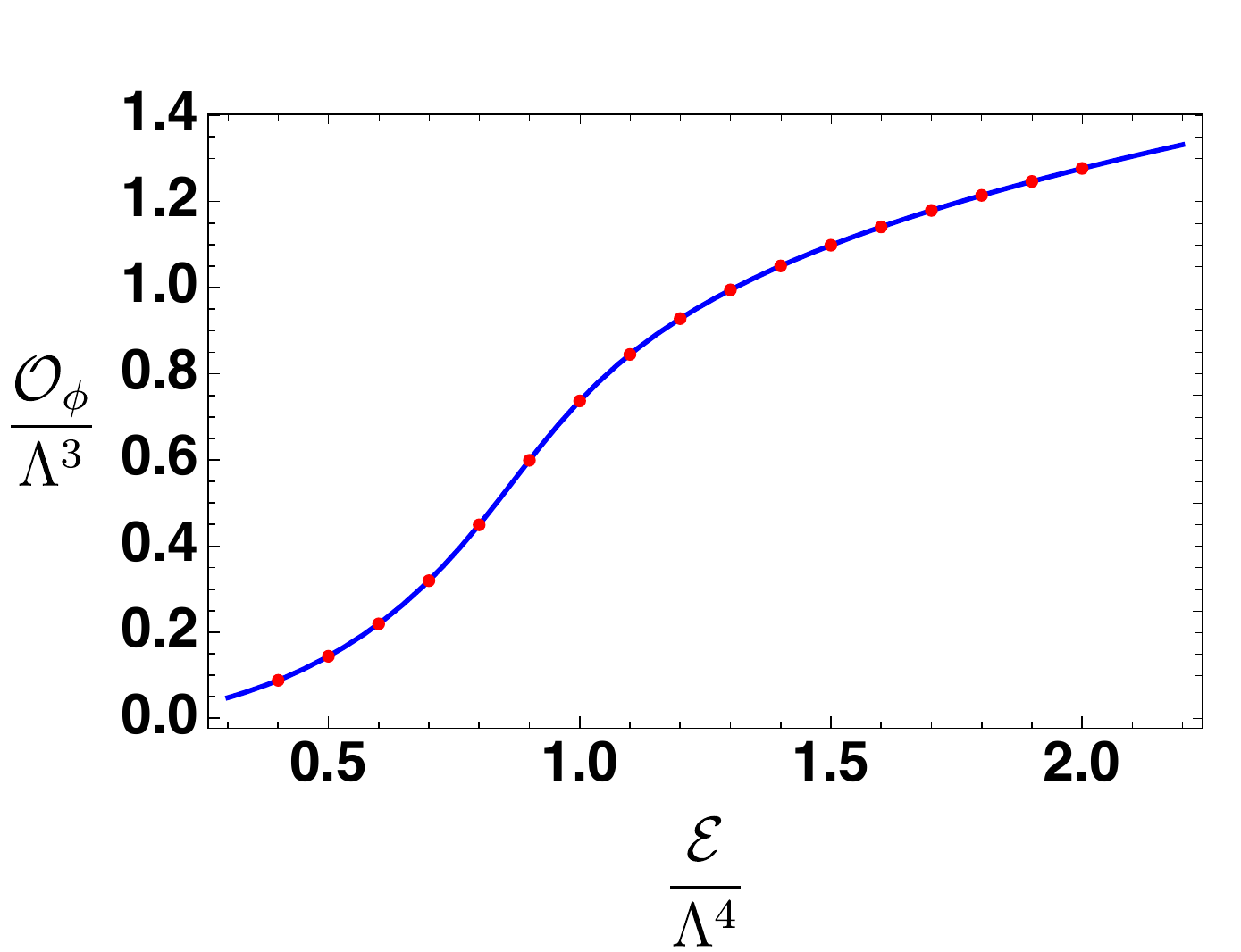}}
  \caption{(Up-left) Pressure, (Up-right) temperature, (Bottom-left) entropy density and (Bottom-right) the scalar VEV, computed solving the static Einstein's equations (blue line) compared against the results obtained from  \texttt{Jecco} after an evolution until $t\Lambda=60$ (red dots).
    \label{Fig:non_conformal_branes}}
\end{figure}
We performed a total of 16 runs with energies evenly distributed in the interval $\mathcal{E}/\Lambda^4\in [0.4,2.0]$ and compared the results with those obtained from directly integrating the static solution of Einstein's equations for the same physical configuration. Each run was performed using a single core in a  16GB memory machine, the runtime being a few minutes. The chosen range of energies is of most relevance because it completely contains the first order thermal phase transition exhibited for this value of $\lambda_4$ and $\lambda_6$ (see \fig{Fig:non_conformal_branes}). For higher and lower energies the theory tends to the conformal case, which was explored previously. \fig{Fig:non_conformal_branes} shows that the results obtained by both methods lie on top of each other. The pressures along the 3 boundary directions are equal to each other (there are no anisotropies), and the behaviour of the energy density as a function of the temperature (up-right panel) shows the typical behaviour of a theory with a first order phase transition, see~\cite{Bea:2018whf,Bea:2020ees}. Notice that the off-diagonal pressure, $\mathcal{P}_{xy}$, and the energy fluxes, $\mathcal{J}_x$ and $\mathcal{J}_y$, are not shown as they are vanishing for these solutions.

\begin{figure}[t]
  {\includegraphics[width=0.475\textwidth]{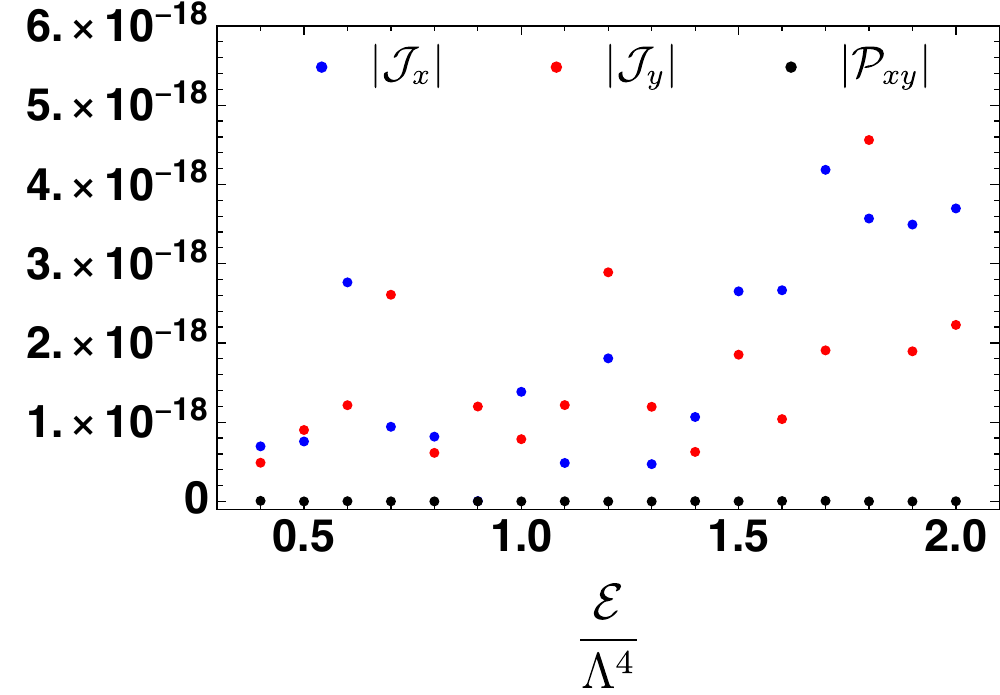}} \quad
  {\includegraphics[width=0.475\textwidth]{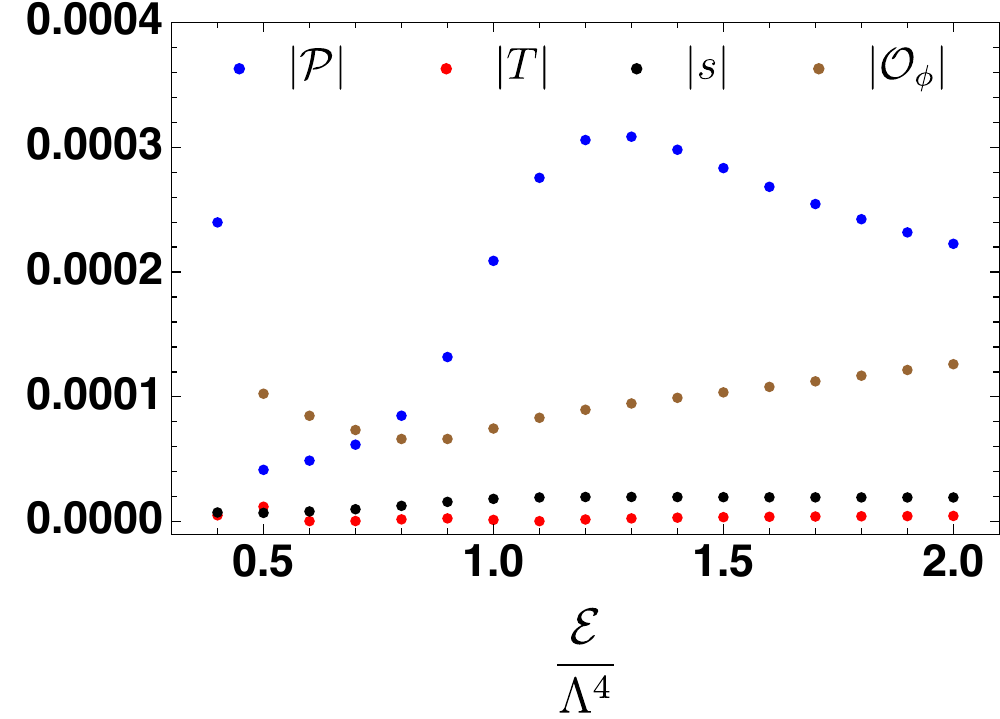}}
  \caption{(Left) Absolute error of the vanishing quantities and (Right) relative error of the non vanishing ones. We took the value obtained by solving the static Einstein's equations as the reference for the relative error. \label{Fig:non_conformal_branes_error}}
\end{figure}
%
In \fig{Fig:non_conformal_branes_error} we show the differences between the quantities obtained with these two methods; 
we plot absolute differences for those quantities that vanish and relative differences 
for the non-trivial ones. As can be seen, the off-diagonal pressure and the energy fluxes have vanishingly small values; for the non-trivial quantities, the pressure presents the largest relative error, which is smaller than $0.03\%$.
We thus see that \texttt{Jecco} is returning the expected properties of these solutions with very good accuracy.

\subsection{Quasi-normal mode tests}
\label{subsec:QNM}


We now show results for a time-dependent test, where we recover expected
quasi-normal mode frequencies.
This test is a replica of the one
performed for~\texttt{SWEC} in~\cite{Attems:2017zam}, and was performed using a single core in a 16GB memory machine running for a few minutes.

We fix $\phi_0=1$, $\lambda_4=0.0025$ and $\lambda_6=0$. We set as
initial conditions $G=0$ together with
$f_{x2}=f_{y2}=0$. $a_4$ is obtained from equation~(\ref{eq:a4_initial_NCBB})
with an average energy density
$\mathcal{E}=0.379686$ and $\phi_2=0.0868357$, which
corresponds to the equilibrium value of the non-conformal uniform
black brane with that same energy density.
For this test $\xi=0$ (initially) is good enough a choice. The vanishing wave number ($k=0$) perturbation is inserted by
activating the anisotropy together with a non equilibrium scalar field
profile:
\begin{equation}
\begin{aligned}
\phi(u)&=\phi_0 u+\phi_2u^3 \,, \\
B_1(u)&=0.15u^8 \,, \\
B_2(u)&=0.05u^8 \,.
\end{aligned}
\end{equation}
Notice that, since the perturbation is independent of both $x$ and $y$, $B_1$ and $B_2$ will behave identically up to an overall constant factor.


The system will relax to the equilibrium state through damped oscillations whose parameters were extracted in~\cite{Attems:2016ugt} for different values of $\lambda_4$. In particular, the boundary variables $\phi_2$, $b_{14}$ and $b_{24}$ will evolve in time according to
\begin{equation}
\begin{aligned}
f(t)&=f_{\rm eq}+f^{(1)}(t)+f^{(2)}(t)+\cdots,\\
f^{(n)}(t)&=A_ne^{-\omega_i^{(n)}t}\cos(\omega_r^{(n)}t+\delta_n),\\
\omega^{(n)}&=\omega_r^{(n)}+i\omega_i^{(n)},
\end{aligned}
\label{eq:QNM_modes}
\end{equation}
with the mode 1, $f^{(1)}(t)$, being the longest lived  (smallest $\omega_i$).



\begin{figure}[tp]
  {\includegraphics[width=0.475\textwidth]{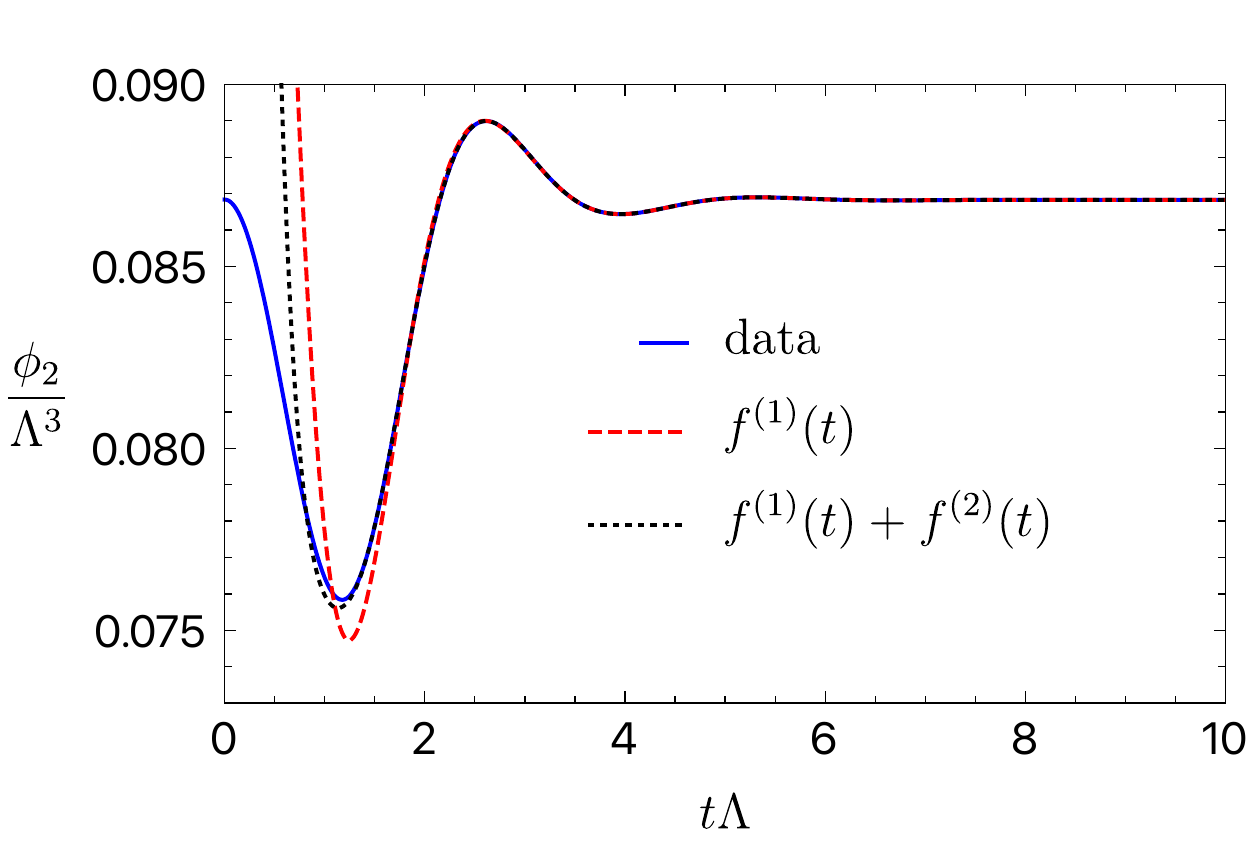}} \hfill
  {\includegraphics[width=0.51\textwidth]{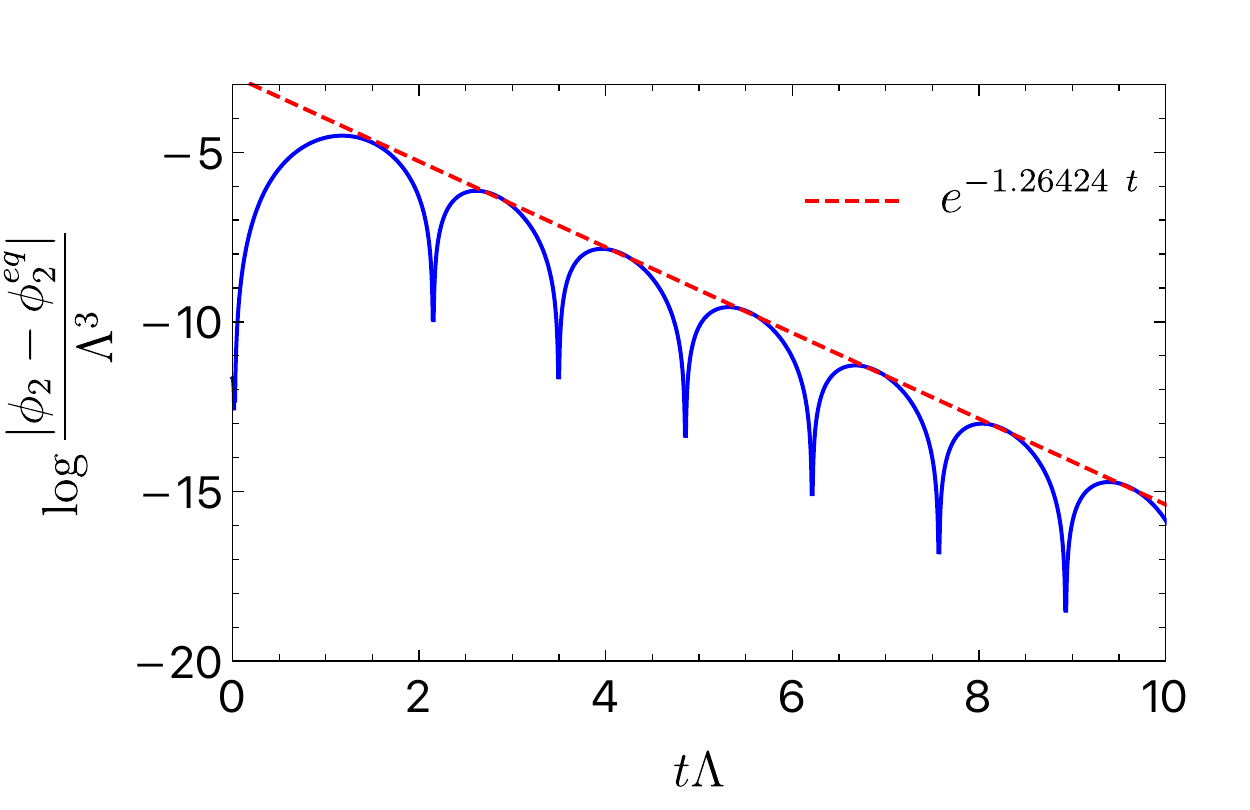}} \\
  {\includegraphics[width=0.475\textwidth]{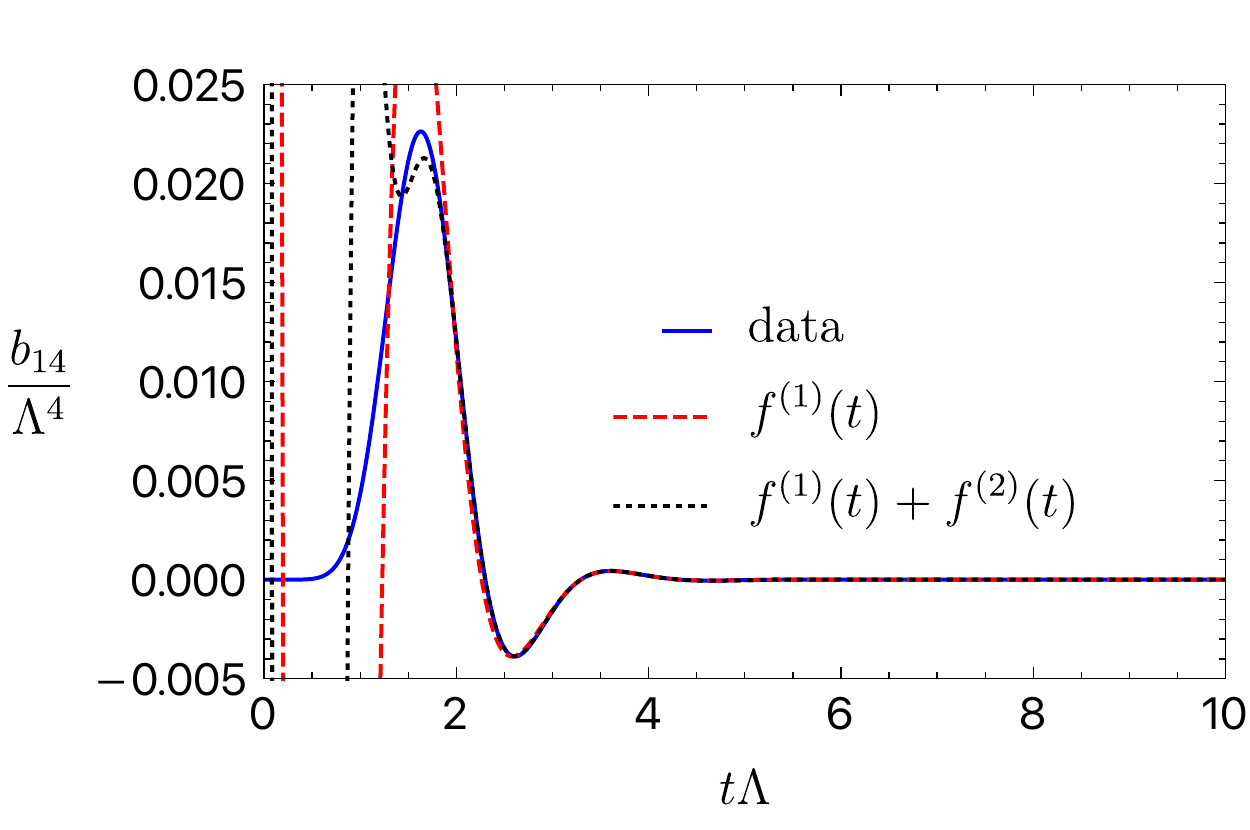}} \hfill
  {\includegraphics[width=0.5\textwidth]{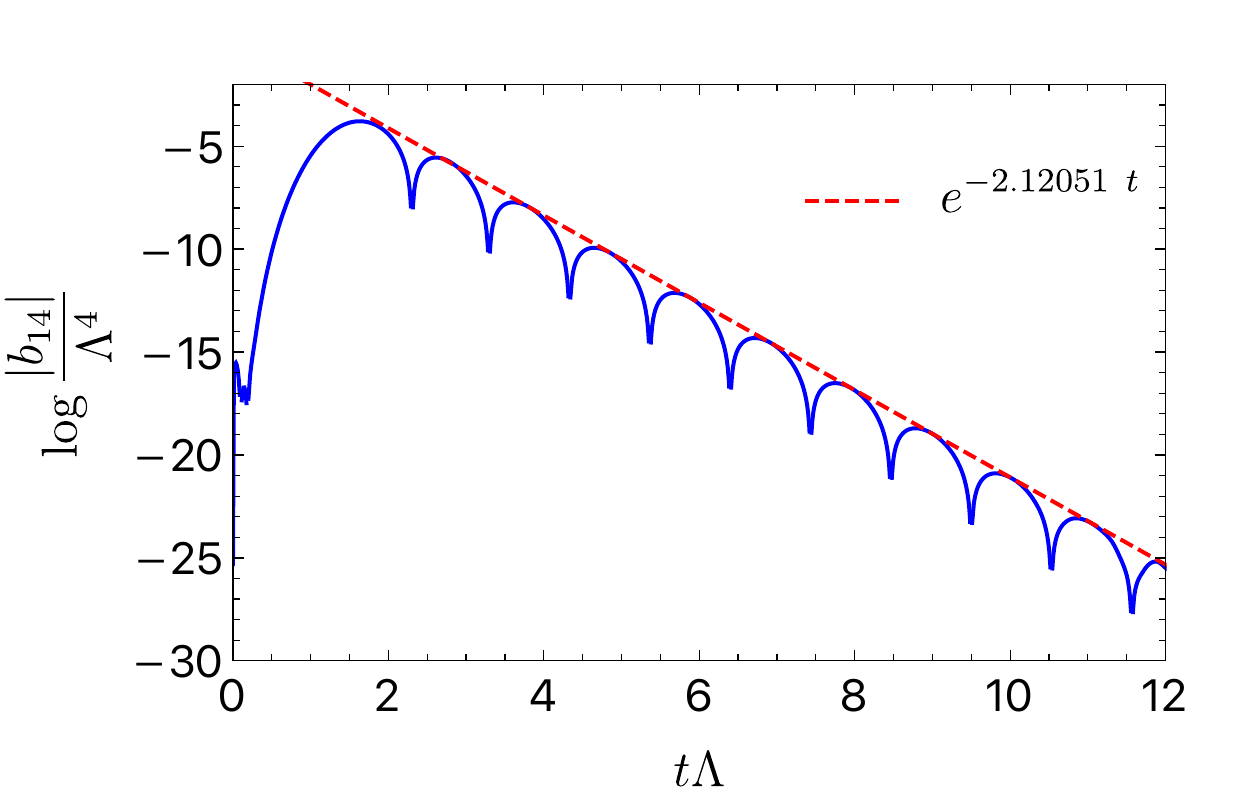}} \\
  {\includegraphics[width=0.475\textwidth]{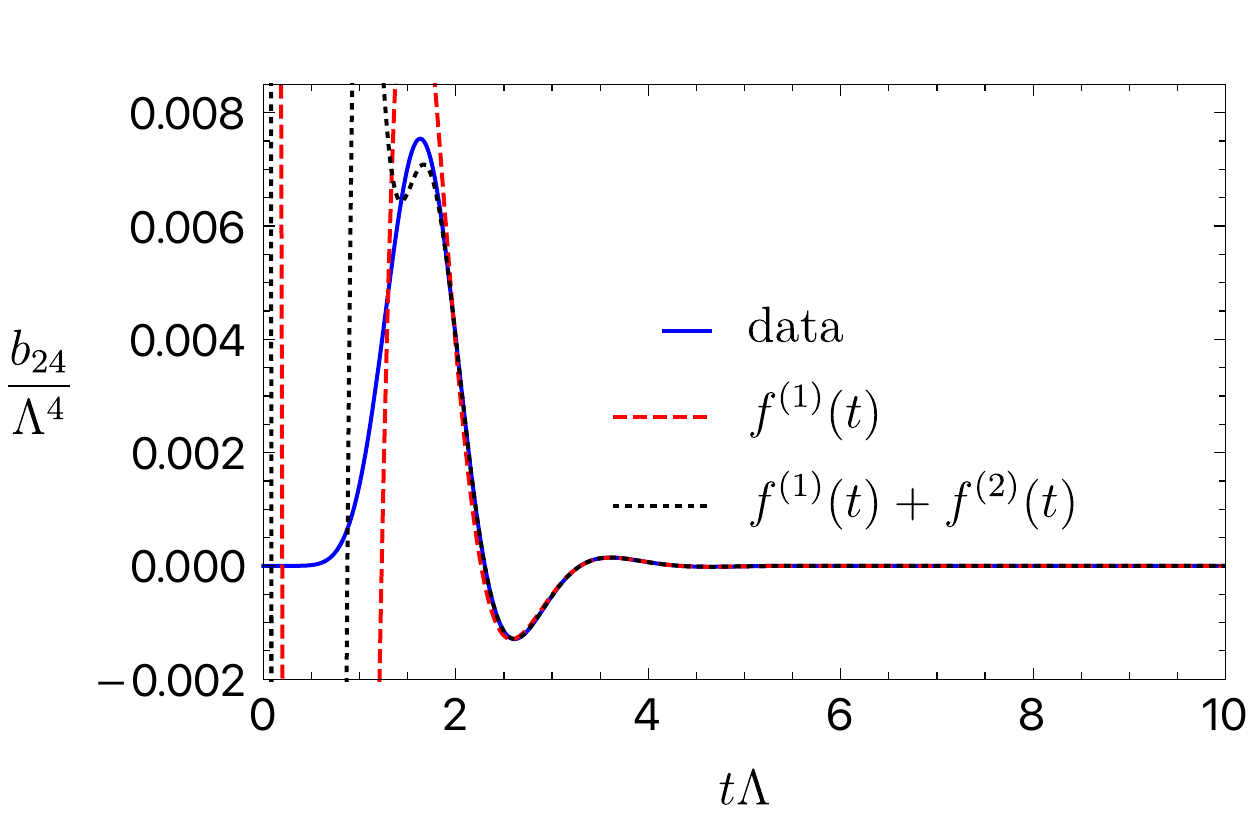}} \hfill
  {\includegraphics[width=0.5\textwidth]{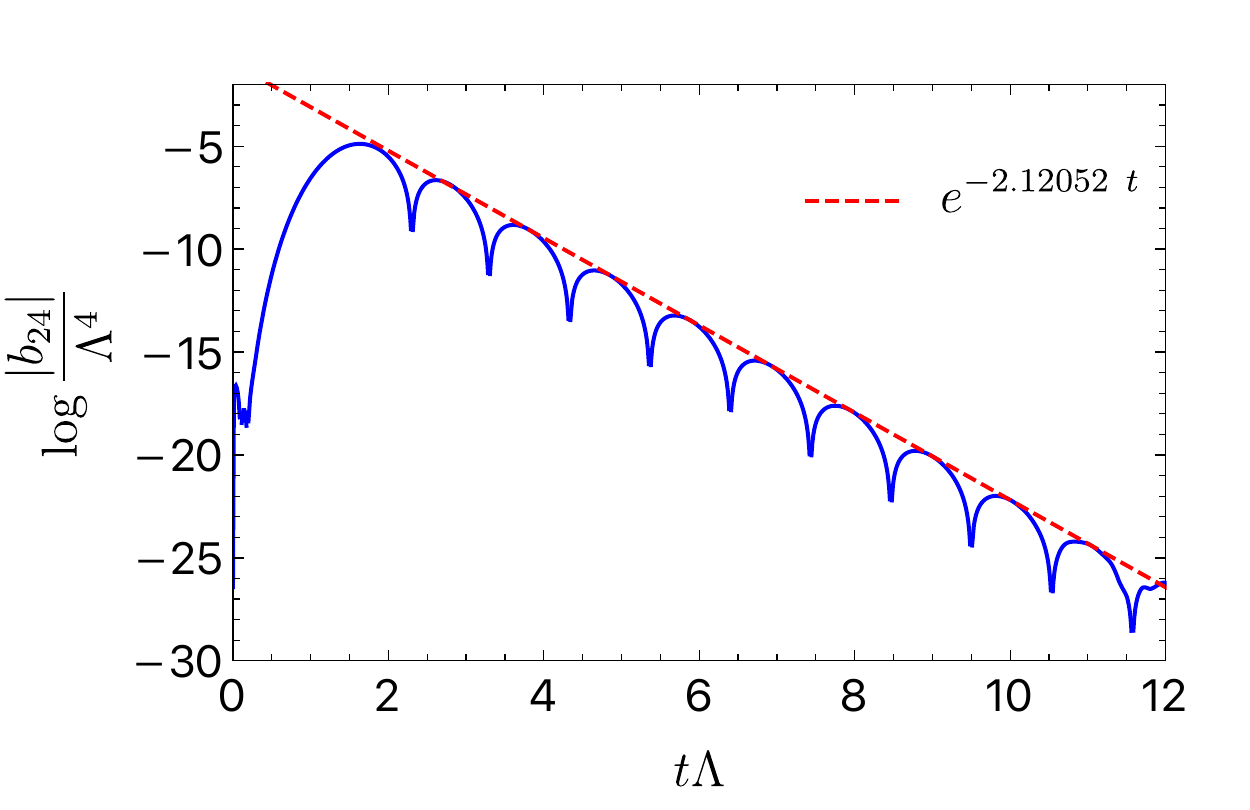}}
  \caption{
    (Left column) Time evolution of the boundary variables.
    Red dotted lines correspond to the longest lived mode, while the black dotted
    ones are the full combination given by equation~\eqref{eq:QNM_modes}.
    (Right column) Log plots of $f-f_{\rm eq}$ for the boundary variables.
    Red dotted lines correspond to the exponential decay that best fits the data at
    late times.
    \label{Fig:QNM_log_all}}
\end{figure}
We can obtain the different parameters from the data for $f-f_{\rm eq}$, whose log-plots are shown in \fig{Fig:QNM_log_all}. For late times, the longest-lived mode dominates and the data clearly behaves as a damped oscillation. We use this fact to fit the $f^{(1)}(t)$ to the data. Once we have the parameters for mode~1, the shorter lived mode can be obtained by fitting $f(t)$ at early times, where its presence is still important. As a consequence we get an improvement in the description, specially at early times, as can be seen in the left column of the figure.

We find the following values for the frequencies
\begin{equation}
\begin{alignedat}{3}
\phi_2:  & \qquad & \omega^{(1)} & =2.313004+i1.264244, \qquad &  \omega^{(2)} & =4.091354+i2.944895,\\
 b_{14}: & \qquad & \omega^{(1)} & =3.039641+i2.120511, \qquad &  \omega^{(2)} & =4.953161+i3.737359,\\
 b_{24}: & \qquad & \omega^{(1)} & =3.039649+i2.120516, \qquad &  \omega^{(2)} & =4.952340+i3.741593,
\end{alignedat}
\end{equation}

The frequencies can also be obtained directly from the equilibrium solution by solving linear perturbation equations.
Taking the final equilibrium solution of this evolution and computing the linearized fluctuations around this background using the Mathematica package~\texttt{QNMspectral} \cite{Jansen:2017oag}, we obtain
\begin{equation}
\begin{alignedat}{3}
\phi_2:  & \qquad & \omega^{(1)} & =2.313080+i1.264337, \qquad &  \omega^{(2)} & =4.108219+i2.931352,\\
 b_{14}: & \qquad & \omega^{(1)} & =3.039399+i2.120359, \qquad &  \omega^{(2)} & =4.934072+i3.739264,\\
\end{alignedat}
\end{equation}
and for $b_{24}$ identical results to those for  $b_{14}$ as they are both tensor fluctuations.

%
%
%
The agreement among the values obtained by both methods is excellent for the lowest-frequency modes, the easiest to extract, with a relative error under $0.01\%$ in all cases. For the shorter-lived modes, as expected, the relative error is higher, but the agreement is still very good, always below $0.5\%$.
\subsection{Fluid/gravity tests}
\label{subsec:fluid/grav_tests}

The fluid/gravity duality establishes a precise map between the equations of relativistic hydrodynamics in $d$ dimensions and the Einstein equations with negative cosmological constant in $d+1$ dimensions in a specific regime (see~\cite{Hubeny:2011hd,Rangamani:2009xk} for comprehensive reviews). This is a map between non-linear equations, and solutions on one side map to solutions on the other side. Even if originally derived from holography, fluid/gravity is an independent statement and constitutes a duality between two classical theories.  This represents a complementary test to the one of \Sec{subsec:QNM}.

We will now use this fluid/gravity mapping to test the code. The idea of this test is to consider a microscopic holographic evolution, which by construction is in the regime of hydrodynamics, and then compare this microscopic evolution against the constitutive relations of hydrodynamics at every spacetime point. 

The constitutive relations of hydrodynamics (see e.g.~\cite{Romatschke:2009kr}) truncated at first order in the hydrodynamic gradient expansion take the form
\begin{equation}
T^{\rm hydro}_{\mu\nu}=T^{\rm ideal}_{\mu\nu}+T^{\rm 1st}_{\mu\nu}+\cdots ~,
\end{equation} 
with
\begin{align}
T^{\rm ideal}_{\mu\nu}&=\mathcal{E}_{\rm loc} u_{\mu}u_{\nu} +P(\mathcal{E}_{\rm loc}) \Delta_{\mu\nu} ~,\\
T^{\rm 1st}_{\mu\nu}&=-\eta(\mathcal{E}_{\rm loc}) \sigma_{\mu\nu} - \zeta(\mathcal{E}_{\rm loc}) (\nabla \cdot u) \Delta_{\mu\nu} ~,
\end{align}
where $\mathcal{E}_{\rm loc}$ is the energy density in the local rest frame of the fluid, $u_{\mu}$ is the local fluid velocity, $\Delta_{\mu\nu}:=\bar{\eta}_{\mu\nu}+u_{\mu}u_{\nu}$ is the projector, $P(\mathcal{E}_{\rm loc})$ is the equation of state, and $\eta$ and $\zeta$ are the shear and bulk viscosities, respectively. Moreover, we define  $\sigma_{\mu\nu}:=2\nabla_{\langle\mu} u_{\nu\rangle}$, where $\langle\cdot\rangle$ indicates symmetrization, tracelessness and orthogonality to the velocity. The equation of state and the viscosities are determined by the specific microscopic theory under consideration, and in our case we obtain them by constructing the set of homogeneous black branes and by using Kubo formulas (see~\cite{Bea:2020ees,Attems:2016ugt}).

We consider as initial state a homogeneous black brane solution with small sinusoidal perturbations along $x$ and $y$.  
For this test we choose $\lambda_4=-0.25$, $\lambda_6=0.1$, $x,y\in[-100,100)$ with $N_x=N_y=100$ grid points in the $x$ and $y$ directions. The average energy density is fixed to $\bar{\mathcal{E}}=10 \Lambda^4$ and the $a_4$ function is chosen to be
\begin{equation}
a_4(0,x,y)=\bar{a}_4\left[1+0.001\left(\cos\left(\frac{\pi x}{100}\right)+\cos\left(\frac{\pi y}{100}\right)\right)\right],
\end{equation}
where the average value, $\bar{a}_4$, is determined by equation~(\ref{eq:a4_initial_NCBB}) (left) with our chosen value of $\bar{\mathcal{E}}$. The initial gauge parameter $\xi$ is obtained from equation (\ref{eq:a4_initial_NCBB}) (right) replacing $a_4$ by $\bar{a}_4$. The scalar bulk profile is again chosen to be given by equation (\ref{eq:phi_initial_NCBB}), with $\phi_0=1.0(=\Lambda)$ and $\phi_2=0.29819$.
If the momentum of the perturbation $k$ is small compared to the temperature of the black brane $T$, the system will be within the regime of hydrodynamics. For this simulation the ratio is $k/T \simeq 0.051$.


We evolve this initial configuration with \texttt{Jecco}
and compare the obtained boundary stress tensor as a function of time
with the constitutive relations of hydrodynamics. As we will see below, we find very good agreement for all  components of the stress tensor.

\begin{figure}[t]
	{\includegraphics[width=0.475\textwidth]{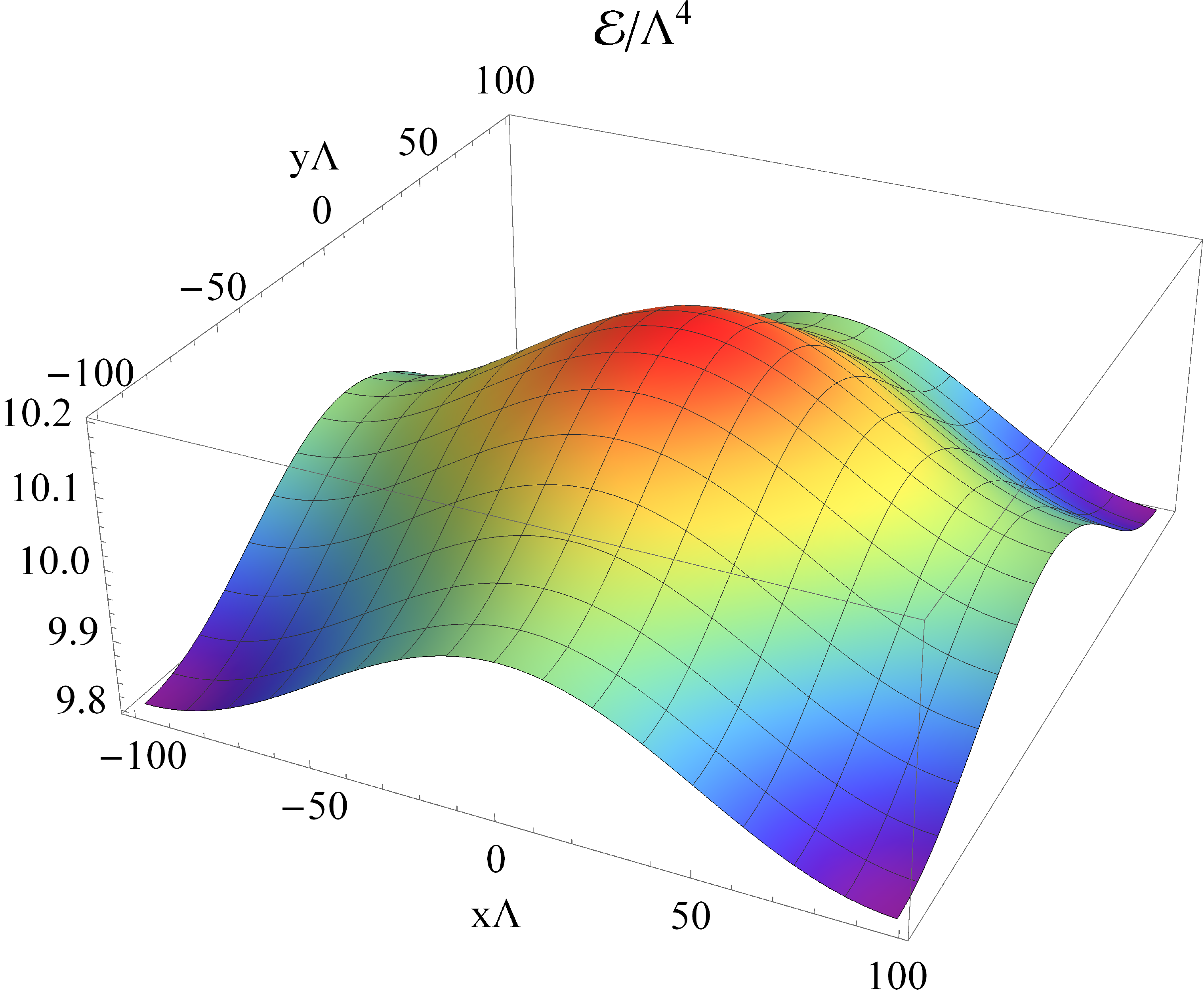}}  \hfill
	{\includegraphics[width=0.475\textwidth]{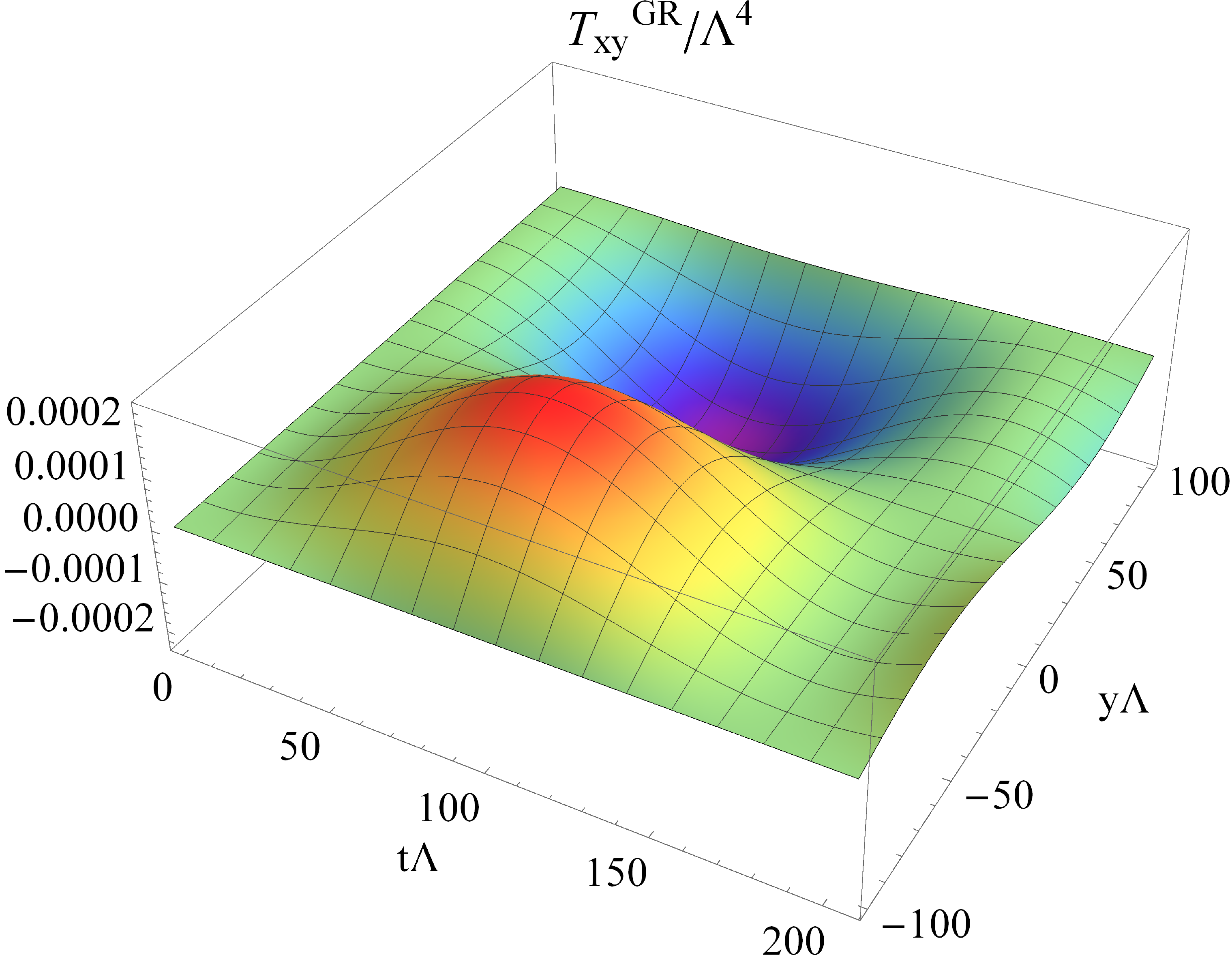}}  \\
	{\includegraphics[width=0.475\textwidth]{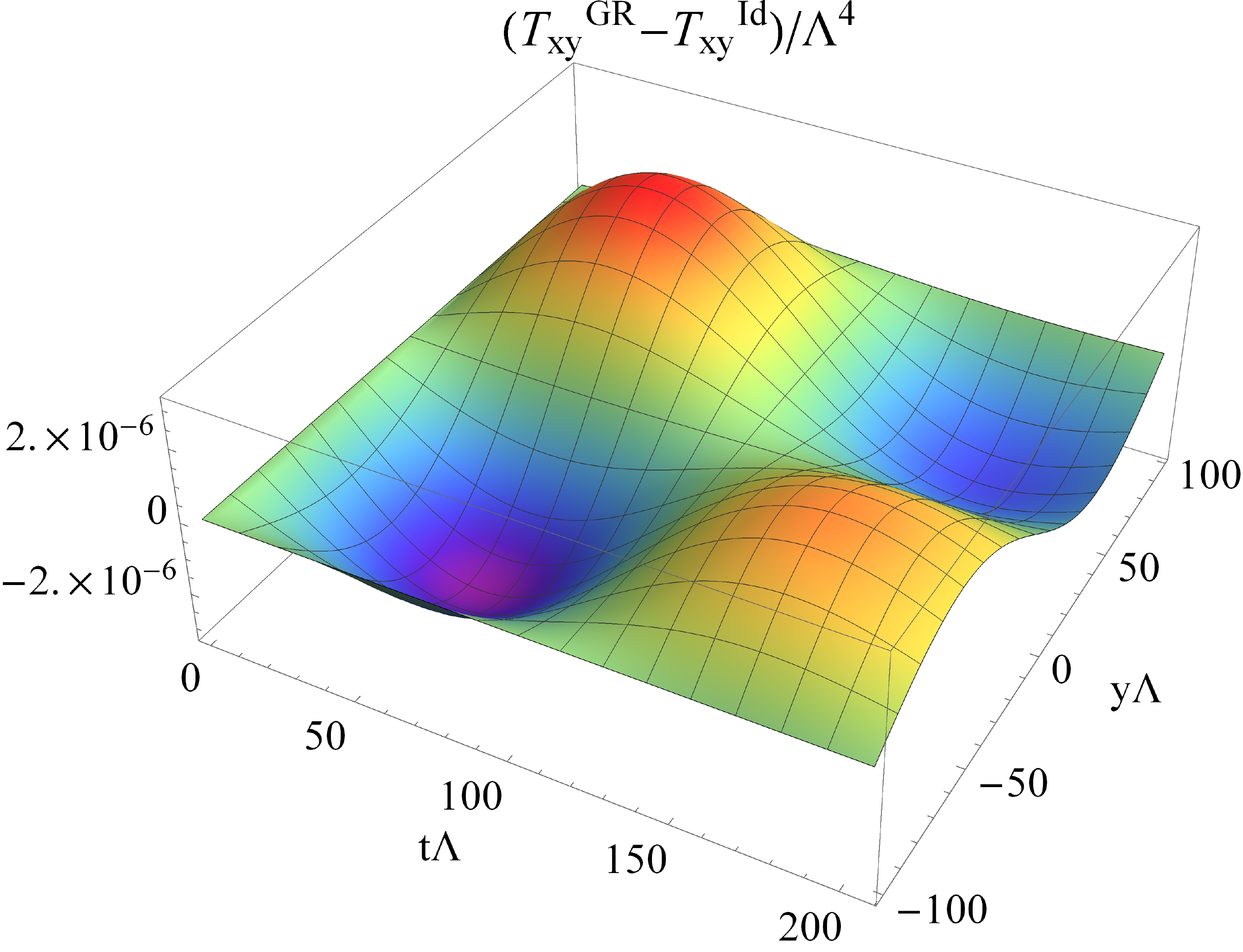}} \hfill
	{\includegraphics[width=0.475\textwidth]{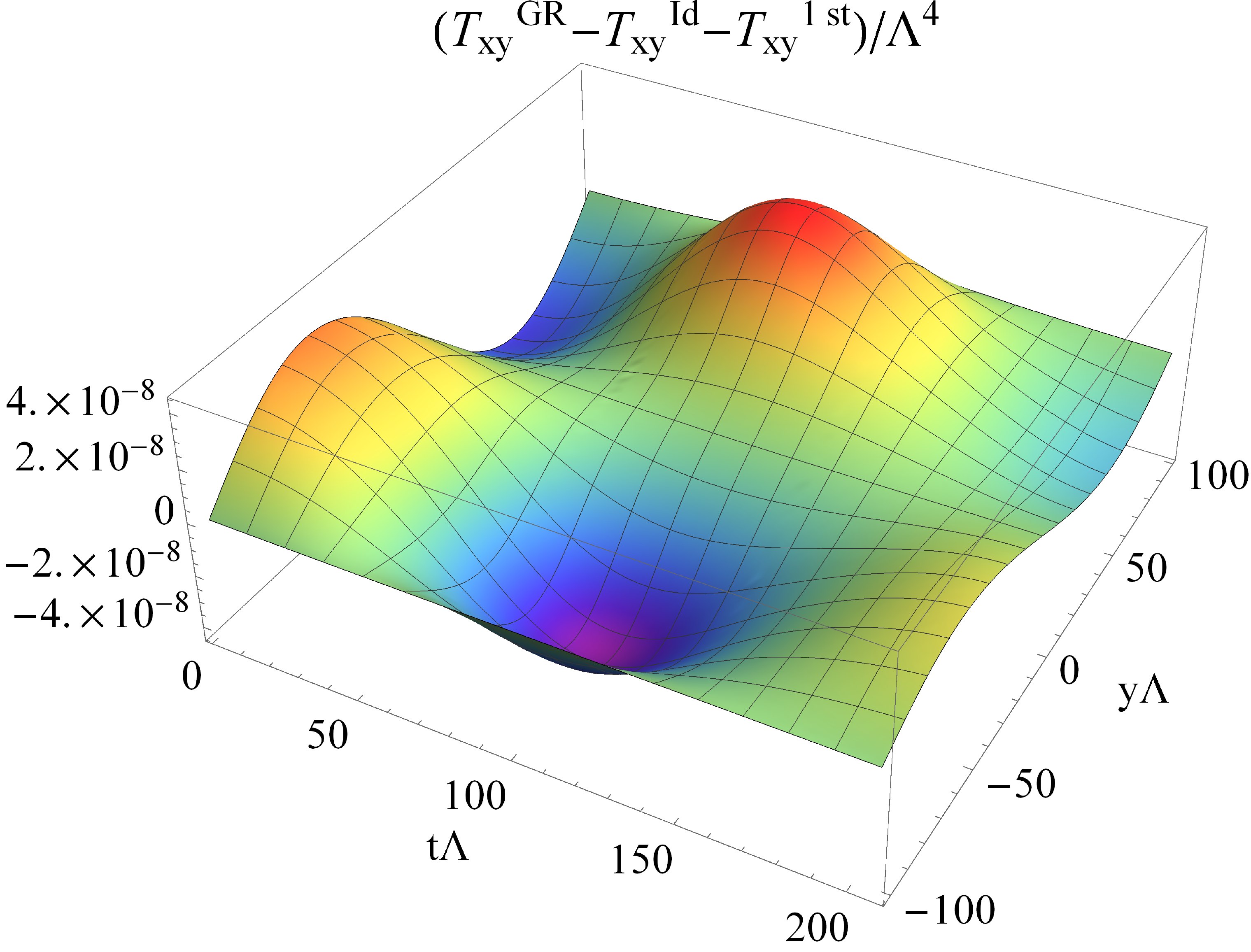}}
	\caption{Real-time evolution from \texttt{Jecco} and comparison with the constitutive relations of hydrodynamics. (Top-left) Initial energy density configuration: a homogeneous state plus small sinusoidal perturbations along $x$ and $y$. (Top-right) $T_{xy}$ component of the stress tensor as a function of time and $y$ (extracted at $x=50 / \Lambda$) obtained from the \texttt{Jecco}.  (Bottom-left) Difference between the $T_{xy}$ obtained from the numerical evolution and $T^{\rm Id}_{xy}$ given by the constitutive relations of ideal hydrodynamics. This difference is very small compared to $T_{xy}$, indicating that ideal hydrodynamics provides a good description. (Bottom-right) We further include the first order terms of hydrodynamics in the previous subtraction, obtaining an even better description.  \label{Fig:Hydro1}}
\end{figure}

See \fig{Fig:Hydro1} (top-left) for the initial energy density configuration. The system has vanishing initial velocity. The fact that we are not initializing~\texttt{Jecco} with the equilibrium $\phi(u)$ does not affect the results, as the time that this takes to decay (through quasi-normal modes) to the equilibrium profile for the specified average energy density is much shorter than the time scale of the dynamics triggered by the sinusoidal perturbation.

In the following, we present the results for the $T_{xy}$ component of the stress tensor. For all other components the results are similar. The component $T_{xy}$ is particularly interesting because it allows to test proper dynamics in 2+1 dimensions on the boundary. 
Moreover, the constitutive relations of hydrodynamics for this component 
are purely non-linear (the linearized expression vanishes),
so this provides also a truly non-linear test.

Figure~\ref{Fig:Hydro1} (top-right) shows the values of $T_{xy}$ obtained from \texttt{Jecco}, at constant $x=50$ as a function of time,
and the bottom-left panel shows the difference between $T_{xy}$ and $T^{\rm Ideal}_{xy}$ -- given by the constitutive relations of ideal hydrodynamics.
This difference is very small compared to $T_{xy}$, indicating that ideal hydrodynamics provides a very good description (within $0.01\%$).
In \fig{Fig:Hydro1} (bottom-right) we further include the first order terms of hydrodynamics in the previous subtraction, obtaining an even better description (within $0.0001\%$). Presumably, this difference would be well described by second order hydrodynamics, but we lack the corresponding coefficients to do this check.

We conclude
that hydrodynamics provides a very good description of the system, in consonance with the fluid/gravity mapping. In particular, we observe that first order hydrodynamics further improves the ideal description, as expected from the hydrodynamic gradient expansion. 
%
We emphasise that this test constitutes a truly non-linear precision test of both the code and the fluid/gravity correspondence in a real-time dynamical configuration.

\section{Radial equations}
\label{sec:full-eqs}

For completeness,  here we list the radial equations obtained from
ansatz~(\ref{eq:metric}).
It is useful to introduce the following operators to make all expressions more compact
\begin{equation}
\begin{aligned}
f^{\prime} &=\partial_rf\,, \\
\dot{f} &=\left(\partial_t+\frac{A}{2}\partial_r\right)f\,, \\
\tilde{f}&=\left(\partial_x-F_x\partial_r\right)f\,, \\
\hat{f}&=\left(\partial_y-F_y\partial_r\right)f\,, \\
\bar{f}&=\left(\partial^2_x-2F_x\partial_r\partial_x+F_x^2\partial^2_r\right)f\,, \\
f^{\star}&=\left(\partial^2_y-2F_y\partial_r\partial_y+F_y^2\partial^2_r\right)f\,, \\
f^{\times}&=\left(\partial_x\partial_y-F_x\partial_r\partial_y-F_y\partial_r\partial_x+F_xF_y\partial^2_r\right)f \,.
\end{aligned}
\label{ec:new_der}
\end{equation}

As shown in Table~\ref{tab:system}, by combining Einstein's equations~(\ref{eq:eom}) in a particular way we obtain a
nested system of radial ODEs where one can sequentially solve for the different
variables. This is common to these characteristic approaches, see e.g.~\cite{Winicour:2012znc,Chesler:2013lia}.
In this case, some of these equations will be coupled. 
%
The full set of equations, in the order to be solved, is:
\begin{equation}
6S^{\prime\prime}+S\left(\cosh^2(G)\left(B_1^{\prime}\right)^2+3\left(B_2^{\prime}\right)^2+\left(G^{\prime}\right)^2+\left(\phi^{\prime}\right)^2\right)=0
\label{ec:nested_first}
\end{equation}

\begin{dgroup}
\begin{dmath}
2 e^{B_1} S^2 F_x''+e^{B_1} \left(S^2 \left(-2 \left(\cosh ^2(G) \left(\tilde{B_1}'-B_1' F_x'\right)+B_2' \left(3 \tilde{B_2}-F_x'\right)+\tilde{G} \left(B_1' \sinh (2G)+G'\right)+\tilde{B_2}'
+4 \tilde{\phi } \phi'\right)-2 \tilde{B_1} B_1' \cosh ^2(G)\right)+S \left(-6 \tilde{S} \left(B_1' \cosh^2(G)+B_2'\right)-8 \tilde{S}'+2 S' F_x'\right)+8 \tilde{S} S'\right)
+S^2 \left(-2 G'\left(\hat{B_1}+F_y'\right)+\sinh (2 G) \left(\hat{B_1}'-B_1'\left(\hat{B_1}+F_y'\right)\right)+2 \hat{G} B_1' \cosh (2 G)+2 \hat{G}'\right)
+3 \hat{S} S \left(B_1' \sinh (2 G)+2 G'\right)=0
\end{dmath}
\begin{dmath}
2 S^2 F_y''+e^{B_1} \left(S^2 \left(2 \left(G' \left(\tilde{B_1}-F_x'\right)+\tilde{G}'\right)-\sinh (2 G) \left(B_1'\left(\tilde{B_1}-F_x'\right)+\tilde{B_1}'\right)-2 \tilde{G}B_1' \cosh (2 G)\right)
-3 S \tilde{S} \left(B_1' \sinh (2 G)-2 G'\right)\right)+2 S^2\left(\cosh ^2(G) \left(\hat{B_1}'-B_1' F_y'\right)+B_2' \left(F_y'-3 \hat{B_2}\right)
+\hat{G} \left(B_1' \sinh (2 G)-G'\right)-\hat{B_1} B_1' \cosh^2(G)-\hat{B_2}'-4 \hat{\phi } \phi '\right)+S \left(6 \hat{S} \left(B_1' \cosh ^2(G)-B_2'\right)
+2 S' F_y'-8 \hat{S}'\right)+8 \hat{S}S'=0
\end{dmath}
\end{dgroup}

\begin{dmath}
12 e^{B_1} S^3 \dot{S}'+e^{B_1+B_2} \left(S^2 \left(2 \cosh (G) \left(-\hat{G} \left(\tilde{B_1}+\tilde{B_2}-F_x'\right)+\tilde{G} \left(\hat{B_1}-\hat{B_2}+F_y'\right)+G'
   \left(\tilde{F_y}+\hat{F_x}\right)
   -2G^{\times}\right)+2 \sinh (G) \left(B_2' \left(\tilde{F_y}+\hat{F_x}\right)+F_y'
   \left(\tilde{B_2}-F_x'\right)+\hat{B_2} \left(F_x'-4 \tilde{B_2}\right)-2{B_2}^{\times}+\tilde{F_y}'
   -2 \hat{G} \tilde{G}-4 \hat{\phi }
   \tilde{\phi }+\hat{F_x}'\right)\right)+S \left(2 \sinh (G) \left(\hat{S} \left(F_x'-4 \tilde{B_2}\right)+\tilde{S} \left(F_y'-4 \hat{B_2}\right)
   +4 S'
   \left(\tilde{F_y}+\hat{F_x}\right)-8S^{\times}\right)-8 \cosh (G) \left(\hat{S} \tilde{G}+\hat{G} \tilde{S}\right)\right)+8 \hat{S} \tilde{S}
   \sinh (G)\right)+e^{2 B_1+B_2} \left(S^2 
   \left(2 \sinh (G) \left(\tilde{G} \left(2 \tilde{B_1}+\tilde{B_2}-F_x'\right)-G'
   \tilde{F_x}+\bar{G}\right)+\cosh (G) \left(2 \left(-\left(B_1'+B_2'\right)
   \tilde{F_x}
   +\bar{B_1}+\bar{B_2}-\tilde{F_x}'+\tilde{G}^2+2 \tilde{\phi }^2\right)-2 \left(\tilde{B_1}+\tilde{B_2}\right) F_x'+2
   \left(\tilde{B_1}{}^2+\tilde{B_2} \tilde{B_1}+2 \tilde{B_2}{}^2\right)+\left(F_x'\right){}^2\right)\right)
   +S \left(2 \cosh (G) \left(\tilde{S}
   \left(4 \left(\tilde{B_1}+\tilde{B_2}\right)-F_x'\right)+4 \left(\bar{S}-S' \tilde{F_x}\right)\right)+8 \tilde{G} \tilde{S} \sinh
   (G)\right)
   -4 \tilde{S}^2 \cosh (G)\right)+e^{B_2} \left(S^2 \left(2 \sinh (G) \left(\hat{G} \left(-2 \hat{B_1}+\hat{B_2}-F_y'\right)-\hat{F_y}
   G'+G^{\star}\right)
   +\cosh (G) \left(2 \left(\left(B_1'-B_2'\right) \hat{F_y}-B_1^{\star}+B_2^{\star}-\hat{F_y}'+\hat{G}^2+2 \hat{\phi
   }^2\right)+2 \left(\hat{B_1}-\hat{B_2}\right) F_y'
   +2 \left(\hat{B_1}{}^2-\hat{B_2} \hat{B_1}+2
   \hat{B_2}{}^2\right)+\left(F_y'\right){}^2\right)\right)+S \left(8 \hat{G} \hat{S} \sinh (G)-2 \cosh (G)
    \left(\hat{S} \left(4 \hat{B_1}-4
   \hat{B_2}+F_y'\right)+4 \hat{F_y} S'-4 S^{\star}\right)\right)-4 \hat{S}^2 \cosh (G)\right)+e^{B_1} \left(8 S^4 V(\phi
   )+24 \dot{S} S^2 S'\right) = 0
\end{dmath}

\begin{dgroup}
\begin{dmath}
12 e^{B_1} S^4 \dot{B'_1}+e^{B_1+B_2} \left(6 S^2 \text{sech}(G) \left(\hat{G} \left(F_x'-\tilde{B_2}\right)+\tilde{G} \left(\hat{B_2}-F_y'\right)+G'
   \left(\tilde{F_y}-\hat{F_x}\right)\right)+6 S \text{sech}(G)
    \left(\hat{S} \tilde{G}-\hat{G} \tilde{S}\right)\right)+e^{2 B_1+B_2} \left(-3 S^2
   \text{sech}(G) \left(-2 B_2' \tilde{F_x}-2 \tilde{B_2} F_x'+4 \tilde{B_2}{}^2+2 \bar{B_2}-2 \tilde{F_x}'+4 \tilde{\phi
   }^2
   +\left(F_x'\right){}^2\right)-6 S \text{sech}(G) \left(\tilde{S} \left(\tilde{B_2}+2 F_x'\right)-S' \tilde{F_x}+\bar{S}\right)+12
   \tilde{S}^2 \text{sech}(G)\right)+e^{B_2} 
   \left(3 S^2 \text{sech}(G) \left(-2 B_2' \hat{F_y}-2 \hat{B_2} F_y'+4 \hat{B_2}{}^2+2
   B_2^{\star}+\left(F_y'\right){}^2-2 \hat{F_y}'+4 \hat{\phi }^2\right)+6 S \text{sech}(G)
    \left(\hat{S} \left(\hat{B_2}+2 F_y'\right)-\hat{F_y}
   S'+S^{\star}\right)-12 \hat{S}^2 \text{sech}(G)\right)+e^{B_1} \left(12 S^4 \tanh (G) \left(\dot{B_1} G'+\dot{G} B_1'\right)
   +18 S^3
     \left(\dot{B_1} S'+\dot{S} B_1'\right)\right)=0
 \end{dmath}
 \begin{dmath}
   12 e^{B_1} S^4 \dot{G}'+e^{B_1+B_2} \left(6 S^2 \cosh (G) \left(B_1' \left(\hat{F_x}-\tilde{F_y}\right)-B_2' \left(\tilde{F_y}+\hat{F_x}\right)+\left(\hat{B_1}-F_y'\right)
   \left(\tilde{B_2}-F_x'\right)-\hat{B_2}
   \left(\tilde{B_1}-4 \tilde{B_2}+F_x'\right)+\tilde{B_1} F_y'+2B_2^{\times}-\tilde{F_y}'+4 \hat{\phi }
   \tilde{\phi }-\hat{F_x}'\right)+6 S \cosh (G) \left(\hat{S} \left(-\tilde{B_1}+\tilde{B_2}+2 F_x'\right)
   +\tilde{S} \left(\hat{B_1}+\hat{B_2}+2
   F_y'\right)-S' \left(\tilde{F_y}+\hat{F_x}\right)+2 S^{\times}\right)-24 \hat{S} \tilde{S} \cosh (G)\right)+e^{2 B_1+B_2} \left(-3 S^2 \sinh (G)
   \left(-2 B_2' \tilde{F_x}-2 \tilde{B_2} F_x'+4 \tilde{B_2}{}^2+2 \bar{B_2}-2 \tilde{F_x}'+4 \tilde{\phi }^2+\left(F_x'\right){}^2\right)-6
   S \sinh (G) \left(\tilde{S} \left(\tilde{B_2}+2 F_x'\right)
   -S' \tilde{F_x}+\bar{S}\right)+12 \tilde{S}^2 \sinh (G)\right)+e^{B_2} \left(-3
   S^2 \sinh (G) \left(-2 B_2' \hat{F_y}-2 \hat{B_2} F_y'+4 \hat{B_2}{}^2+2 B_2^{\star}+\left(F_y'\right){}^2
   -2 \hat{F_y}'+4 \hat{\phi }^2\right)-6
   S \sinh (G) \left(\hat{S} \left(\hat{B_2}+2 F_y'\right)-\hat{F_y} S'+S^{\star}\right)+12 \hat{S}^2 \sinh (G)\right)+e^{B_1} \left(18 S^3
   \left(\dot{S} G'+\dot{G} S'\right)-6 \dot{B_1} S^4 B_1' \sinh (2 G)\right)=0
 \end{dmath}
\end{dgroup}

\begin{dmath}
12 e^{B_1} S^4 \dot{B_2}'+e^{B_1+B_2} \left(S^2 \left(2 \cosh (G) \left(\hat{G} \left(\tilde{B_1}-2 \tilde{B_2}-F_x'\right)-\tilde{G} \left(\hat{B_1}+2 \hat{B_2}+F_y'\right)-G'
   \left(\tilde{F_y}+\hat{F_x}\right)
   +2 G^{\times}\right)+2 \sinh (G) \left(2 B_2' \left(\tilde{F_y}+\hat{F_x}\right)+F_y' \left(2
   \tilde{B_2}+F_x'\right)+2 \hat{B_2} \left(F_x'-\tilde{B_2}\right)-4 B_2^{\times}-\tilde{F_y}'+2 \hat{G} \tilde{G}
   +4 \hat{\phi } \tilde{\phi
   }-\hat{F_x}'\right)\right)+S \left(2 \sinh (G) \left(2 \left(\hat{S} \left(F_x'-\tilde{B_2}\right)+\tilde{S}
   \left(F_y'-\hat{B_2}\right)+S^{\times}\right)-S' \left(\tilde{F_y}+\hat{F_x}\right)\right)
    +2 \cosh (G)\left(\hat{S} \tilde{G}+\hat{G}
   \tilde{S}\right)\right)-8 \hat{S} \tilde{S} \sinh (G)\right)+e^{2 B_1+B_2} \left(S^2 \left(2 \sinh (G) \left(\tilde{G} \left(-2 \tilde{B_1}+2
   \tilde{B_2}+F_x'\right)
   +G' \tilde{F_x}-\bar{G}\right)-\cosh (G) \left(2 \left(-\left(B_1'-2 B_2'\right) \tilde{F_x}+\bar{B_1}-2
   \bar{B_2}-\tilde{F_x}'+\tilde{G}^2+2 \tilde{\phi }^2\right)-2 \left(\tilde{B_1}-2 \tilde{B_2}\right) F_x'
   +2 \left(\tilde{B_1}{}^2-2
   \tilde{B_2} \tilde{B_1}-\tilde{B_2}{}^2\right)+\left(F_x'\right){}^2\right)\right)+S \left(-2 \cosh (G) \left(\tilde{S} \left(\tilde{B_1}-2
   \tilde{B_2}+2 F_x'\right)-S' \tilde{F_x}+\bar{S}\right)
   -2 \tilde{G} \tilde{S} \sinh (G)\right)+4 \tilde{S}^2 \cosh (G)\right)+e^{B_2}
   \left(S^2 \left(2 \sinh (G) \left(\hat{G} \left(2 \left(\hat{B_1}+\hat{B_2}\right)+F_y'\right)+\hat{F_y} G'-G^{\star}\right)
   -\cosh (G) \left(2
   \left(\left(B_1'+2 B_2'\right) \hat{F_y}-B_1^{\star}-2 B_2^{\star}-\hat{F_y}'+\hat{G}^2+2 \hat{\phi }^2\right)+2 \left(\hat{B_1}+2
   \hat{B_2}\right) F_y'
   +2 \left(\hat{B_1}{}^2+2 \hat{B_2} \hat{B_1}-\hat{B_2}{}^2\right)+\left(F_y'\right){}^2\right)\right)+S \left(2 \cosh (G)
   \left(\hat{S} \left(\hat{B_1}+2 \hat{B_2}-2 F_y'\right)+\hat{F_y} S'-S^{\star}\right)
   -2 \hat{G} \hat{S} \sinh (G)\right)+4 \hat{S}^2 \cosh
   (G)\right)+18 e^{B_1} S^3 \left(\dot{B_2} S'+\dot{S} B_2'\right)=0
\end{dmath}

\begin{dmath}
8 e^{B_1} S^3 \dot{\phi}'+e^{B_1+B_2} \left(S \left(4 \sinh (G) \left(\hat{\phi } \left(F_x'-\tilde{B_2}\right)+\tilde{\phi } \left(F_y'-\hat{B_2}\right)+\phi '
   \left(\tilde{F_y}+\hat{F_x}\right)-2 \phi ^{\times}\right)-4 \cosh (G)
    \left(\hat{\phi } \tilde{G}+\hat{G} \tilde{\phi }\right)\right)-4 \sinh
   (G) \left(\hat{\phi } \tilde{S}+\hat{S} \tilde{\phi }\right)\right)+e^{2 B_1+B_2} \left(4 S \left(\cosh (G) \left(\tilde{\phi }
   \left(\tilde{B_1}+\tilde{B_2}-F_x'\right)
   -\phi ' \tilde{F_x}+\bar{\phi }\right)+\tilde{G} \tilde{\phi } \sinh (G)\right)+4 \tilde{S}
   \tilde{\phi } \cosh (G)\right)+e^{B_2} \left(S \left(4 \hat{G} \hat{\phi } \sinh (G)-4 \cosh (G)
    \left(\hat{\phi }
   \left(\hat{B_1}-\hat{B_2}+F_y'\right)+\hat{F_y} \phi '-\phi^{\star}\right)\right)+4 \hat{S} \hat{\phi } \cosh (G)\right)+e^{B_1} \left(12 S^2 \left(\dot{\phi } S'+\dot{S} \phi '\right)
   -4 S^3 V'(\phi )\right)=0
\end{dmath}

\begin{dmath}
6 e^{B_1} S^4 A''+e^{B_1+B_2} \left(S^2 \left(6 \cosh (G) \left(\left(\hat{B_2}-\hat{B_1}\right) \tilde{G}+\hat{G}
   \left(\tilde{B_1}+\tilde{B_2}\right)-G' \left(\tilde{F_y}+\hat{F_x}\right)+2
   G^{\times}\right)
   +6 \sinh (G) \left(-B_2' \left(\tilde{F_y}+\hat{F_x}\right)+2
   B_2^{\times}+4 \hat{B_2} \tilde{B_2}+2 \hat{G} \tilde{G}+4 \hat{\phi } \tilde{\phi }-F_x'
   F_y'\right)\right)+24 S \left(\sinh (G)
    \left(\hat{B_2} \tilde{S}+\hat{S} \tilde{B_2}-S'
   \left(\tilde{F_y}+\hat{F_x}\right)+2 S^{\times}\right)+\cosh (G) \left(\hat{S}
   \tilde{G}+\hat{G} \tilde{S}\right)\right)-24 \hat{S} \tilde{S} \sinh (G)\right)
   +e^{2 B_1+B_2}
   \left(S^2 \left(3 \cosh (G) \left(\left(F_x'\right){}^2-2 \left(-\left(B_1'+B_2'\right)
   \tilde{F_x}+\tilde{B_1}{}^2+2 \tilde{B_2}{}^2+\bar{B_1}+\bar{B_2}+\tilde{B_1}
   \tilde{B_2}
   +\tilde{G}^2+2 \tilde{\phi }^2\right)\right)-6 \sinh (G) \left(\left(2
   \tilde{B_1}+\tilde{B_2}\right)\tilde{G}-G' \tilde{F_x}+\bar{G}\right)\right)+S \left(-24
   \cosh (G)
    \left(\left(\tilde{B_1}+\tilde{B_2}\right) \tilde{S}-S'
   \tilde{F_x}+\bar{S}\right)-24 \tilde{G} \tilde{S} \sinh (G)\right)+12 \tilde{S}^2 \cosh
   (G)\right)
   +e^{B_2} \left(S^2 \left(6 \sinh (G) \left(\left(2
   \hat{B_1}-\hat{B_2}\right) \hat{G}+\hat{F_y} G'-G^{\star}\right)+3 \cosh (G)
   \left(\left(F_y'\right){}^2-2 \left(\left(B_1'-B_2'\right) \hat{F_y}+\hat{B_1}{}^2+2
   \hat{B_2}{}^2-B_1^{\star}+B_2^{\star}-\hat{B_1} \hat{B_2}+\hat{G}^2+2 \hat{\phi
   }^2\right)\right)\right)
   +S \left(24 \cosh (G) \left(\left(\hat{B_1}-\hat{B_2}\right)
   \hat{S}+\hat{F_y} S'-S^{\star}\right)-24 \hat{G} \hat{S} \sinh (G)\right)+12 \hat{S}^2 \cosh
   (G)\right)
   +e^{B_1} \left(S^4 \left(6 \left(\dot{B_1} B_1' \cosh ^2(G)+3 \dot{B_2} B_2'+\dot{G} G'+4
   \dot{\phi } \phi '+4\right)-2 (4 V(\phi )+12)\right)-72 S^2 \dot{S} S'\right)
=0
 \label{ec:nested_last}
\end{dmath}

These next equations (to solve for $\ddot{S}$ and $\dot{F}_{x,y}$) are not needed for our evolution scheme, but they are used
in the equation for the gauge condition $\partial_t \xi$:
\begin{dmath}
6 e^{B_1} \ddot{S} S^3+e^{B_1+B_2} \left(S^2 \left(\sinh (G) \left(-A' \left(\tilde{F_y}+\hat{F_x}\right)+\hat{B_2}
   \left(\tilde{A}+2 \dot{F_x}\right)+\tilde{B_2} \left(\hat{A}+2 \dot{F_y}\right)+2 A^{\times}+2
   \tilde{\dot{F_y}}
   +2 \hat{\dot{F_x}}\right)+\cosh (G) \left(\hat{G} \left(\tilde{A}+2
   \dot{F_x}\right)+\tilde{G} \left(\hat{A}+2 \dot{F_y}\right)\right)\right)+S \sinh (G) \left(\hat{S}
   \left(\tilde{A}+2 \dot{F_x}\right)
   +\tilde{S} \left(\hat{A}+2 \dot{F_y}\right)\right)\right)+e^{2
   B_1+B_2} \left(S^2 \left(\tilde{G} \sinh (G) \left(-\left(\tilde{A}+2 \dot{F_x}\right)\right)-\cosh
   (G) \left(-A' \tilde{F_x}
   +\left(\tilde{B_1}+\tilde{B_2}\right) \left(\tilde{A}+2
   \dot{F_x}\right)+\bar{A}+2 \tilde{\dot{F_x}}\right)\right)-S \tilde{S} \cosh (G)
   \left(\tilde{A}+2 \dot{F_x}\right)\right)-e^{B_1+2 B_2}
   \left(\hat{F_x}-\tilde{F_y}\right)^2
   +e^{B_2} \left(S^2 \left(\cosh (G) \left(A'
   \hat{F_y}+\left(\hat{B_1}-\hat{B_2}\right) \left(\hat{A}+2 \dot{F_y}\right)-A^{\star}-2
   \hat{\dot{F_y}}\right)-\hat{G} \sinh (G) \left(\hat{A}+2 \dot{F_y}\right)\right)
   -S \hat{S} \cosh
   (G) \left(\hat{A}+2 \dot{F_y}\right)\right)+e^{B_1} \left(S^4 \left(\dot{B_1}{}^2 \cosh ^2(G)+3
   \dot{B_2}{}^2+\dot{G}^2+4 \dot{\phi }^2\right)-3 S^3 \dot{S} A'\right)=0
\end{dmath}

\begin{dgroup}
\begin{dmath}
4 e^{B_1} S^3 \dot{F_x}'+e^{B_1} \left(2 S^3 \left(\left(\tilde{A}+2 \dot{F_x}\right) \left(B_1' \cosh ^2(G)+B_2'\right)+2
   \tilde{A}'+2 \tilde{\dot{B_1}} \cosh ^2(G)
   +2 \dot{B_1} \left(\tilde{B_1} \cosh ^2(G)+\tilde{G}
   \sinh (2 G)\right)+2 \tilde{\dot{B_2}}+6 \dot{B_2} \tilde{B_2}+2 \dot{G} \tilde{G}+8 \dot{\phi }
   \tilde{\phi }-A' F_x'\right)+4 S^2
    \left(-S' \left(\tilde{A}+2 \dot{F_x}\right)+3 \tilde{S}
   \left(\dot{B_1} \cosh ^2(G)+\dot{B_2}\right)+4 \tilde{\dot{S}}+3 \dot{S} F_x'\right)-16 \dot{S} S
   \tilde{S}\right)+e^{B_1+B_2}
    \left(4 S \sinh (G) \left(\hat{F_x} \left(F_x'-2 \tilde{B_2}\right)+2
   \tilde{B_2} \tilde{F_y}-\tilde{F_x} F_y'-F_x^{\times}+\bar{F_y}\right)+4 \tilde{S}
   \sinh (G) \left(\hat{F_x}-\tilde{F_y}\right)\right)
   +e^{B_2} \left(4 S \cosh (G) \left(2 \hat{B_2}
   \left(\hat{F_x}-\tilde{F_y}\right)+\tilde{F_y} F_y'-F_y^{\times}-\hat{F_y}
   F_x'+F_x^{\star}\right)+4 \hat{S} \cosh (G) \left(\tilde{F_y}-\hat{F_x}\right)\right)
   +S^3
   \left(-\left(\hat{A}+2 \dot{F_y}\right) \left(B_1' \sinh (2 G)+2 G'\right)+4 \hat{B_1} \dot{G}-2
   \hat{\dot{B_1}} \sinh (2 G)+2 \dot{B_1} \left(\hat{B_1} \sinh (2 G)
   -2 \hat{G} \cosh (2 G)\right)-4
   \hat{\dot{G}}\right)-6 \hat{S} S^2 \left(\dot{B_1} \sinh (2 G)+2
   \dot{G}\right)=0
\end{dmath}
\begin{dmath}
4 S^3
   \dot{F_y}'+e^{B_1} \left(S^3 \left(\left(\tilde{A}+2 \dot{F_x}\right) \left(B_1' \sinh (2 G)-2 G'\right)-4
   \dot{G} \tilde{B_1}+2 \tilde{\dot{B_1}} \sinh (2 G)+2 \dot{B_1} \left(\tilde{B_1} \sinh (2 G)
   +2
   \tilde{G} \cosh (2 G)\right)-4 \tilde{\dot{G}}\right)+6 S^2 \tilde{S} \left(\dot{B_1} \sinh (2 G)-2
   \dot{G}\right)\right)+e^{B_2} \left(4 S \sinh (G) \left(2 \hat{B_2}
   \left(\hat{F_x}-\tilde{F_y}\right)
   +\tilde{F_y} F_y'-F_y^{\times}-\hat{F_y}
   F_x'+F_x^{\star}\right)+4 \hat{S} \sinh (G)
   \left(\tilde{F_y}-\hat{F_x}\right)\right)+e^{B_1+B_2} \left(4 S \cosh (G) \left(\hat{F_x}
   \left(F_x'-2 \tilde{B_2}\right)
   +2 \tilde{B_2} \tilde{F_y}-\tilde{F_x}
   F_y'-F_x^{\times}+\bar{F_y}\right)+4 \tilde{S} \cosh (G)
   \left(\hat{F_x}-\tilde{F_y}\right)\right)+2 S^3 \left(-A' F_y'-\left(\hat{A}+2 \dot{F_y}\right)  
    \left(B_1' \cosh ^2(G)-B_2'\right)+2 \hat{A}'-2 \hat{\dot{B_1}} \cosh ^2(G)+2 \dot{B_1}
   \left(\hat{B_1} \cosh ^2(G)-\hat{G} \sinh (2 G)\right)+2 \hat{\dot{B_2}}+6 \dot{B_2} \hat{B_2}
   +2
   \dot{G} \hat{G}+8 \dot{\phi } \hat{\phi }\right)+4 S^2 \left(-S' \left(\hat{A}+2 \dot{F_y}\right)+3
   \hat{S} \left(\dot{B_2}-\dot{B_1} \cosh ^2(G)\right)+3 \dot{S} F_y'+4 \hat{\dot{S}}\right)-16 \dot{S} \hat{S} S
   =0
 \end{dmath}
\end{dgroup}

\section{Apparent horizon finder}
\label{sec:AH}

In order to find the AH we need to compute the expansion of the outgoing null rays. We can construct the tangent vector to such outgoing rays using the ingoing null rays, $n$, together with the form perpendicular to the AH, $s$,
\begin{equation}
\begin{aligned}
s&=N_s\left(-\partial_t\sigma dt-\partial_y\sigma dy-\partial_y\sigma dy+dr\right),\\
n&=-N_n\partial_r,
\end{aligned}
\end{equation}
from where we can compute the vector $s$ by simply raising the indices. The normalisation factors, $N_s$ and $N_n$, can be computed by imposing $s^2=1$ and $s\cdot n=-1/\sqrt{2}$. Combining these two vectors we can construct another vector tangent to outgoing trajectories,
\begin{equation}
l^{\mu}=\sqrt{2}s^{\mu}+n^{\mu},
\end{equation}
so that it is null, $l^2=0$, and properly normalised, $l\cdot n=-1$. The expansion of these rays can be computed as
\begin{equation}
\theta_l=h^{\mu\nu}\nabla_{\mu}l_{\nu},
\end{equation}
where
\begin{equation}
h_{\mu\nu}=g_{\mu\nu}+l_{\mu}n_{\nu}+l_{\nu}n_{\mu}
\end{equation}
is the induced metric over hypersurfaces normal to both in- and out-going null rays. The AH location is given by the condition $\theta_l=0$. Imposing it at a generic surface, $r=\sigma(x,y)$, we obtain the following equation,
\begin{dmath}
2 e^{B_2} \left(F_y+\partial_y\sigma\right) \left(S \left(e^{B_1} \cosh (G) \left(\tilde{G}+G' \left(F_x+\partial_x\sigma\right)\right)+e^{B_1} \sinh (G) \left(\tilde{B_2}+B_2' \left(F_x+\partial_x\sigma\right)\right)
   +\cosh (G) \left(B_1' \left(F_y+\partial_y\sigma\right)+\hat{B_1}\right)-\cosh (G) \left(B_2' \left(F_y+\partial_y\sigma\right)+\hat{B_2}\right)-\sinh (G) \left(G' \left(F_y+\partial_y\sigma
\right)+\hat{G}\right)\right)
   +e^{B_1} \sinh (G) \left(\tilde{S}-2 S' \left(F_x+\partial_x\sigma\right)\right)-\cosh (G) \left(S'
   \left(F_y+\partial_y\sigma\right)+\hat{S}\right)\right)-2 e^{B_1+B_2} \left(F_x+\partial_x\sigma\right)
    \left(S \left(e^{B_1} \left(\cosh (G)
   \left(\tilde{B_1}+B_1' \left(F_x+\partial_x\sigma\right)\right)+\cosh (G) \left(\tilde{B_2}+B_2' \left(F_x+\partial_x\sigma\right)\right)
   +\sinh (G)
   \left(\tilde{G}+G' \left(F_x+\partial_x\sigma\right)\right)\right)-\sinh (G) \left(B_2' \left(F_y+\partial_y\sigma\right)+\hat{B_2}\right)-\cosh
   (G) \left(G' \left(F_y+\partial_y\sigma\right)+\hat{G}\right)\right)
   +e^{B_1} \cosh (G) \left(\tilde{S}+S' \left(F_x+\partial_x\sigma
\right)\right)-\sinh (G) \left(S' \left(F_y+\partial_y\sigma\right)+\hat{S}\right)\right)+S \left(e^{B_1} \left(2 e^{B_2} \sinh (G)
   \left(\tilde{F_y}+F_y' \left(F_x+\partial_x\sigma\right)+\partial_{xy}\sigma\right)-2 e^{B_1+B_2} \cosh (G) \left(\tilde{F_x}+F_x' \left(F_x+\partial_x\sigma
\right)+\partial_{xx}\sigma\right)+6 S \dot{S}\right)
+2 e^{B_1+B_2} \sinh (G) \left(F_x' \left(F_y+\partial_y\sigma\right)+\hat{F_x}+\partial_{xy}\sigma\right)-2 e^{B_2} \cosh (G) \left(F_y' \left(F_y+\partial_y\sigma\right)+\hat{F_y}+\partial_{yy}\sigma
\right)\right)
+3 e^{2 B_1+B_2} \cosh (G) S' \left(F_x+\partial_x\sigma\right){}^2+3 e^{B_2} \cosh (G) S' \left(F_y+\partial_y\sigma\right)^2=0,
\label{eq:expansion-gen}
\end{dmath}
where every function is evaluated at the $r=\sigma(x,y)$ surface defining the AH.
When the AH is located at constant radial surfaces, i.e.\
$\sigma(t,x,y)=r=\mathrm{constant}$ -- which is what we impose to find the
evolution equation for the gauge function $\xi$ -- equation~\eqref{eq:expansion-gen} reduces to
\begin{dmath}
  \Theta \equiv -2 e^{B_1+B_2} F_x \left(S \left(e^{B_1} \left(\cosh (G)
\left(\tilde{B_1}+B_1' F_x\right)+\cosh (G) \left(\tilde{B_2}+B_2'
F_x\right)+\sinh (G) \left(\tilde{G}+F_x G'\right)\right)
-\sinh (G) \left(B_2' F_y+\hat{B_2}\right)-\cosh (G) \left(F_y
G'+\hat{G}\right)\right)+e^{B_1} \cosh (G) \left(\tilde{S}+F_x S'\right)-\sinh
(G) \left(F_y S'+\hat{S}\right)\right) +2 e^{B_2} F_y \left(S \left(e^{B_1}
\cosh (G) \left(\tilde{G}+F_x G'\right)+e^{B_1} \sinh (G) \left(\tilde{B_2}+B_2'
F_x\right)+\cosh (G) \left(B_1' F_y+\hat{B_1}\right)
-\cosh (G) \left(B_2' F_y+\hat{B_2}\right)-\sinh (G) \left(F_y
G'+\hat{G}\right)\right)+e^{B_1} \sinh (G) \left(\tilde{S}-2 F_x S'\right)-\cosh
(G) \left(F_y S'+\hat{S}\right)\right) +S \left(e^{B_1} \left(2 e^{B_2} \sinh
(G) \left(\tilde{F_y}+F_x F_y'\right)-2 e^{B_1+B_2} \cosh (G)
\left(\tilde{F_x}+F_x F_x'\right)+6 S \dot{S}\right) +2
e^{B_1+B_2} \sinh (G) \left(F_y F_x'+\hat{F_x}\right)-2 e^{B_2} \cosh (G)
\left(F_y F_y'+\hat{F_y}\right)\right)+3 e^{2 B_1+B_2} F_x^2 \cosh (G) S' +3
e^{B_2} F_y^2 \cosh (G) S' = 0
\label{ec:expansion}
\end{dmath}
%
One can check that when going to the 2+1 case, by imposing conditions (\ref{ec:to_2+1}), equation~(3.17) of~\cite{Attems:2017zam} is recovered.

To start with initial data that satisfies $\Theta|_{r=\mathrm{const}} = 0$ we first need to find the AH and adjust $\xi$ accordingly.
Solving the differential equation~\eqref{eq:expansion-gen} gives us the location of the AH at a given time slice $t$. Contrary to what we have found so far this equation is non-linear, with the form
\begin{equation}
\begin{aligned}
\mathcal{L}\left(\sigma,\partial\sigma,\partial^2\sigma\right)&=\alpha_{xx}(t,\sigma,x,y)\partial_{xx}\sigma+\alpha_{xy}(t,\sigma,x,y)\partial_{xy}\sigma+\alpha_{yy}(t,\sigma,x,y)\partial_{yy}\sigma\\
&\quad+\beta_{xx}(t,\sigma,x,y)\left(\partial_x\sigma\right)^2+\beta_{xy}(t,\sigma,x,y)\partial_x\sigma\partial_y\sigma+\beta_{yy}(t,\sigma,x,y)\left(\partial_y\sigma\right)^2\\
&\quad+\gamma_x(t,\sigma,x,y)\partial_x\sigma+\gamma_y(t,\sigma,x,y)\partial_y\sigma+\delta(t,\sigma,x,y)=0,
\end{aligned}
\label{eq:AH-nonlin}
\end{equation}
where
%
\begingroup
\allowdisplaybreaks
\begin{align*}
\alpha_{xx}&=-e^{B_1+B_2} S \cosh (G),\\
\alpha_{xy}&=2 e^{B_2} S \sinh (G),\\
\alpha_{yy}&=-e^{B_2-B_1} S \cosh (G),\\
\beta_{xx}&=\frac{1}{2} e^{B_1+B_2} \left(\cosh (G) S'-2 S \left(B_1' \cosh (G)+B_2' \cosh (G)+G' \sinh (G)\right)\right),\\
\beta_{xy}&=e^{B_2} \left(2 S \left(B_2' \sinh (G)+G' \cosh (G)\right)-\sinh (G) S'\right),\\
\beta_{yy}&=\frac{1}{2} e^{B_2-B_1} \left(2 S \left(B_1' \cosh (G)-B_2' \cosh (G)+G' (-\sinh (G))\right)+\cosh (G) S'\right),\\
\gamma_x&=e^{B_2} \left(S \left(-e^{B_1} \tilde{G} \sinh (G)-e^{B_1} \tilde{B_1} \cosh (G)-e^{B_1} \tilde{B_2} \cosh (G)-2 e^{B_1} F_x G' \sinh (G)\right.\right.\\
&\left.\left.-2 e^{B_1} B_1'
   F_x \cosh (G)-2 e^{B_1} B_2' F_x \cosh (G)-e^{B_1} \cosh (G) F_x'+2 B_2' F_y \sinh (G)+\hat{B_2} \sinh (G)\right.\right.\\
   &\left.\left.+2 F_y G' \cosh (G)+\sinh (G) F_y'+\hat{G}
   \cosh (G)\right)-e^{B_1} \tilde{S} \cosh (G)+e^{B_1} F_x \cosh (G) S'\right.\\
   &\left.-F_y \sinh (G) S'+\hat{S} \sinh (G)\right),\\
\gamma_y&=e^{B_2-B_1} \left(S \left(e^{B_1} \tilde{B_2} \sinh (G)+e^{B_1} \tilde{G} \cosh (G)+2 e^{B_1} F_x G' \cosh (G)+2 e^{B_1} B_2' F_x \sinh (G)\right.\right.\\
&\left.\left.+e^{B_1} \sinh
   (G) F_x'+2 B_1' F_y \cosh (G)-2 B_2' F_y \cosh (G)+\hat{B_1} \cosh (G)-\hat{B_2} \cosh (G)\right.\right.\\
   &\left.\left.-2 F_y G' \sinh (G)-\cosh (G) F_y'-\hat{G} \sinh
   (G)\right)+e^{B_1} \tilde{S} \sinh (G)-e^{B_1} F_x \sinh (G) S'\right.\\
   &\left.+F_y \cosh (G) S'+\hat{S} (-\cosh (G))\right),\\
\delta&=-e^{B_2-B_1} S \left(-F_y \left(e^{B_1} \tilde{B_2} \sinh (G)+e^{B_1} \tilde{G} \cosh (G)+2 e^{B_1} F_x G' \cosh (G)+2 e^{B_1} B_2' F_x \sinh (G)\right.\right.\\
&\left.\left.+e^{B_1}
   \sinh (G) F_x'+\hat{B_1} \cosh (G)-\hat{B_2} \cosh (G)-\cosh (G) F_y'-\hat{G} \sinh (G)\right)+e^{B_1} F_x \left(e^{B_1} \tilde{G} \sinh (G)\right.\right.\\
   &\left.\left.+e^{B_1}
   \tilde{B_1} \cosh (G)+e^{B_1} \tilde{B_2} \cosh (G)+e^{B_1} \cosh (G) F_x'-\hat{B_2} \sinh (G)-\sinh (G) F_y'-\hat{G} \cosh (G)\right)\right.\\
   &\left.+e^{2 B_1}
   \cosh (G) \tilde{F_x}-e^{B_1} \sinh (G) \tilde{F_y}+e^{2 B_1} F_x^2 \left(B_1' \cosh (G)+B_2' \cosh (G)+G' \sinh (G)\right)\right.\\
   &\left.+F_y^2 \left(-B_1' \cosh
   (G)+B_2' \cosh (G)+G' \sinh (G)\right)-e^{B_1} \hat{F_x} \sinh (G)+\hat{F_y} \cosh (G)\right)\\
   &+\frac{1}{2} e^{B_2-B_1} \left(-2 F_y \left(\hat{S} \cosh
   (G)-e^{B_1} \sinh (G) \left(\tilde{S}-F_x S'\right)\right)+e^{B_1} F_x \left(2 \hat{S} \sinh (G)\right.\right.\\
   &\left.\left.-e^{B_1} \cosh (G) \left(2 \tilde{S}-F_x
   S'\right)\right)+F_y^2 \cosh (G) S'\right)+3 \dot{S} S^2,
\end{align*}
\endgroup


We solve equation~\eqref{eq:AH-nonlin} with the Newton-Kantorovich method by
linearising the equation around a guess solution $\sigma_0(x,y)$.
Expanding the operator $\mathcal{L}$ we obtain
\begin{equation}
\begin{aligned}
\mathcal{L}\left(\sigma,\partial\sigma,\partial^2\sigma\right)&=\left(\mathcal{L}+\frac{\partial\mathcal{L}}{\partial\sigma}+\frac{\partial\mathcal{L}}{\partial(\partial_x\sigma)}\partial_x+\frac{\partial\mathcal{L}}{\partial(\partial_y\sigma)}\partial_y+\frac{\partial\mathcal{L}}{\partial(\partial_{xx}\sigma)}\partial_{xx}+\frac{\partial\mathcal{L}}{\partial(\partial_{xy}\sigma)}\partial_{xy}\right.\\
&\qquad\left.+\frac{\partial\mathcal{L}}{\partial(\partial_{yy}\sigma)}\partial_{yy}\right)_{\sigma=\sigma_0}\delta\sigma+\mathcal{O}\left(\delta\sigma^2\right)=0,\\
\end{aligned}
\end{equation}
where $\delta\sigma=\sigma(x,y)-\sigma_0(x,y)$.
The associated linear problem for the correction $\delta \sigma$ is then
\begin{equation}
\begin{aligned}
&\left[\alpha_{xx}(\sigma_0)\partial_{xx}+\alpha_{xy}(\sigma_0)\partial_{xy}+\alpha_{yy}(\sigma_0)\partial_{yy}+\left(\gamma_x(\sigma_0)+2\beta_{xx}(\sigma_0)\partial_x\sigma_0+\beta_{xy}(\sigma_0)\partial_y\sigma_0\right)\partial_x\right.\\
&\quad\left.+\left(\gamma_y(\sigma_0)+2\beta_{yy}(\sigma_0)\partial_y\sigma_0+\beta_{xy}(\sigma_0)\partial_x\sigma_0\right)\partial_y+\partial_\sigma\mathcal{L}(\sigma_0)\right]\delta\sigma=-\mathcal{L}(\sigma_0,\partial\sigma_0,\partial^2\sigma_0),
\end{aligned}
\end{equation}
which has the same functional form as that of equation~(\ref{ec:xi_evol}) and
which we solve in the same fashion. For the purpose of implementing this in the
code, the only remaining thing to be done is rewriting the coefficients of this
linearized equation in terms of the outer grid redefinitions, $g2$.

\end{appendix}

\bibliography{Holographic_Bubbles_v3}{}
\bibliographystyle{JHEP}
\end{document}